\documentstyle[times,emulateapj]{article}

\def\vvmax{$\langle V/V_{max} \rangle$}
\def\avgz{$\langle z \rangle$}
\def\fxfr{log($f_x/f_r$)}

\def\avgPext{$\langle P_{ext} \rangle$}
\def\wlam{$W_{\lambda}$}
  
\def\lya{Ly$\alpha$}

\lefthead{Rector \& Stocke}
\righthead{The Properties of 1Jy RBLs}

\begin{document}

\title{The Properties of the Radio-Selected 1Jy Sample of BL Lacertae Objects}
\author{Travis A. Rector\altaffilmark{1,2,3,4} and John T.
Stocke\altaffilmark{1,2,3}}
\affil{Center for Astrophysics and Space Astronomy, University of Colorado,
Boulder, Colorado 80309-0389}
\authoremail{rector@casa.Colorado.EDU}

\altaffiltext{1}{Visiting Astronomer, Kitt Peak National Observatory, National
Optical Astronomy Observatories, which is operated by the Association of
Universities for Research in Astronomy, Inc. (AURA) under cooperative agreement
with the National Science Foundation.}
\altaffiltext{2}{Observations reported here were obtained at the MMT Observatory, a
joint facility of the Smithsonian Institution and the University of Arizona.}
\altaffiltext{3}{Visiting Astronomer, National Radio Astronomy Observatory.  NRAO
is a facility of the National Science Foundation operated under cooperative
agreement by Associated Universities, Inc.}
\altaffiltext{4}{Current address:  National Optical Astronomy Observatories, 950 N.
Cherry Avenue, Tucson, AZ  85719}

\begin{abstract}

We present new optical and near-IR spectroscopy as well as new high dynamic range,
arcsecond-resolution VLA radio maps of BL Lacs from the complete radio-selected ``1
Jansky'' (1Jy) sample (RBLs) for which such data were not previously available.  Redshift
information is now available for all but six of the 37 BL Lacs in the 1Jy sample.  Of
the 31 with redshift information, four redshifts are only minimum values based upon
absorption lines and four other objects have uncertain redshifts based upon the
detection of only a single emission line.  Unlike  BL Lacs from the complete
X-ray-selected {\it Einstein} Medium Sensitivity Survey (EMSS) sample (XBLs), most RBLs
possess weak but moderately luminous emission lines.  The emission-line luminosities of
RBLs are several orders of magnitude lower than flat-spectrum radio quasars (FSRQs);
however, there is significant overlap in the luminosity distributions of the two
classes.  All but one object in the 1Jy sample has now been observed with the
VLA; and extended flux was detected for all but three of the
observed objects.  Whereas nearly all XBLs have extended power levels consistent with
FR--1s, more than half of the RBLs have extended radio power levels too luminous to be
beamed FR--1 radio galaxies.   In fact, we find evidence for and examples of three
distinct mechanisms for creating the BL Lac phenomenon in the 1Jy sample: beamed FR--1s,
beamed FR--2s and possibly a few gravitationally-lensed quasars.  The \vvmax\ determined
for the 1Jy sample is $0.614\pm0.047$, which is markedly different from the
negative evolution seen in the EMSS and other XBL samples.  A correlation between
logarithmic X-ray to radio flux ratio and \vvmax\ is observed across the EMSS and 1Jy
samples, from negative evolution in the more extreme XBLs to positive evolution in
the more extreme RBLs.  There is evidence that the
selection criteria chosen by Stickel et al. eliminates some BL Lac objects
from the 1Jy sample, although how many is unknown.  And several objects currently in the
sample have exhibited strong emission lines in one or more epochs, suggesting they
should be reclassified as FSRQs.  However these selection effects cannot account for the
observed discrepancy in XBL and RBL properties.  From these observational
properties we conclude that RBLs and XBLs cannot be related by viewing angle alone,
and that RBLs are more closely related to FSRQs.

\end{abstract}


\section{Introduction}

BL Lacertae objects are an extreme type of Active Galactic Nuclei (AGN), whose
hallmark is their ``featureless" optical spectra.  They are a member of the blazar
class;
and
like other blazars, their observed properties are due largely to bulk relativistic
outflows.   The characteristic broadband spectral energy
distribution (SED) of BL Lac objects and other blazars is double-peaked, with
synchrotron radiation dominating at lower energies and a second peak at hard
X-ray and $\gamma$-ray energies.  The second peak is likely due to inverse-Compton
scattering, however other mechanisms are possible (e.g., Mannheim 1998).   The overall
properties of BL Lacs are still not well known, due in part to the difficulty in
obtaining redshifts for many of these objects, but also because detailed studies of
complete BL Lac samples have been sparse.   
For many years there were only two statistically-complete
samples of BL Lacs: the X-ray-selected {\it Einstein} Medium Sensitivity Survey (EMSS)
sample (Morris et al. 1991, hereafter M91; Rector et al. 2000, hereafter R00) and the
radio-selected 1Jy sample (Stickel et al. 1991, hereafter S91), each sample
containing $\sim$40 objects.  Since the writing of S91 three objects have been
added to the 1Jy sample.  New radio and X-ray surveys, e.g., the VLA Faint
Images of the Radio Sky at Twenty centimeters (FIRST; Laurent-Muehleisen et al. 1998)
and the NRAO VLA Sky Survey (NVSS; Condon et al. 1998), the ROSAT All-Sky Survey (RASS;
Nass et al. 1996) and the Deep X-Ray Blazar Survey (DXRBS; Perlman et al. 1998) are now
discovering BL Lacs in prodigious numbers.  These new surveys are invigorating BL Lac
research, however they have not yet been studied in as great detail as the 1Jy and EMSS
samples.  

The dominant paradigm for BL Lac objects is that they are ``highly-beamed" FR--1
(Fanaroff \& Riley 1974) radio galaxies which are seen nearly along the line of sight
of the relativistic jet (Blandford \& Rees 1978; Urry \& Padovani 1995).  This
hypothesis, known as the ``unification model," accounts for many of the observed
properties of BL Lacs, e.g., rapidly-varying polarized continuum emission seen at all
wavelengths (Angel \& Stockman 1980); highly core-dominated kpc-scale radio structure
(e.g., Antonucci \& Ulvestad 1985; Perlman \& Stocke 1994); and compact parsec-scale
radio core structure, often exhibiting superluminal motion (Zensus 1989).

Alternatively, Ostriker \& Vietri (Ostriker 1989; Ostriker \& Vietri 1985, 1990)
suggested that BL Lacs may be gravitationally-lensed background quasars.  
Stars associated with a
foreground galaxy could selectively amplify the continuum emission from the compact
nucleus relative to the extended emission line regions, thus creating the low-\wlam\
emission-line spectra which is a BL Lac hallmark.
This hypothesis has been ruled out as an explanation for all BL Lacs, but it may be
an explanation for a few.   So far
only one proven example of a lensed BL Lac is known, the ``smallest Einstein ring"
object S4 0218+357 (O'Dea et al. 1992).  However, several examples have been noted
of BL Lacs which are potentially gravitationally lensed quasars (e.g., Scarpa et al.
1999, Stocke, Wurtz \& Perlman 1995, Stickel et al. 1988a,b).
A preliminary accounting of the present work given in Stocke \& Rector (1997;
hereafter SR97) finds evidence for a substantial excess of intervening MgII
absorption systems in the 1Jy BL Lac sample when compared to quasars.  This result
requires a correlation between the nearly featureless spectra of BL Lacs (i.e.,
low-\wlam\ emission lines) and the presence of foreground absorbing gas.  The
most obvious interpretation of such a correlation is gravitational lensing, thereby
raising the question of the importance of this mechanism for producing BL Lacs in the
1Jy sample.

Historically BL Lacs have been divided into radio-selected (RBLs) and X-ray-selected
(XBLs) based upon the method of discovery.  Interestingly, optical surveys have not
been efficient in discovering BL Lacs; e.g., only 6 have been found in the optical PG
survey (Fleming et al. 1993), and other attempts have not been successful (Jannuzi
1990).  In recent years this terminology has given way to classification based upon
the overall SED.  In ``low-energy-peaked" blazars (LBLs), the peak of the synchrotron
radiation occurs at IR wavelengths, whereas in ``high-energy-peaked" blazars (HBLs)
this peak occurs in the UV/X-ray, with an approximate dividing line defined by
\fxfr $ = -5.5$, the logarithmic flux density ratio from 1 keV to 5 GHz (e.g., Wurtz et
al. 1996).  For the most part XBLs are HBLs and RBLs are LBLs, although exceptions do
exist (e.g., Mkn 501 is an HBL in the 1Jy sample).  
New surveys (e.g., the ROSAT-Green Bank sample; Laurent-Muehleisen et al. 1999)
are finding BL Lacs intermediate between HBLs and LBLs, indicating a continuum of SEDs in
BL Lacs and likely rendering the LBL/HBL terminology obsolete.
We choose to retain the term RBL to describe the 1Jy
sample because it is a radio-selected sample; neither X-ray emission nor the overall
SED is considered in the selection process.

Previous studies of XBLs have shown that their properties are less extreme than RBLs;
e.g., XBLs show less extreme optical variability (Jannuzi, Smith \& Elston 1993,
1994), lower maximum percentages of optical polarization and less polarization
variability with preferred position angles (Jannuzi et al. 1994), less core dominated
radio structure (Perlman \& Stocke 1993) and significant fractions of starlight in
their optical spectra (M91).  The unified model therefore was generalized to suggest
that XBLs are viewed further from the jet axis than RBLs (e.g., Jannuzi et al. 1994;
Perlman \& Stocke 1993).  Thus, in the unification model, RBLs, XBLs and FR--1s
represent the low-luminosity population of radio-loud AGN, with each class seen
increasingly further off the jet axis.  Similarly, radio-loud quasars are believed to
be beamed, high-luminosity FR--2 radio galaxies (e.g., Barthel 1989), wherein
steep-spectrum radio quasars (SSRQs) are seen further off-axis than flat-spectrum
radio quasars (FSRQs).
Alternatively, it has been suggested that the difference lies not in orientation
but in the high-energy cutoff in their energy distributions, where XBLs and RBLs
represent a single family of objects with a smooth energy distribution followed by
a sharp cutoff.  For RBLs this cutoff occurs in the near-IR/optical and for XBLs
this cutoff is at UV/X-ray energies (Giommi \& Padovani 1994; Padovani
\& Giommi 1995; Sambruna, Maraschi \& Urry 1996).  

There are important observations that bear upon these hypotheses.  First, the \vvmax\
values for the 1Jy and EMSS BL Lac samples are different.  XBLs show
``negative" evolution whereas RBLs evolve ``positively" (i.e., greater numbers
and/or luminosities in the past), suggesting that RBLs and XBLs are two distinct classes
of AGN.  The
\vvmax\ values for XBLs are similar to FR--1s; and the \vvmax\ values for RBLs are more
like FR--2 galaxies and quasars.  However for both samples the \vvmax\ value is
consistent with a no-evolution result to 2$\sigma$; and prior measurements of the
\vvmax\ for the 1Jy sample (S91) were based upon incomplete redshift information,
raising some doubt as to the validity of the discrepancy.  Second, the extended radio
structure of some 1Jy RBLs were known to be too luminous to be that of an FR--1 radio
galaxy, suggesting that some RBLs are instead beamed FR--2s (Kollgaard et al. 1992).  A
detailed understanding of the 1Jy RBL sample has been hampered by two major impediments:
(1) a lack of high-quality optical spectra for many 1Jy BL Lacs prevented the detection
of weak spectral lines and the determination of redshifts, hence their absolute
properties; and (2) dynamic range limited VLA observations were unable to detect faint,
extended radio flux near the bright core.  Without complete sample data both of the
aforementioned problems with BL Lac unification remained unverified.

Here we present high-SNR optical and moderate-SNR near-IR spectroscopy
as well as high dynamic-range 20cm, A-array VLA maps of BL Lacs from the 1Jy sample of
S91 for which emission-line redshifts and faint extended radio structure had not been
detected in previous observations.  In \S 2 we present our observations and discuss
the data reduction.  We also discuss the overall status of
observations of the 1Jy sample.  In \S 3 we discuss individual sources and how
our observations affect other results for these sources.  In \S 4 we discuss
reasons why the 1Jy sample may be incomplete.  In \S 5 we discuss the overall
properties of the 1Jy BL Lac sample and compare them to the complete EMSS XBL samples
(R00), discussing the impact on unified schemes.  In
\S 6 we summarize the conclusions of our work. Throughout this paper we assume a
standard cosmology with $H_0 = 50$ km s$^{-1}$ Mpc$^{-1}$ and $q_0 = 0.0$.

\section{Observations and Reduction}

\subsection{Optical and Near-IR Spectroscopy}

The optical spectroscopy presented here was completed during four observing runs:
two at the KPNO 2.1m during 4--7 April 1995 and 18--21 November 1995 and one each at
the Multiple Mirror Telescope Observatory (MMTO) 4.5m during 15--16 August 1996 and
the upgraded MMTO 6.5m during 22--23 November 2000.  The log of observations are shown
in Table~\ref{tbl-1}.  The columns are: [1] the 1Jy object name; [2] dates of
observation; and [3] the SNR (continuum flux density/ $1\sigma$ RMS) for each spectrum
at $\sim$5000\AA.  At the 2.1m we used the ``Goldcam" spectrograph with a 300 line
mm$^{-1}$ grating to give a dispersion of 2.4\AA\ pixel$^{-1}$ and a
resolution (FWHM) of 8.4\AA\ with a 2\arcsec\ slit over the spectral range of
4000-8000\AA.  At the MMTO 4.5m we used the blue-channel spectrograph, also with a
300 line mm$^{-1}$ grating, to give a 2.4 \AA\ pixel$^{-1}$ dispersion and a
resolution of 7.4 \AA\ with a 1.25\arcsec\ slit for a spectral coverage over
3200-7500\AA.  At the MMTO 6.5m we used the blue-channel spectrograph with the 500
line mm$^{-1}$ and 800 line mm$^{-1}$ gratings to achieve higher resolution in the
near-UV.  The 500 line mm$^{-1}$ grating gave a 1.2\AA\ pixel$^{-1}$ dispersion and
3.6\AA\ resolution with a 1.5\arcsec\ slit for spectral coverage over 3500-6500\AA.
The 800 line mm$^{-1}$ grating gave a 0.75\AA\ pixel$^{-1}$ dispersion and 2.2\AA\
resolution with a 1.5\arcsec\ slit for spectral coverage over 3200-5200\AA.
Unfortunately, due to contamination of the MMTO 6.5m primary
mirror coating, sensitivity in the near-UV was poor, with a $>$60\% loss of
light blueward of $\sim$4500\AA (Foltz, private communication).

In an effort to improve the flat-fielding and sky subtraction as well as to remove the
effects of detector defects, we observed each target at two locations along the
slit.  Although we achieved excellent removal of telluric absorption features, weak
Na ``D" absorption is present in most spectra as well as the oxygen absorption bands
at varying strengths depending on the object observed.  Also, weak ``emission" bumps
just blueward of telluric absorption features are common.  Features caused by
imperfect telluric correction are marked.

The spectra were extracted with IRAF\footnote{IRAF is distributed by NOAO, which is
operated by AURA, Inc., under contract to the National Science Foundation.} using the
optimal extraction routines.  The error in the wavelength calibration is $\leq 0.2$\AA\
from 4000\AA\ to 9000\AA.  The solution is slightly poorer at shorter wavelengths.  Flux
calibration was performed using observations of flux standards at similar airmass. 
No correction was made for the loss of light at the slit; thus the derived flux
density values are probably uncertain by a factor of
$\sim$2.  Most objects were observed on multiple runs.  These spectra were not
combined but rather were used to confirm any features detected.  

\begin{figure}
\plottwo{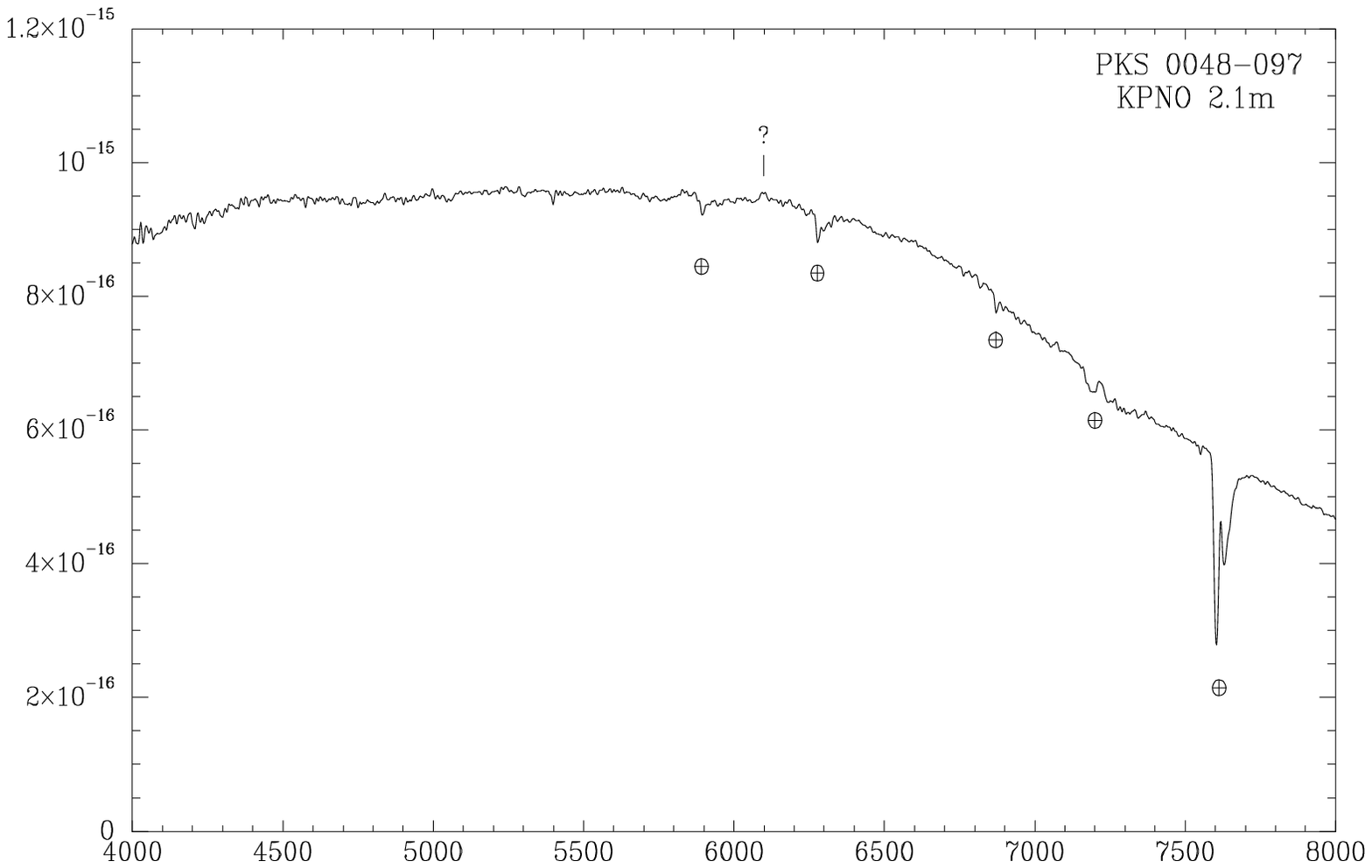}{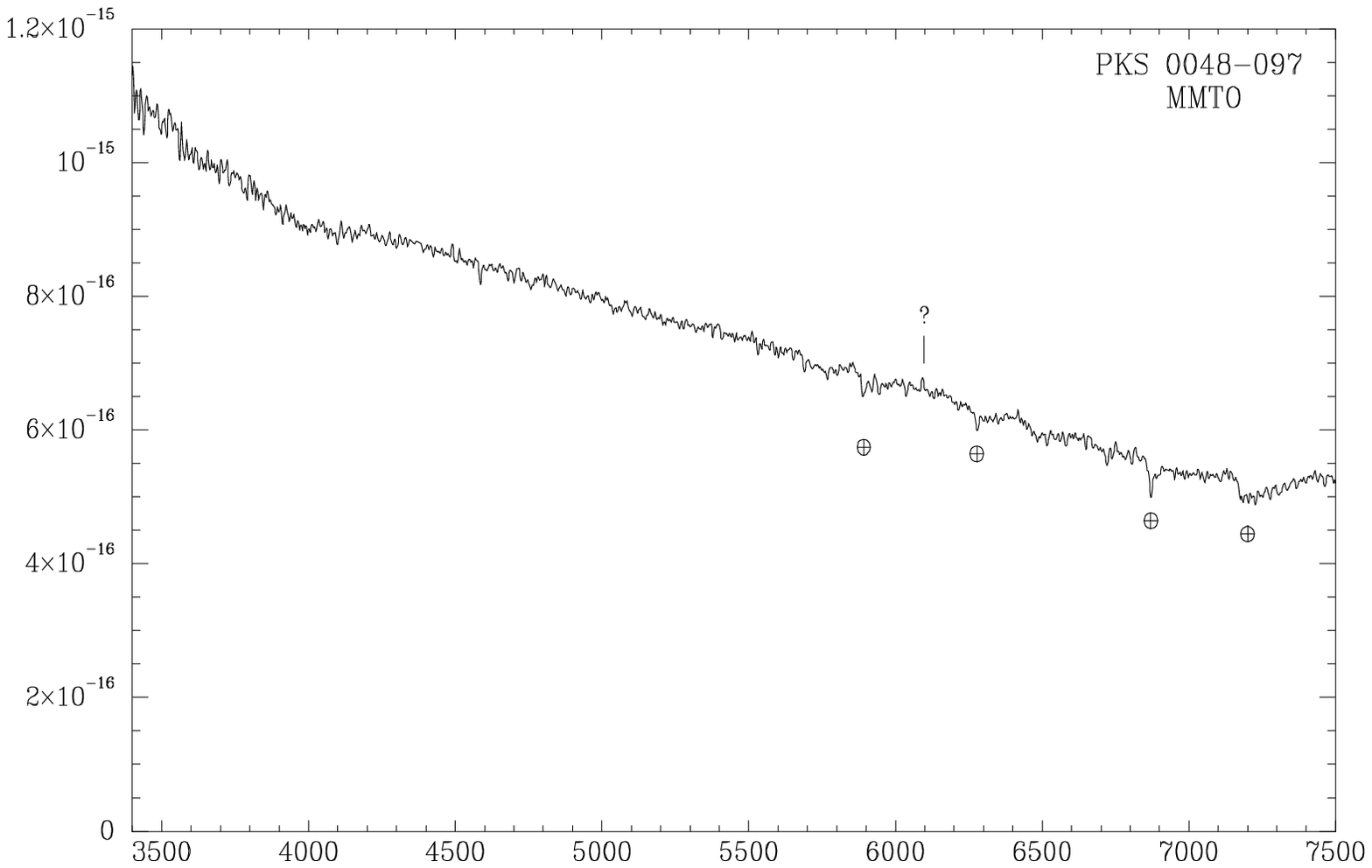}
\caption{The flux scale is $F_{\lambda}$ in ergs s$^{-1}$ cm$^{-2}$; the x-axis is wavelength
in \AA.  The symbol $\earth$ identifies features due to the Earth's atmosphere. 
Possible features are labeled with a question mark.  A scaled-up version of the
6900\AA\ to 9000\AA\ region of S5 1803+784 is shown in the upper right corner to
show the weak detection of H$\beta$ at $z=0.684$.  Due to the small slit width
($2\arcsec$) these flux values are only approximate, although the relative flux
values blueward of 6500\AA\ are reasonably correct.  The spectral energy
distributions of the 2.1m spectra should not be trusted redward of 6500\AA\ due to
second-order overlap.
\label{fig-1}}
\end{figure}

The flux calibration of the KPNO 2.1m spectra suffer from the presence of
second-order blue light redward of $\sim$7000\AA, causing an incorrect continuum
slope redward of $\sim$6500\AA.  Our attempts to correct for the order overlap
introduced small-scale artifacts that could have been misinterpreted as real spectral
features; thus we present the uncorrected spectra in Figure~\ref{fig-1}.  Many of
these spectra also suffer from poor flux calibration blueward of $\sim 5000$\AA. 
Thus the detailed spectral slopes of the 2.1m spectra should not be trusted.  The
MMTO 6.5m spectra also suffered from minor flux calibration problems.  Spectra taken
with the 500  line mm$^{-1}$ grating exhibits an artificial drop in flux redward of
6000\AA.  Spectra taken with the 800  line mm$^{-1}$ grating show a weak
$\sim$400\AA\ oscillation in the continuum shape.  To correct for this oscillation,
each spectrum was divided by a smoothed and normalized spectrum of 0716+714, a
high-SNR, featureless spectrum.  The spectrum of 0716+714 itself is shown in its
uncorrected state.

Near-IR spectroscopy in the $I$ ($0.90-1.30 \mu$m), $J$ ($1.14-1.34 \mu$m), $H$
($1.47-1.80 \mu$m) and $K$ ($1.94-2.46 \mu$m) bands was also attempted on several
objects which remained featureless after extensive optical spectroscopy.  Two
observing runs were completed on the KPNO 2.1m during 24-26 November 1996 and 15-16
February 1997.  We used the IR Cryogenic Spectrometer (CRSP) with a low-resolution 75
line mm$^{-1}$ grating to give a dispersion of 18-35\AA\ per CCD pixel and a
resolution (FWHM) of 50-100\AA\ (depending on the band) with a 1.7\arcsec\ slit. 
Each object was observed in as many bands as possible, to moderate SNRs resulting in
minimum equivalent width limits of
$W_{\lambda} =$ 10--50\AA\ for each band.  Atmospheric corrections were made in a
manner similar to those described in Maiolino et al. (1996).  Unfortunately due to
numerous hardware and weather problems we were unable to obtain many high-SNR
spectra; and no emission lines were detected above the minimum \wlam\ values quoted
in Table~\ref{tbl-5}.
Table~\ref{tbl-5} includes by column: [1]
the object name; [2] dates of observation; [3] the observed IR band(s); [4] the
typical SNR (continuum flux / $1\sigma$ RMS) for each spectra; and [5] the
minimum \wlam\ limit for any emission line (assuming a $3\sigma$
detection limit).  Most of the near-IR spectra are of poor quality for the
purposes of this work and put only very loose constraints on any emission lines
present; in all cases the weak ($W_{\lambda} \leq 5$\AA) emission lines often
seen in BL Lacs would be undetectable.

With our new spectra emission-line redshifts are now determined for 27 of the 37 BL
Lacs; four of which are in need of confirmation due to the detection of only one
line.  Redshift lower limits for four more sources have been determined by 
absorption systems (e.g. SR97).  The remaining five sources show no certain emission
nor absorption features, however a lower limit of $z \geq 0.2$ can be inferred because
all five are optically unresolved in ground-based images (Stickel, Fried \& K\"uhr
1993; hereafter SFK93).  Redshifts for the entire 1Jy sample are given in
Table~\ref{tbl-2}.  The columns are: [1] the 1Jy object name; [2] and [3] the
emission and absorption redshifts.  An absorption redshift is listed only if it is
different from the emission redshift.  Redshifts followed by a colon are not
confirmed; and [4] the reference from which the redshift was obtained; RS01 refers
to this paper.  Table~\ref{tbl-4} lists the observed spectral line properties for the
objects observed.  Average values are reported when more than one spectrum is
available.  The columns are: [1] the 1Jy object name; [2] the average redshift of the
lines observed.  An average is given for the emission lines; absorption lines are
listed separately if they are at different redshifts; [3], [4], [5] and [6] are the
line identifications, their rest and observed wavelengths (in \AA) and the redshift
of the line; [7] and [8] are the observed full-width half-maximum and observed
equivalent
\wlam\ of each line (in \AA); for emission lines [9] is the observed flux
density of the line in 10$^{-16}$ erg s$^{-1}$ cm$^{-2}$ and [10] is the line
luminosity in 10$^{41}$ erg s$^{-1}$ in the object rest frame.

With our new spectra as well as with a review of spectra available in the
literature, the classification of several BL Lacs must be called into question due
to the presence of emission lines with substantial \wlam.  Stocke et. al
(1991) proposed a
$W_{\lambda} \leq 5$\AA\ classification limit for BL Lacs; S91 later modified this
definition to accommodate higher redshift candidates by requiring that the rest
$W_{\lambda} \leq 5$\AA.  Neither of these criteria have any physical basis but are
supported by the observed paucity of quasars with rest $W_{\lambda} \leq 50$\AA\
(Stocke et al. 1991). Further, the spectral range over which an object must be
featureless has not been specified; e.g., our spectrum of PKS 2029+121 taken with
the MMT blue-channel spectrograph discovered strong CIV emission at
$\sim 3500$\AA, a line which was not detectable in our 2.1m spectra because the
Goldcam spectrograph is not sensitive blueward of $\sim 4000$\AA.  We mark with an
asterisk in Table~\ref{tbl-2} all objects whose spectra have been observed to have
had at any time emission lines with rest $W_{\lambda} \geq 5$\AA.  For most of
these objects the high--\wlam\ emission lines are transient (e.g., BL Lac
itself; Vermeulen et al. 1995, OJ 287; Sitko \& Junkkarinen 1985 and B2 1308+326;
S91).  In addition, strong ($W_{\lambda} > 5$\AA) emission lines have also been
detected on at least one occasion in PKS 0537-441 (Wilkes 1986), B2 1308+326 (SFK93),
PKS 1749+096 (Scarpa
\& Falomo 1997) and PKS 2029+121 (this paper), so the BL Lac classification for
these objects is questioned.

\subsection{Radio Continuum Imaging}

Ten 1Jy BL Lacs were observed with the NRAO\footnote{The National Radio
Astronomy Observatory is operated by Associated Universities, Inc., under
cooperative agreement with the National Science Foundation.} VLA on 5 August
1995.  We chose to observe with the A-array at 20cm with a 50MHz bandwidth to
maximize sensitivity to extended, steep-spectrum structure while achieving
$\sim$1.5\arcsec\ resolution.  Approximately eight
6-minute scans, each bracketed by a 90--second scan on a primary VLA flux
calibrator, were made to obtain $\sim$100,000 ``visibilities" for each source. 
Scans were widely spaced to optimize coverage in the $(u,v)$ plane; allowing dynamic
ranges (peak flux density to $1\sigma$ RMS) of $\sim$10,000 : 1 to be achieved for
most objects.

Epoch 1995.2 VLA values were used to flux calibrate the maps using multiple
observations of 3C 286.  Since these sources are highly core dominated, a point
source model was assumed to start the self-calibration process, followed by clean
component models.  Phase-only self-calibration in decreasing solution intervals
was used for the first four iterations.  Amplitude and phase self-calibration were
then used until the maximum dynamic range was achieved, usually requiring only one or
two more iterations.  The AIPS task IMAGR was used to generate the maps and clean
components.  Robust weighting (ROBUST = 0.5; see Briggs 1995 for an explanation) was
used to achieve a smaller beam FWHM with only a 10--12\% increase in noise over
natural weighting.  The maps are shown in Figure~\ref{fig-2}.

The core flux densities were measured by fitting the core with a single Gaussian with
the synthesized beam's parameters.  The extended flux was determined by measuring
the total flux density with a box enclosing the entire source and then subtracting the
core flux density.  The luminosity is calculated with the standard cosmology, with
$K$- and bandpass corrections applied.  A power-law continuum of the form
$F_{\nu} \propto \nu^{-\alpha}$ is assumed where $\alpha = 0.8$ is assumed for
extended flux densities and $\alpha = 0.3$ is assumed for the core (values typical of
BL Lacs; PS93).

We do not account for cosmological surface-brightness dimming and beam dilution
effects.  These effects decrease the amount of extended flux observed,
especially so for edge-dimmed FR--1 sources at redshift $z > 0.2$. FR--2 sources
are not greatly affected because their extended flux is often dominated by high
surface-brightness hotspots.  PS93 describe a method for correcting for these
biases, but the corrections are strongly dependent on the source structure and
redshift.  It is not clear how to account for the unusual, highly distorted
morphologies seen in the 1Jy sample because many are not clearly FR--1 or FR--2 in
morphology.  Thus we do not attempt to correct for these effects.

A concern with observing with the A-array is that extended structure will be
``overresolved" and that the extended flux will be underestimated.  We do not
believe this to be a significant concern because arcminute-scale radio emission in
BL Lacs is rare (Ulvestad \& Johnston 1984; Kollgaard et al. 1992).  3C 371 is a
notable exception (Wrobel \& Lind 1990) but it is at a much lower redshift than
most of the sources presented herein.

With these observations all but two 1Jy BL Lacs 
have 
been imaged with the VLA A-array at 20cm to a dynamic range $\geq 1000:1$. 
Curiously, we were unable to locate a deep 20cm A-array map of BL Lac itself;
thus we used the 20cm B-array map from Antonucci (1986).  PKS 2005--489 is too far
south to be observed with the VLA.  We are now able
to look at the radio properties of the 1Jy sample at a single frequency and array
configuration, as summarized in Table~\ref{tbl-3}.  Columns include: [1] object name;
[2] and [3] the uncorrected core and extended flux densities at 20cm (mJy); [4] and
[5] the
$K$- and bandpass-corrected core and extended logarithmic radio power in W Hz$^{-1}$;
[6] the ratio of core to extended flux; [7] the physical size of the source (see
discussion below); [8], [9] and [10] the Fanaroff-Riley (1974) classification based
upon the extended radio structure (R), optical emission line properties (O)
and a final (F) classification based upon both (see \S 4.1 for discussion); and
[11] is the reference for the radio map.  Absolute quantities assume either the
emission line redshift from Table~\ref{tbl-2} or a lower limit on the redshift from
either the presence of an absorption system in the spectrum or an optically
unresolved image.  Tentative values are marked with a colon.  Values in parenthesis
are highly uncertain because they are based upon an upper limit on extended flux
density  and a lower limit or redshift.

We had difficulty choosing an appropriate means for measuring the largest-angular
size for each source.  For example, Murphy et al. (1993; hereafter M93) chose to
measure from the core to a single peak, or ``hotspot," in the extended emission
because many of the objects in their sample show only one-sided emission.  This
definition is not precise because many 1Jy BL Lacs have diffuse extended emission
without clear hotspots.  For consistency with PS93 and Laurent-Muehleisen et al.
(1993) we choose to define the largest angular size as the sum of two straight
lines from the core to the outermost 3$\sigma$ contours, with each line
intersecting the brightest hotspot, if any, in the extended structure.  By doing so
we are assuming that BL Lacs are intrinsically ``triple" sources consisting of a
core and two outer lobes, which may not be the case.  For sources without a clear
``triple" morphology a best guess is made; these values are thus questionable and
are marked with a colon in Table~\ref{tbl-3}.  S5 0454+844 and PKS
1749+096 remain unresolved at $\sim$1 arcsecond resolution and at the 0.1
mJy beam$^{-1}$ level; and PKS 1519-273 is unresolved at the 0.5 mJy beam$^{-1}$
level (Cassaro et al. 1999, hereafter C99).  Extended flux is tentatively measured in
PKS 1144-379 (C99) and PKS 0823+033 (a weak eastern component shown in the map of M93
is most likely not real).  
Unresolved
sources are given an upper flux density limit based upon the limit on extended flux
shown in Table~\ref{tbl-3}.  The estimates for S5 0454+844 and PKS 1519-273 are very
uncertain because they are unresolved and have only a lower limit on redshift.

\subsection{Observational Status of the 1Jy Sample}

All of our optical spectra are of sufficient quality to determine whether or not these
objects should be classified as BL Lacs based upon the rest $W_{\lambda} \leq 5$\AA\
criterion.  However several faint objects (S4 0828+493, PKS 1519-273 and PKS 2150+173)
are in need of higher-SNR spectra to determine if weak ($W_{\lambda} \approx
0.5$\AA) emission lines, often found in BL Lac spectra, are present.  The spectra
of four of these objects (PKS 0118--272, S4 0218+357, S5 0454+844 and PKS 0735+178)
detect only absorption systems, thus giving a lower redshift
limit, and are also in need of additional observations to search for emission
features from the AGN or absorption features from the host galaxy.  Four bright BL
Lacs (PKS 0048-097, S5 0716+714, S4 0814+425 and B2 1147+245) have no confirmed
spectral features despite numerous high-SNR observations by us and by others.

All but three 
of the 1Jy BL Lacs
have been mapped with the VLA to a dynamic range of $\geq$1000:1.  PKS 2005--489 is
too far south to be observed with the VLA; it is in need of synthesis-array
observations.  The 20cm B-array map from Antonucci (1986) marginally resolves BL Lac
(S4 2200+420); therefore it should be deeply mapped with the
A-array.  PKS 2254+074 is also in need of high-dynamic range mapping to confirm weak
extended structure detected by Antonucci \& Ulvestad (1985).  

\section{Notes on Individual Sources}

PKS 0048--097: A very weak emission line may be detected at 6092\AA\ in both our
MMTO 4.5m and 2.1m spectra.  It is plausibly [OIII] $\lambda$5007 at $z=0.216$
because Ly$\alpha$ may be present in the IUE spectra at this redshift (Lanzetta,
Turnshek
\& Sandoval 1993).  
Alternatively, there is evidence that the redshift for this
source is higher.  Falomo (1996) reports that it is optically unresolved and,
assuming a host elliptical of $M_R = -23.5$, sets a lower limit to the redshift of
$z > 0.5$.  
The presence of a hotspot near the leading edge makes this source a
plausible ``dogleg" FR--2 similar to 3C 270.1 and 3C 275.1 (Stocke, Burns \&
Christiansen 1985).  The radio structure sets no firm limits on the distance to
this object, however its FR--2-like radio structure does suggest that [OII]
$\lambda$3727 at $z=0.634$ is a more likely identification for the weak emission
line at 6092\AA.  At this redshift PKS 0048-097 would have an extended radio
luminosity more commensurate with an FR--2 (log
$P_{ext} = 26.23$).  This redshift is also in agreement with Falomo (1996).   However,
because both lines are so weak, we choose not to adopt either redshift. 
Because it is unresolved by ground-based, optical observations (SFK93), 
we assume
$z>0.2$ for luminosity and linear size calculations rather than either tentative
redshift.

\begin{figure}
\plottwo{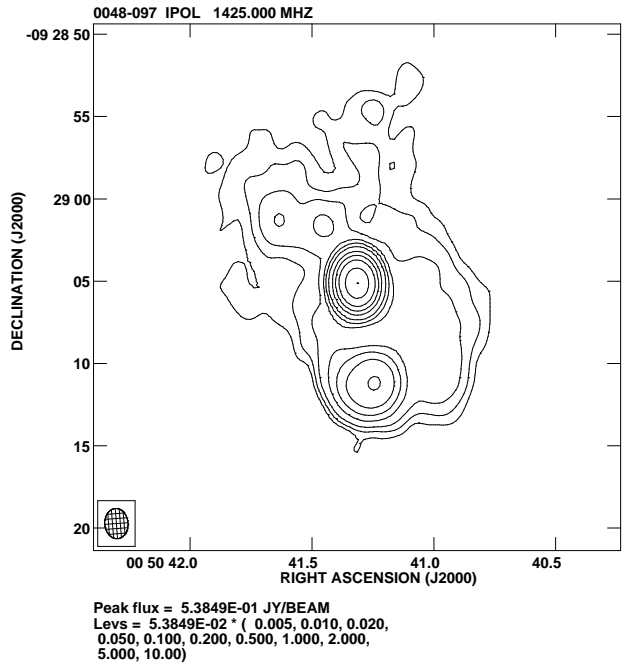}{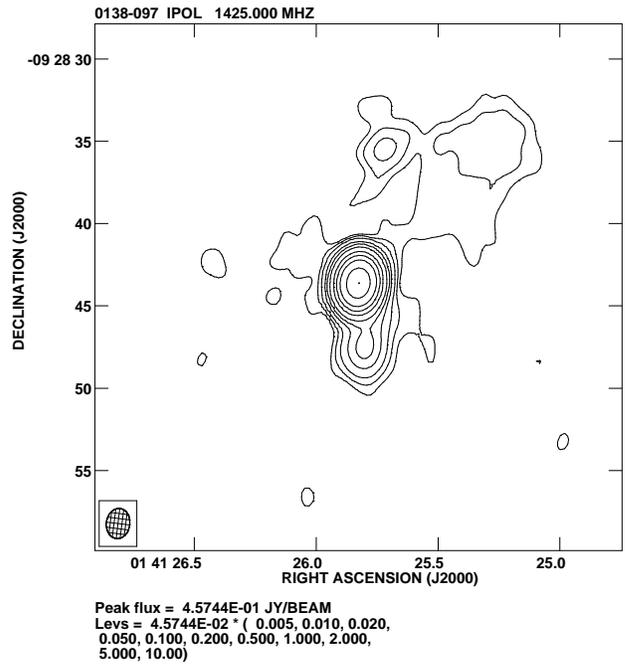} 
\caption{Deep VLA 20cm A-array radio maps of 1Jy BL Lacs.  The base level of each
map is set at the 2$\sigma$ RMS noise level.  The beam is shown in the lower left
corner.
\label{fig-2}}
\end{figure}

PKS 0138--097: SFK93 report an absorption feature at 4200\AA, identified as MgII
$\lambda$2798 at $z = 0.501$, which we confirm.  We also detect MgII $\lambda$2798,
[NeV] $\lambda$3426 and [OII] $\lambda$3727 in emission and CaII $\lambda
\lambda$3933,3968 in absorption at $z = 0.733$.  An MMTO spectrum of this object is
also presented in SR97. Deep R- and K'-band imaging reveal four nearby nonstellar
objects, one or more of which may be the absorbing galaxy (Heidt et al. 1996).  At
$z=0.733$ the extended radio structure is too luminous for an FR--1 (see
Figure~\ref{fig-2}).  The morphology appears to be that of a ``dogleg" FR--2 source
(Stocke et al. 1985); and is confirmed by C99.

S5 0454+844: Lawrence et al. (1996; hereafter L96) first identified an absorption
feature at $\sim 6550$\AA\ to be Na ``D" at $z=0.112$.  However our 2.1m
observations resolve this feature into the MgII doublet $\lambda\lambda$2796,2802
at $z = 1.340$.  A discussion of this source is presented in SR97. S5 0454+844 is
unresolved at radio and optical wavelengths, which is consistent with a high
redshift.  This new redshift lower limit makes this the most distant object in
the 1Jy sample.  The MMTO 6.5m spectrum shows evidence of a broad hump at $\sim$4855\AA. 
If real, this may be CIII] $\lambda 1909$ at $z=1.54$, although CIV$\lambda 1549$
is not detected at this redshift.  Alternatively, the hump at $\sim$4855\AA\ may be
CIV$\lambda 1549$ at $z=2.12$.  A potential feature at 3790\AA may be identified with
\lya\ at this redshift, although there is no evidence for CIII] $\lambda 1909$.  The
hump is unlikely to be \lya\ at $z\sim3$ as no depression from a \lya\ forest is
present, nor is CIV$\lambda 1549$ present at that redshift.

S5 0716+714 (DA 237):  This object is remarkably featureless in our 2.1m and MMTO 6.5m
spectra.  This well-studied source (e.g. Vermeulen \& Taylor 1995; SFK93) has yet
to show any spectral features.  The weak $\sim$400\AA\ oscillation in the continuum
shape and the small absorption feature at $\sim$4380\AA\ in our MMTO 6.5m spectrum are
due to flux calibration errors.  Because its has a high-SNR and
is featureless, the MMTO 6.5m spectrum of 0716+714 was used to correct for the same
calibration errors in other MMTO 6.5m spectra taken with the 800 line mm$^{-1}$
grating.  0716+714 is optically unresolved (SFK93), suggesting $z > 0.2$. 

PKS 0735+178: We confirm the strong absorption feature at 3985\AA\ first identified
as MgII $\lambda$2798 at $z = 0.424$ by Carswell et al. (1974).  Our MMTO 6.5m
spectra resolves the MgII doublet as well as detects MgI $\lambda$2852 and FeII
$\lambda$2344,
$\lambda$2383,
$\lambda$2587 and $\lambda$2600 at this redshift.  Galactic CaII
$\lambda\lambda$3933,3968 is also detected.  We do not confirm a second, tentative
absorption system suggested by Carswell et al. (1974).  No other features are clearly
apparent.  The radio source is marginally resolved (M93, Perlman \& Stocke 1994),
suggesting a high redshift.

S4 0814+425 (OJ 425): Wills \& Wills (1976) suggested a highly uncertain redshift of
$z = 0.258$ based upon two weak emission lines identified as MgII $\lambda$2798 and
[OII] $\lambda$3727.   However, our observations and others (SFK93, Dunlop et al.
1989) find no evidence for spectral features at this or any other redshift.  Falomo
et al. (1997) report this object to be unresolved by HST/WFPC-2 observations and,
assuming a host elliptical of $M_R = -23.3$, suggest $z \geq 0.6$.  L96 detect very
weak features which they identify as either MgII at $z = 0.245$ or CIV
$\lambda$1549, CIII] $\lambda$1909 and [OII] $\lambda$3727 at $z = 1.2516$.  Both
are very speculative; and the former redshift is probably ruled out by Falomo et
al. (1997).  We note the possible detection of a weak line at 4262\AA, which is
potentially seen in both our KPNO 2.1m spectra but not in our MMTO 6.5m spectrum. 
A better spectrum blueward of 4500\AA\ is required for this object. 

S4 0828+493: This object was observed twice at the 2.1m.  Both spectra confirm the
narrow emission line seen at 5770\AA\ by SFK93, which they identify as [OII]
$\lambda$3727 at $z = 0.548$.  [NeV] $\lambda$3426 is also possibly present at
this redshift in our spectrum.  However, given that only one emission line is
definitely detected, there is no compelling argument for this redshift.  Indeed, at
$z=0.548$ MgII emission should be visible in our spectrum and that of SFK93 and it
is not observed.  
The radio structure shown in Figure~\ref{fig-2} appears to be that of an
edge-darkened twin jet source, typical of FR--1s; and yet at $z = 0.548$ the extended
power is too luminous to be an FR--1, suggesting a lower redshift.  
Therefore the redshift of this object remains in doubt.  A better spectrum blueward
of 5000\AA\ is required for this object.

S4 0954+658:  L96 measure a firm redshift of $z = 0.3668$ with the detection of [OII]
$\lambda$3727, [OIII] $\lambda$5007 and H$\alpha$.  SFK93 and our spectra confirm this
redshift by detecting CaII $\lambda\lambda$3933,3968 in absorption and [OII] in
emission. The one-sided radio emission resembles an FR--1 morphology and has an extended
radio power consistent with that morphology at this redshift.

B2 1147+245: This object remains stubbornly featureless, in agreement with SFK93. 
It is optically unresolved (SFK93), suggesting $z > 0.2$. Its FR--2 radio morphology
(Antonucci \& Ulvestad 1985; C99) suggests it is a high-$z$ object.

PKS 1519--273: No spectral features are present, in agreement with SFK93 and White
et al. (1988).  However, due to its low declination, we were unable to obtain a
high-quality spectrum.  Further observations are needed.  It is optically unresolved
(SFK93), suggesting $z > 0.2$.

S4 1749+701: SFK93 report an emission feature which they have detected on two
occasions and which they identify as [OII] $\lambda$3727 at $z = 0.770$.  L96
confirm this redshift with the detection of [OII] as well as CIII] $\lambda$1909. 
Our spectra do not confirm the [OII] line and CIII] is outside of our wavelength
range.  The VLA map is only marginally resolved, consistent with a high-$z$. 
At $z=0.770$ the extended radio power level is only consistent with this source
being an FR--2.

PKS 1749+096 (4C 09.57): SFK93 report $z=0.320$ as confirmed independently by
Stickel et al. (1988a) and White et al. (1988) by detecting [OII], H$\beta$ and
[OIII] emission.  The host galaxy of this BL Lac is easily resolved on the CFHT
image of Wurtz et al. (1996) and has an absolute magnitude consistent with BL Lac
hosts at that redshift.  However, despite its moderate redshift the radio emission
is completely unresolved at a dynamic range of $\sim10,000:1$.  The absence of
extended flux for this relatively nearby source is unusual, although very faint,
diffuse extended radio cannot be completely ruled out.  Nonetheless the extended
radio power for this source is very low, and most likely lower than any other 1Jy
BL Lac.  This anomaly is similar to PKS 1413+135 (Perlman et al. 1996) and should
be investigated further.

S5 1803+784: Our spectrum confirms the broad emission feature identified by SFK93
as MgII $\lambda$2798 at $z = 0.680$.  We also detect H$\beta$ at this redshift.
L96 confirm this redshift.

S4 1823+568 (4C 56.27): L96 detect MgII $\lambda$2798, [OII] $\lambda$3727,
H$\beta$ and [OIII] $\lambda$5007 at $z = 0.6634$.  Our MMTO spectrum confirms the
MgII and [OII] lines as well as detects [OIII] $\lambda$2321 and CaII H\&K in
absorption at that redshift.

PKS 2029+121: Stickel \& K\"uhr (1993) first found the intervening MgII system at
$z=1.117$ and suggested an emission redshift of $z=1.223$ based upon broad MgII.  SR97
present our spectrum which confirms both redshifts (although at a slightly different
emission redshift of $z=1.215$ based upon the detection of CIV$\lambda 1549$, CIII]
$\lambda 1909$, MgII $\lambda$2798 and [OII] $\lambda$3727).  Our VLA map shows a
one-sided, edge-brightened source consistent with an FR--2 level of extended radio
power.  The rest equivalent width of the CIV$\lambda 1549$ emission line (\wlam\ $=
15.7$\AA) exceeds the criterion proposed by S91.

PKS 2131--021 (4C -02.81): Wills \& Lynds (1978) suggested a redshift of $z =
0.557$; however, our spectrum shows [OII] $\lambda$3727, MgII $\lambda$2798 and  CIII]
$\lambda$1909 in emission at $z = 1.285$.  These emission lines are at the
upper limit in \wlam\ for BL Lac objects. 
Drinkwater et al. (1997) independently confirm our redshift.  In addition, there are
very weak absorption features at $z
\sim 0.36$ which are more apparent in a subset of our 2.1m spectra.  These very
tentative detections require confirmation and suggest that MgII absorption should
be searched for at $\sim 3800$\AA.  Surprisingly the extended radio structure
resembles a wide-angle tail but with the extended radio power and leading-edge
hotspots like an FR--2.  This unusual radio morphology is suggestive of a
gravitationally-lensed source; however $\sim$0.1 arcsecond-resolution maps reveal a jet
to the southeast of the core which bisects the lobes to the south and east (Rector \&
Stocke 2002). 

PKS 2150+173: Our 2.1m and MMTO spectra show no evidence for any lines, in agreement
with Stickel, K\"uhr \& Fried (1993).  The very extended narrow-angle tail radio
morphology suggests a low to moderate redshift (i.e., assuming an upper limit on
extended radio power for an FR--1 of log $P_{ext} \leq 25.5$ W Hz$^{-1}$ requires
that $z < 0.5$ for this source).

\section{Completeness of the 1Jy Sample}

Browne and March\~a (1993; hereafter BM93) and March\~a and Browne (1995; hereafter
MB95) describe optical selection effects which can make BL Lacs
difficult to recognize, thus introducing the possibility that such samples are
incomplete.  A ``low-luminosity BL Lac" (i.e., a BL Lac object which is either
intrinsically less luminous or not as strongly Doppler boosted) would be difficult
to identify optically because the weak nonthermal continuum from the AGN could be
strongly diluted by starlight from the luminous host elliptical galaxy, thus
causing the object to be misidentified as either a radio galaxy or,
in the X-ray case, as a cluster of galaxies (Rector, Stocke \& Perlman 1999), rather
than as a BL Lac.  The magnitude cutoff of the 1Jy sample ($V \leq 20$) will
bias against these objects, as they are intrinsically faint optically.  MB95
predict that the percentage of missing objects in the 1Jy sample is 22\%, 18\% of
which are missing due to galaxy dilution and 4\% due to the magnitude cutoff.  MB95
estimate that most of these  misidentified BL Lacs are at low $z$ ($z < 0.2$). 
Although significant, this effect probably cannot explain the \vvmax\ discrepancy
between the EMSS and 1Jy samples (MB95).

The predictions of BM93 and MB95 are based only upon selection effects due
to the optical selection criteria for BL Lacs.  It is also possible that the 1Jy sample
is incomplete due to selection effects at radio wavelengths.  In particular, the flat
radio-spectrum criterion ($\alpha_r \leq 0.5$ between 11 and 6cm; S91)
for BL Lacs effectively biases the 1Jy sample against lobe-dominated radio sources,
as the spectral index for the extended radio structure is typically steeper than the
core.  A BL Lac whose core is not strongly Doppler boosted may thus be eliminated.  

A study of eight XBLs from the EMSS and HEAO-1 samples by Stocke et al. (1985) found
that three of the objects showed a steep radio spectrum which exceed this criterion,
and two more were marginally steep.  Furthermore, a spectral analysis  of the EMSS
sample using the discovery 6cm flux and the NVSS 20cm flux find that about 30\% of
the objects show a spectrum steeper than $\alpha_r = 0.5$.  However, it is certainly
possible that XBLs have radio spectral indices that are significantly different than
RBLs.  Further, the flux density measurements were not simultaneous, and therefore
variability may affect the results.  Simultaneous, multi-frequency observations of EMSS
XBLs are underway to address this issue (Cavallotti et al. 2001).

\begin{figure}
\plotone{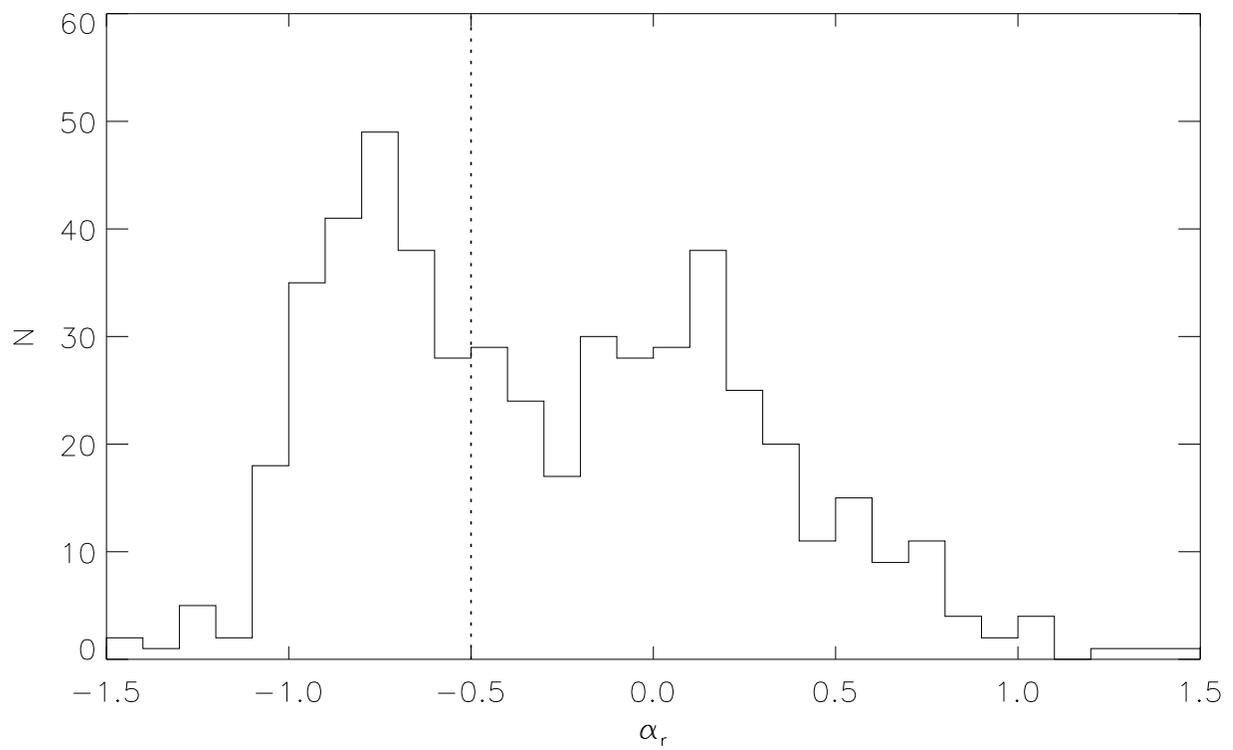}
\caption{The distribution of radio spectral indices $\alpha_r$ for objects in the 1Jy
survey (K\"uhr et al. 1981).  The two-point (11 and 6cm) $\alpha_r$ values are
determined from non-simultaneous observations.  The dashed line marks the flat-radio
spectrum criterion imposed upon the 1Jy BL Lac sample (S91).  This criterion does not
correspond to an observed break in the distribution.
\label{fig-7}}
\end{figure}

However, the primary point to be made here is that flat radio-spectrum criterion is
arbitrary.  The distribution of spectral indices in the entire 1Jy survey can be 
described as marginally bimodal, with a minimum between the peaks at roughly $\alpha_r =
-0.25$ (Figure~\ref{fig-7}).  There is no indication that $\alpha_r = -0.5$ should be
considered as an intrinsic break between flat and steep spectral objects in the 1Jy
survey.  Almost half (216 of 518) of the objects in the 1Jy survey have steep radio
spectra ($\alpha_r < -0.5$); and  
115 of these objects have  spectral indices which are near the flat
radio-spectrum criterion ($-0.8 < \alpha_r < -0.5$).  Of these, 66 of these objects are
galaxies and 32 are quasars.
As spectral variability of $\Delta\alpha_r \sim 0.2$ is not unusual for RBLs (e.g.,
0716+714; Wagner et al. 1996), many of these objects may be RBLs which were observed
in the wrong epoch to be classified as such.

What number of 1Jy objects have been misclassified because of the flat radio
spectrum criterion, or because of the optical selections effects described in BM93, is
unknown.  However, there is already evidence that these criteria do eliminate some BL
Lac objects.  Rector et al. (1999) identify 3C 264, a narrow-angle tail radio galaxy
in Abell 1367 and a member of the 1Jy survey (1Jy 1142+198), as a low-luminosity BL
Lac object.  Its marginally-steep radio spectrum ($\alpha_r = 0.65\pm0.08$; K\"uhr et
al. 1981) eliminated it from consideration for the 1Jy BL Lac sample.  It is at a low
redshift ($z = 0.0215$; Zabludoff et al. 1993) and has a relatively high radio flux
($\langle S_{5\ {\rm GHz}} \rangle = 2.19$ Jy; K\"uhr et al. 1981), giving it a
$V/V_{max} = 0.31$.  The flat radio-spectrum criterion biases the 1Jy sample towards
BL Lacs which are more strongly Doppler boosted, hence the BL Lacs which are most
extreme in their properties.  If objects such as 3C 264 are abundant in the 1Jy
sample, the range of observed properties for 1Jy BL Lacs,
including the sample \vvmax, could be different.

\section{Properties of the 1Jy Sample and Impact on Unified Schemes}

In recent years the unification picture for BL Lac objects has become increasingly
complex.  In contrast to the unified model, several authors have suggested (e.g., 
Brinkmann et al. 1996; Scarpa \& Falomo 1997) that RBLs may be more closely related
to other blazars, i.e. FSRQs, than to XBLs.  Here we compare the properties of 1Jy
RBLs to the complete XBL sample of M91 as well as to complete samples of radio
galaxies (FR--1s and FR--2s) and FSRQs drawn from the 2Jy sample of Wall \& Peacock
(1985). 

\subsection{Radio Power Levels}

\begin{figure}
\plotone{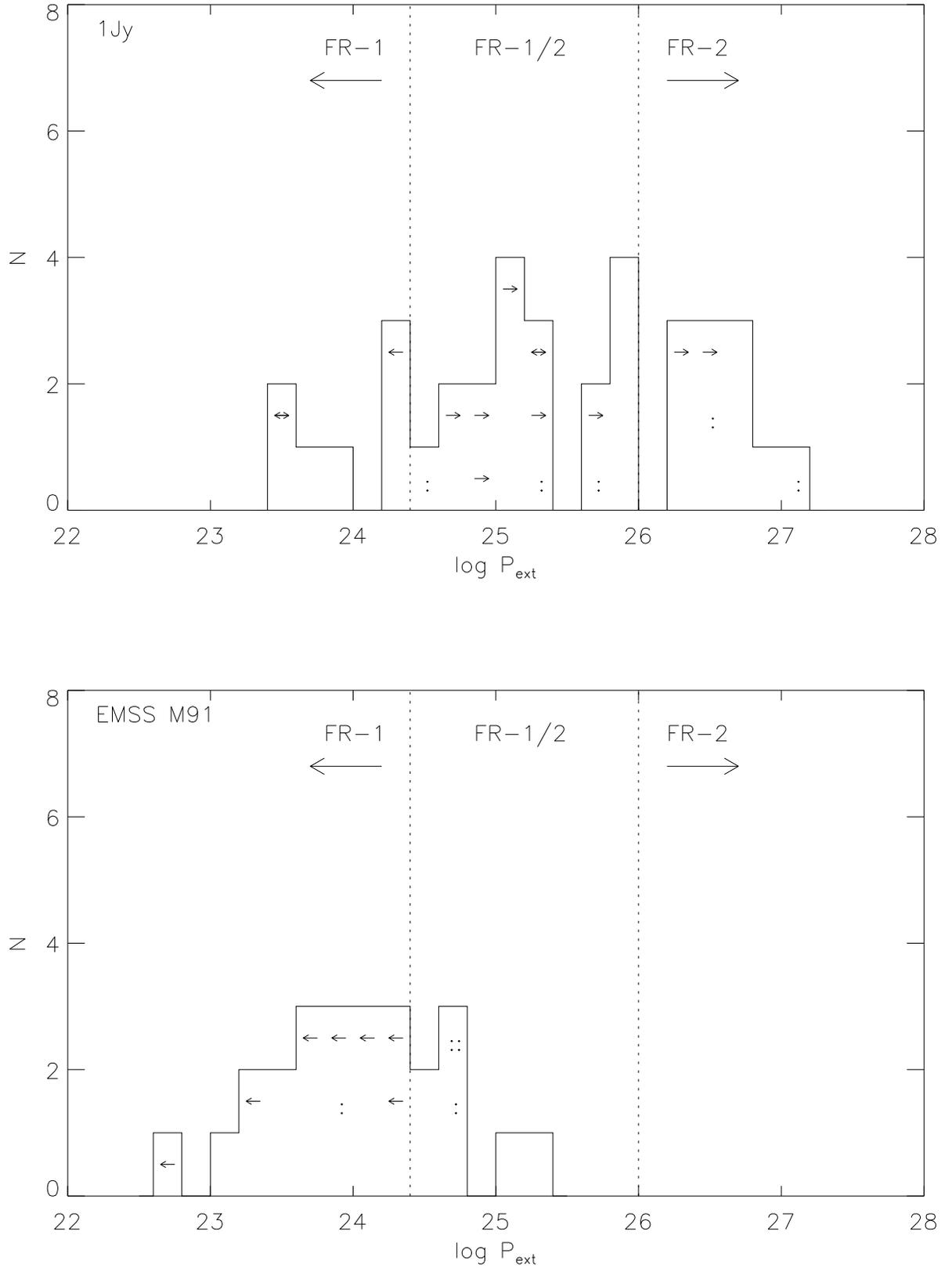}
\caption{Distribution of extended 20cm radio power for the entire 1Jy BL Lacs
sample (above) and the EMSS M91 sample (below; R00) in units of log W
Hz$^{-1}$. Objects marked with a colon have a luminosity based upon a tentative
redshift; a double colon indicates a spectulative redshift.  Objects marked with an
``$\rightarrow$" have a lower limit on their extended radio power based upon an
absorption system in the spectrum.  Objects marked with a ``$\leftarrow$" are
unresolved; therefore the quoted value is an upper limit.  The luminosity ranges
marked for Fanaroff-Riley classification are adopted from Owen
\& Laing (1989).
\label{fig-4}}
\end{figure}

If BL Lacs are simply a highly-beamed population of FR--1 radio galaxies their,
presumably unbeamed, extended radio flux densities should be similar.  R00 find that
BL Lacs from the EMSS have extended radio powers almost completely consistent with
that of FR--1s.  Based upon the VLA maps and updated redshifts presented here and in
R00, a comparison of the extended luminosities of the complete 1Jy
and EMSS samples reveal that they are very different (Figure~\ref{fig-4}).  The 1Jy
luminosities are nearly two orders of magnitudes more luminous: log \avgPext\ $>
26.23$ W Hz$^{-1}$ for the 1Jy sample, as compared to log
\avgPext\ $< 24.43$ W Hz$^{-1}$ for the complete EMSS ``M91" sample (R00).  Note that
the former value is considered a lower limit as many 1Jy BL Lacs have only a lower
limit in redshift; and the latter value is an upper limit as extended radio
flux was not detected in roughly half of the objects in the M91 sample.  A
Kolmogorov-Smirnov (K-S) test on the distributions of extended radio power of the 1Jy
and EMSS samples finds that the probability these two samples are drawn from the same
parent population is $< 0.1$\%.  While the extended radio luminosities for 1Jy RBLs
are more luminous than XBLs, they are however significantly lower than 2Jy FSRQs
(Padovani 1992).

If RBLs are highly-beamed radio galaxies we expect them to have extended radio
structures that are more compact and distorted due to projection effects.  As can be
seen in Figure~\ref{fig-2} the extended radio morphologies of RBLs do tend to be
compact and highly distorted, often with a ``halo" structure around the core.  Their
compact, distorted morphology makes it difficult, if not impossible, to classify a BL
Lac object as being FR--1 or FR--2-like based solely upon its morphology.  Thus we
have relied primarily upon the extended radio luminosity to make this classification.
Figure~\ref{fig-4} shows the distribution of extended radio powers for RBLs in the
1Jy sample and XBLs in the EMSS (R00) which have been mapped by the
VLA to a high dynamic range.  Owen \& Laing (1989) find that almost all radio
galaxies with extended radio powers $P_{ext} \leq$ log 24.5 W Hz$^{-1}$ are
morphologically FR--1s; and those with log $P_{ext} \geq 26.0$ W Hz$^{-1}$ are
morphologically FR--2s; thus we use these dividing lines for our classification. 
RBLs with intermediate powers ($24.5 < $ log $P_{ext} < 26.0$  W Hz$^{-1}$) are given
a Fanaroff-Riley classification based upon the radio morphology where possible. 
Uncertain classifications are marked with a colon in Table~\ref{tbl-3}.  PKS
2005--489 has no radio FR-classification due to the lack of a sufficient radio
map.  We find that many RBLs have extended radio powers which are too luminous to be
FR--1s; and several have extended radio morphologies similar to FR--2s (e.g., PKS
0048--097 and PKS 2131--021; Figure~\ref{fig-2}).  Our classifications are roughly in
agreement with the radio luminosity-based classifications of Kollgaard et al.
(1992).  

We also classify each object by its optical spectral properties.  Objects with broad
(FWHM $\geq 1000$ km s$^{-1}$), luminous emission lines typical of FR--2s (e.g.,
Urry \& Padovani 1995) are classified as such.  Objects with only narrow (FWHM $\leq
1000$ km s$^{-1}$), weak emission lines or stellar absorption lines (i.e., not
intervening absorption systems) are classified as FR--1s.  Six objects with no
detectable emission or absorption lines are not classified in this manner.  These two
classification schemes are in agreement for most objects (Table~\ref{tbl-3}); seven
objects have only one classification available and two have
conflicting classifications that are not tentative.  We classify PKS 0851+202 as a
tentative FR--1 based upon its radio structure, despite the detection of broad
H$\alpha$ in its spectra (Sitko \& Junkkarinen 1985); and we classify SR 1749+701
as a tentative FR--1 despite its relatively luminous extended radio power because
only weak, narrow emission lines are seen in its spectrum (L96).  Based upon these
criteria we classify 11 RBLs as being FR--1-like and 15 as FR--2-like; 5 and 6 more
are tentatively classified as FR--1 and FR--2 respectively.  

Thus we conclude that the 1Jy sample contains roughly equal amounts of FR--1 and
FR--2-like sources.  Why this is the case depends on several factors: (1) the radio
luminosity function (RLF) at 5 GHz for FR--1s and FR--2s over $0<z<1.5$; (2) selection
effects induced by Doppler boosting, which depend on the jet Lorentz factors and
physical opening angles for the two radio galaxy classes; and (3) the importance of
gravitational microlensing in the 1Jy (SR97).  Unfortunately, none of these factors are
well known.  The RLFs for FR--1s and FR--2s are well-known for $z<0.2$ (Urry et al. 1991;
Padovani \& Urry 1992), but their evolution is poorly constrained at higher redshift and
virtually unknown at $z \sim 1$.  Further, the Lorentz factors in radio galaxy jets
have been measured in only a few instances (e.g., Lara et al. 1997).  And a
problem with the unification model is that radio galaxies do not show apparent
superluminal motion as often as expected for a relatively unbeamed BL Lac population
(Urry
\& Padovani 1995, and references therein), implying that they have smaller Lorentz
factors than BL Lacs.  A possible explanation for this discrepancy is that
the motion of reverse shocks observed within the jet can be much slower than the
actual flow velocity (Bicknell 1994).  Thus, the Lorentz factors for radio galaxies
are not necessarily inconsistent with BL Lacs, but they are also not
well-constrained.  Finally, as is discussed below, the number of
gravitationally-lensed BL Lacs in the 1Jy sample is not known; nor do we know the
source amplification, which is necessary to correctly determine $V/V_{max}$ for each
lensed source.

Can the number of
FR--1- and FR--2-like BL Lacs found the 1Jy sample be understood as a subset of 
the beamed counterparts of FR--1 and FR--2 radio galaxies, a subset which also includes
SSRQs and FSRQs?  Or is another mechanism (e.g., gravitational lensing) necessary
to account for the surprisingly large number of FR--2-like BL Lacs?  By using the
observed RLFs of FR--1s and FR--2s as well as their evolutionary
constraints (albeit with large errors; Urry \& Padovani 1995), it is possible to
construct RLFs for Doppler-boosted FR--1s and FR--2s using the formalism originally
developed by Urry \& Shafer (1984).  By assuming reasonable values for the jet
outflow velocity ($\gamma =$ 5-10; Urry \& Padovani 1995, Gabuzda 1992, PS93) and
unbeamed core to lobe radio power ratios for FR--1s and FR--2s ($f=$ 0.01-0.1;
Ulrich 1989; Urry et al. 1991; Padovani \& Urry 1992), BL Lac samples can be
constructed which meet the flux density and flat radio spectrum criteria of the 1Jy
BL Lac sample.  With nominal values for $\gamma$ and $f$, the observed numbers and
redshift distributions of FR--1 and FR--2-like BL Lacs are readily produced (see
Rector 1998 for specific examples), indicating that, at least based upon the radio
selection criteria, the 1Jy BL Lac sample is not necessarily unusual.  But the loose
constraints of the model parameters are such that significantly different FR--1/FR--2
ratios are also possible.  Tighter constraints on FR--1 and FR--2 evolution and
Doppler boosting parameters are necessary before the FR--1/FR--2 content of the 1Jy
sample may be used as a detailed constraint on BL Lac populations.  We also note that
this approach does not consider the other 1Jy selection criteria which also affect
the observed sample properties (e.g., optical properties, as discussed in \S 5.3.)

\subsection{Radio Core-Dominance Values}

The radio core dominance value $f$ must be, at least to some extent, an indicator of
beaming angle.  The higher core-dominance values seen in RBLs have been known for
some time; and have been invoked as a primary piece of evidence in support of
the unified model that BL Lacs are beamed FR--1s.  While nearly all XBLs are
FR--1-like (R00), we have shown above that the 1Jy sample consists of roughly equal
numbers of FR--1- and FR--2-like RBLs.  An important question is the relationship
between the low-luminosity, FR--1-like RBLs in the 1Jy and the low-luminosity XBLs in
the EMSS.  

\begin{figure}
\plotone{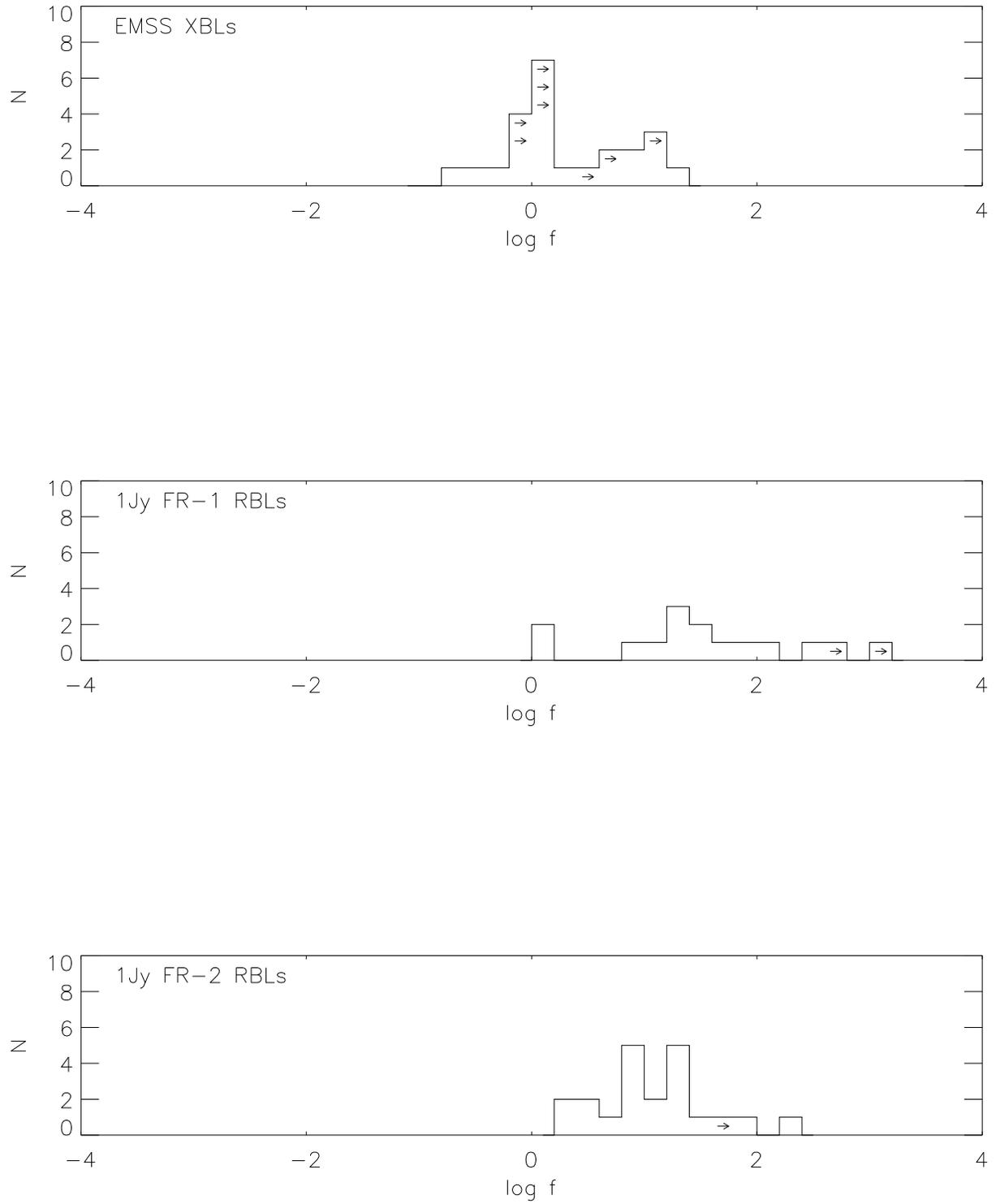}
\caption{The distribution of core-to-extended flux ratios ($f$) for the M91 XBL
sample (R00) as well as FR--1-like RBLs and FR--2-like RBLs from the 1Jy sample.  Objects
marked with an arrow were unresolved; the values indicate a lower limit based upon
the upper limits on the extended radio power.  While the distribution of core-domince
values for FR--1- and FR--2-like RBLs are consistent, EMSS XBLs have significantly
lower core-dominance values than FR--1-like RBLs, suggesting that they are less
beamed.
\label{fig-6}}
\end{figure}

The distribution of core-dominance values for RBLs is significantly higher
than for XBLs (Figure~\ref{fig-6}), supporting the unified hypothesis that XBLs are
indeed seen, on average, at larger angles.  Further, the distribution of $f$
values for FR--1-like RBLs are similar to FR--2-like RBLs but not XBLs.  Thus the
1Jy sample is not simply an admixture of low-luminosity, XBL-like BL Lacs and
high-luminosity, FRSQ-like BL Lacs.  A plausible explanation as to why the
1Jy sample contains only the most highly beamed sources is that the radio jet has
either a larger Lorentz factor (hence a smaller beam pattern) or a smaller opening
angle $\theta_{jet}$ than the X-ray jet, such that a shallow radio survey such as the
1Jy will detect only the most highly beamed objects.  However, an explanation is
still required for the discrepancy in \vvmax\ values for XBLs and FR--1-like RBLs,
in order for the two samples to be part of the same population (see \S 5.4).  A possible
explanation is that the 1Jy RBL sample is incomplete due to its flat-spectrum
selection criterion.  If an object is seen at an orientation angle ``too far" off
axis such that it is seen outside the beam pattern or physical opening angle of
the jet ($\theta > 1/\gamma$ or $\theta > \theta_{jet}$), it will become
lobe-dominated and have too steep a radio spectrum to be included into the 1Jy.  3C
264 is an example of such an object (Rector, Stocke \& Perlman 1999; see
\S 4 for further discussion).  Further, surface-brightness dimming and beam dilution
effects (PS93), which strongly affect FR--1s, will bias the 1Jy sample {\it against}
nearby, off-axis objects; i.e., more distant objects will appear more core-dominated
and hence are more likely to be included into the 1Jy based upon their flat spectrum,
thus artificially raising the \vvmax\ value for the 1Jy.  However, roughly three times
the currently known FR--1-like BL Lac population would need to be identified in the
1Jy survey in order to bring the \vvmax\ statistic for the 1Jy sample into agreement
with the EMSS.

\subsection{Emission Line Properties}

While traditionally known for their linelessness, emission lines have been detected
in most (27 of 37) 1Jy BL Lacs.  However, the difficulty in detecting lines in BL
Lacs does pose a problem for statistical studies of emission-line properties.  And the
wide range of redshifts in XBLs, RBLs and FSRQs prevents a comprehensive study of a
single emission line at optical wavelengths; e.g., the [OIII] emission line is detected
in only eight 1Jy RBLs, all of which are $z < 0.4$.  Our observations did not 
discover previously undetected [OIII] emission in any RBLs; however, we did discover
moderately luminous MgII emission in PKS 0138--097, PKS 2131--021 and PKS 2029+121.  

\begin{figure}
\plotone{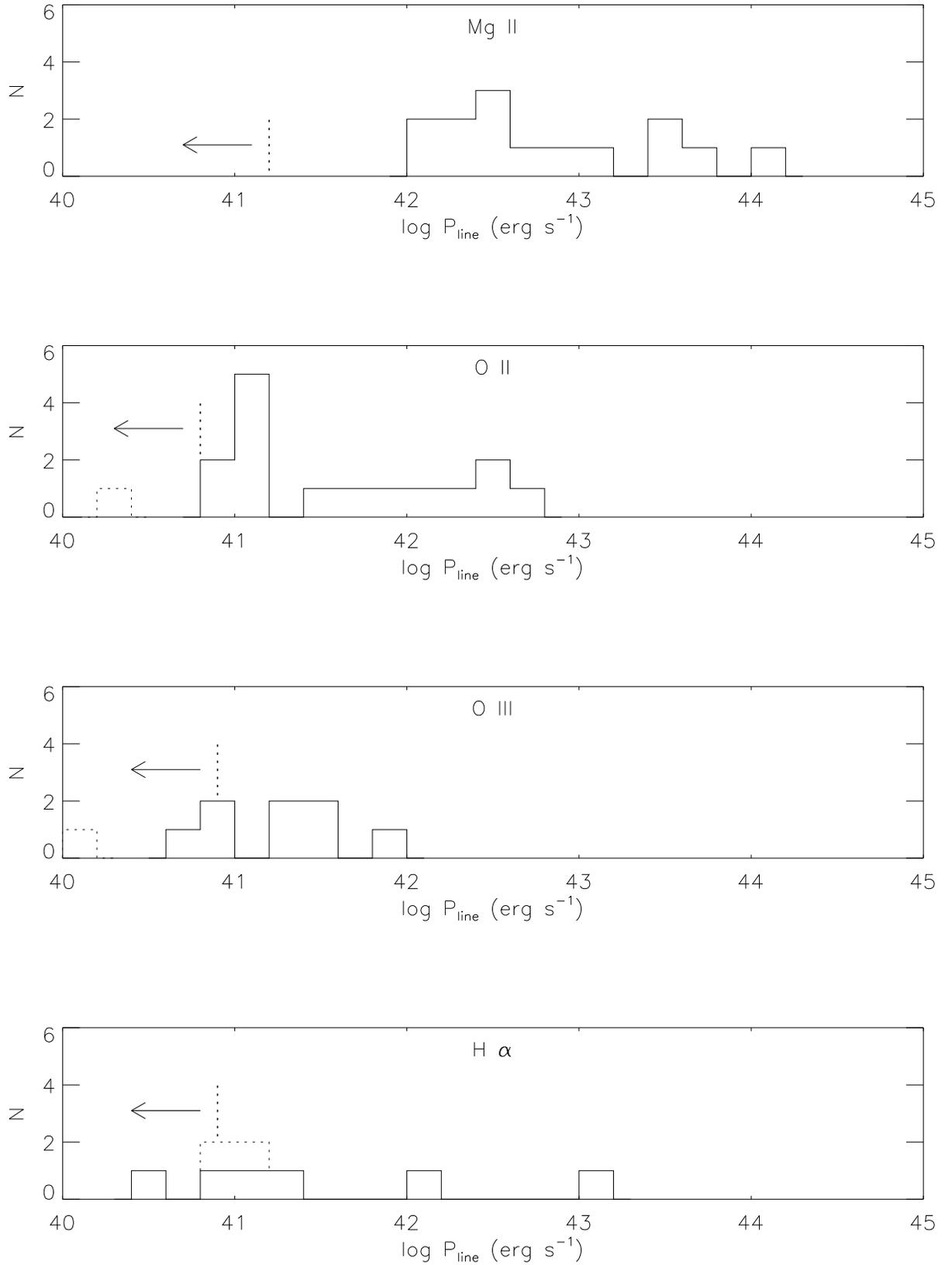}
\caption{Distribution of the optical emission line luminosities of MgII
$\lambda$2798, [OII] $\lambda$3727, [OIII] $\lambda$5007 and H$\alpha$ for 1Jy
and EMSS BL Lacs (R00).  RBLs are shown with a solid line, XBLs with a dashed line.
While most RBLs show weak but luminous emission lines only three XBLs in the EMSS
sample show any emission lines, all of which are very weak.  A vertical dashed line
with an arrow to the left marks the approximate median $5\sigma$ upper limits of
detectability for both RBLs and XBLs.
\label{fig-5}}\end{figure}

Evidence indicates that the central region in BL Lacs is not obscured (e.g.,
Chiaberge et al. 1999 and references therein); thus, according to the unified model,
RBLs and XBLs should possess identical distributions of emission-line luminosities.
While most RBLs have moderately luminous (log $41 \leq L \leq 44$ erg s$^{-1}$)
emission lines, only three EMSS BL Lacs show any emission lines at all and they are
very weak (log $L \leq 40$ erg s$^{-1}$; Figure~\ref{fig-5}).  The spectra for both
the 1Jy and EMSS samples are of sufficient quality to detect lines of this strength
or weaker.  Thus this discrepancy is not due to a systematic difference in the
quality of spectra obtained, although it may be due to a difference in the redshift
distribution for the two samples.  EMSS BL Lacs are predominantly at lower redshifts
($z<0.5$), thus many potentially strong emission lines (MgII, CIII] $\lambda$1909 and
CIV $\lambda$1549) are not redshifted into the optical window.

While the sampling is poor, especially at high-$z$ (only 3 RBLs at $z > 1$), the MgII
and [OIII] emission-line luminosities for RBLs are several orders of magnitude lower
than for 2Jy FSRQs.  However, there is significant overlap between the two classes,
forming a continuum of line luminosities among blazars (e.g., Scarpa \& Falomo
1997).  This suggests that the limiting equivalent-width criterion ($W_{\lambda} <
5$\AA) which differentiates RBLs and FSRQs is an arbitrary division. 
However, it should be noted that studies of blazars with VLBI techniques have revealed
two distinct groups of blazars, as determined by the orientations of the magnetic
fields within the radio jets.  While not definitive, sources with strong emission
lines tend to have primarily longitudinal magnetic fields in their jets, while in
weak-lined sources the magnetic field is usually transverse to the jet axis (e.g.,
Gabuzda et al. 1992).  

The lower emission-line luminosities in RBLs compared to FSRQs is may be exaggerated
because RBLs lie at systematically lower redshifts than FSRQs.  RBLs are rarely seen at
high redshifts ($z > 1$) because the strong CIII] $\lambda 1909$, CIV$\lambda 1549$ and
\lya\ lines appear in the optical at these redshifts, thereby preventing a BL Lac
classification in most cases.  PKS 2029+121 is an example of such an RBL.

\subsection{The Spatial Distribution of BL Lacs}

\begin{figure}
\plotone{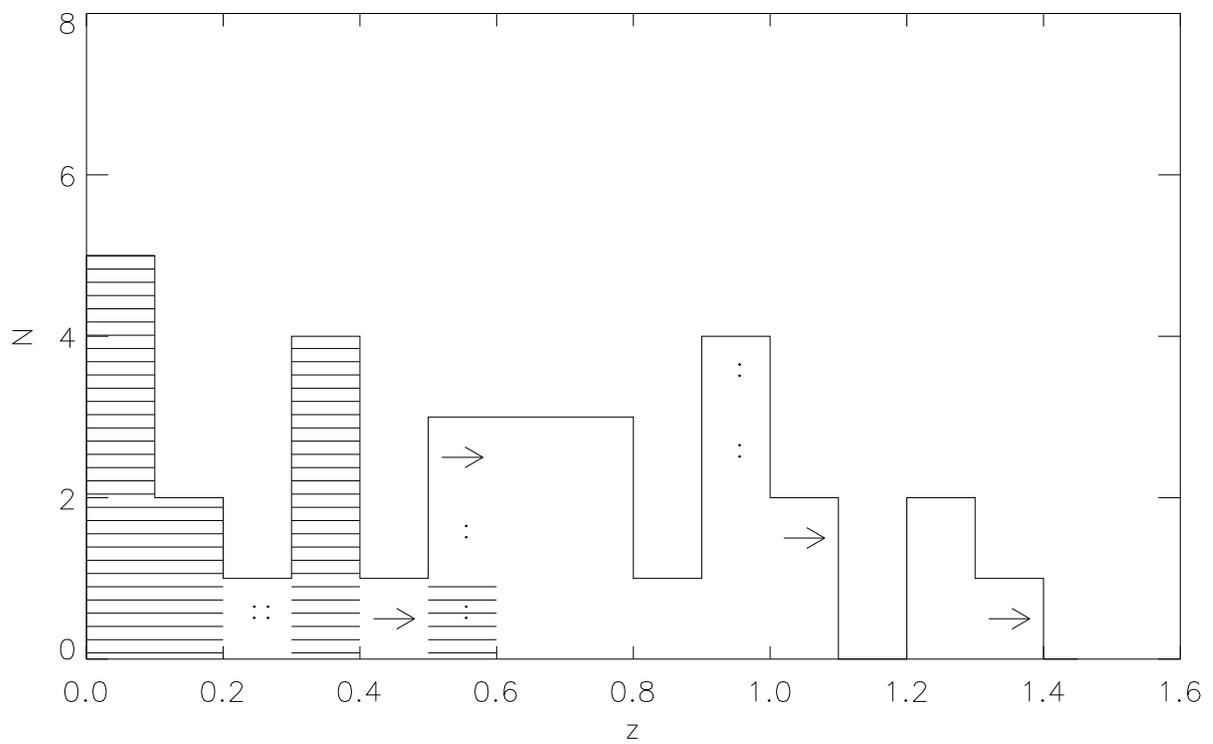}
\caption{Distribution of redshifts for 1Jy BL Lacs.  Tentative redshifts are marked
with a colon.  Redshift lower limits based upon observed absorption system(s) are
marked with an ``$\rightarrow$".  The hatched area represents objects which are
FR--1-like.  Note that the distribution is roughly flat up to $z \sim 1.5$.
\label{fig-3}}
\end{figure}

Six new redshifts for 1Jy BL Lacs are considered here: four are revisions to
redshifts reported in S91 (PKS 0138-097, S5 0454+844, S4 0814+425 and PKS 2131-021)
and two are for BL Lacs new to the sample (S4 0218+357 and PKS 2029+121).  The
distribution of redshifts is roughly flat up to $z \sim 1.5$ (Figure~\ref{fig-3}). 
We do not see a preference towards detection of low-$z$ BL Lacs, as was previously
reported for incomplete samples of RBLs (e.g., Browne 1989).  The discussion in \S 3
describes how the new redshifts impact previous results on individual sources. 
These revisions include more high-redshift objects in the 1Jy, increasing the
average redshift of the sample from \avgz\ $= 0.506$ to \avgz\
$= 0.595$.

With these new redshift values, \vvmax\ for the 1Jy sample increases slightly
from $0.60\pm0.05$ (S91) to $0.614\pm0.047$.  We include objects without firm redshifts
in the following manner: (1) the tentative redshifts of four objects (listed in
Table~\ref{tbl-2}) are assumed to be correct; (2) the four objects
which only have lower limits on redshift based upon absorption lines in their spectra
are set to these lower limits; and (3) the redshifts for the six objects with no
redshift information are assumed to be
$z = 0.595$, the \avgz\ of the 1Jy sample. Despite the increase in \avgz\ the moderate
change in \vvmax\ is not surprising as $V/V_{max}$ is primarily dependent on the
sample flux limit and increases only weakly with redshift. The inclusion of
objects with uncertain redshifts does not significantly affect \vvmax, although the
inclusion of objects with redshift lower limits does cause \vvmax\ to be slightly
underestimated.  Now that redshift information is available for all but six of the 37
BL Lacs in the 1Jy sample, it is clear that new redshift information will change
the sample \vvmax\ value only slightly.  The
\vvmax\ discrepancy between 1Jy RBLs and EMSS XBLs is 4$\sigma$.  

As discussed in \S 4.1, about half of the BL Lacs in 1Jy sample which were
observed with the VLA have extended radio powers more consistent with that of
FR--2s.  Thus an explanation to consider for the \vvmax\ discrepancy between XBLs
and RBLs is that the 1Jy survey draws from two populations of BL Lacs:
low-luminosity, XBL-like objects and high-luminosity, FSRQ-like objects.
However dividing the 1Jy sample by their extended radio luminosities into FR--1 and
FR--2 groups does not explain the discrepancy, as the \vvmax\ for both
subsamples are indistinguishable (\vvmax\ $= 0.609\pm0.07$ and $0.601\pm0.07$
respectively).  So, not only does this fail to explain the \vvmax\ discrepancy
between the two samples, but it also suggests that low-luminosity, FR--1-like RBLs do
not share the same parent population as XBLs. We do note a discrepancy between the
\vvmax\ values of low-$z$ and high-$z$ RBLs: \vvmax\ $= 0.529	\pm 0.08$ for $z<0.5$
and $0.644\pm 0.06$ for $z >0.5$, although this is insufficient to reconcile the
\vvmax\ for RBLs and XBLs with $z<0.5$.

The \vvmax\ for RBLs does agree with FSRQs in the 2Jy sample (\vvmax\ $=
0.64\pm0.04$; Urry \& Padovani 1995), further suggesting that RBLs are more closely
related to FSRQs than XBLs; although 2Jy FSRQs are at significantly higher
redshifts (\avgz\ = 0.91).  The \vvmax\ for 2Jy FR--2s with $z \leq 0.7$ (\vvmax\
$=0.55\pm0.04$) is slightly lower than for RBLs; and 2Jy FR--1s have a significantly
lower \vvmax\ value (\vvmax\ $=0.42\pm0.05$) which is consistent with the \vvmax\ for
XBLs.  Thus the \vvmax\ for the 1Jy sample cannot be explained as an
admixture of FR--1s and FR--2s; although this comparison is difficult as the 2Jy radio
galaxy samples are at much lower redshifts (\avgz\ = 0.02 for FR--1s and \avgz\ =
0.15 for FR--2s).  The possible incompleteness of the 1Jy sample suggested in \S 4 is
unlikely to explain these discrepancies.

R00 find evidence of a correlation between \fxfr\ and \vvmax\ among
EMSS XBLs, wherein the more extreme HBLs (\fxfr\ $> -4.5$) clearly show negative
evolution and the less extreme HBLs ($-6.0 < $ \fxfr\ $< -4.5$) are compatible
with a no-evolution result.  Other HBL samples also show this characteristic
(e.g., Bade et al. 1998; Giommi, Menna \& Padovani 1999).  Table~\ref{tbl-7}
presents the results of the \vvmax\ test as it is applied to subsamples of the 1Jy
and EMSS divided by their \fxfr\ ratios.  Data for the EMSS were obtained from R00.  The
\fxfr\ values for the 1Jy sample were determined from 1 keV flux densities in
Urry et al. (1996); the three objects newly classified as RBLs were excluded due to a
lack of sufficient X-ray data.  Column [1] is the BL Lac sample from which the
subsample is derived; column [2] is the
\fxfr\ range for the subsample; column [3] is the number of objects
$N$ in the subsample; and column [4] is the
\vvmax\ and its error; the error is $(1/12N)^{1/2}$.  The correlation between \fxfr\ and
\vvmax\ originally observed in HBL samples spans the entire range of observed \fxfr\
values in BL Lacs.  A trend in
\vvmax\ is observed, from negative evolution in the more extreme HBLs to positive
evolution in the more extreme LBLs.  Georganopoulos \&
Marscher (1998) explain this correlation as a result of positive evolution of the
electron kinetic luminosity of the jet $\Lambda_{kin}$.  They demonstrate that as
$\Lambda_{kin}$ decreases the peak synchrotron frequency increases, thereby
increasing the \fxfr\ value and producing the observed positive evolution in RBLs and
negative evolution in XBLs.

\subsection{Gravitational Microlensing}

As was discussed earlier, the distribution of emission-line luminosities for RBLs are
systematically lower than for FSRQs, but with significant overlap between the two
classes.  Due to the limiting equivalent-width criterion ($W_{\lambda} < 5$\AA), RBLs may
have systematically higher optical continuum levels.  A plausible explanation for the
boosted continuum in {\it some} but not {\it all} BL Lacs is gravitational microlensing
of a quasar by a foreground galaxy.  Microlensing can preferentially amplify the
nonthermal emission from the compact AGN over the extended emission-line regions, thus
``swamping" the emission lines.  However only one known BL Lac object (S4 0218+357)
shows the multiple images expected from associated macrolensing (O'Dea et al. 1992). 
Narayan \& Schneider (1990) point out that if the lensing galaxy has a low surface
brightness or a large scale length the surface mass distribution will be below the
critical amount necessary to produce multiple images.  However such cases may never be
individually confirmed conclusively as it is difficult to separate the effects of
microlensing from the intrinsic variability of the background quasar; although
intrinsic variability may be differentiated from lensing variability by the latter's
achromatic nature.  As was discussed in SR97 we find strong ($W_{\lambda} > 1$\AA) MgII
absorption systems along RBL sightlines at a rate 4--5 times greater than expected from
quasar sightlines (Steidel \& Sargent 1992). And recent work on damped Ly$\alpha$
absorption systems (which are very strong MgII absorbers like those seen in BL Lacs) by 
Turnshek \& Rao (2000) show that most absorbing galaxies are of low luminosity.  These
galaxy types would not necessarily produce multiple macrolensed images (Narayan \&
Schneider 1990).  Further, at least six RBLs show possible evidence of lensing by a
foreground galaxy (PKS 0138--097; Heidt et al. 1996, S4 0218+357; O'Dea et al. 1992, AO
0235+164; Stickel et al. 1988a, PKS 0426--380; SFK93, PKS 0537--441; Stickel et al.
1998b and B2 1308+326; Stickel 1991); and three RBLs have extended radio morphologies
that are unusual enough to be suggestive of gravitational lensing (S4 0814+425; M93, B2
1308+326; M93 and PKS 2131--021; this paper).  However, for some of these sources there
also is evidence against the microlensing scenario (e.g. AO 0235+164; Abraham et al.
1993, Chu et al. 1996 and PKS 0537--441; Falomo et al. 1992, Lewis \& Ibata 2000).

The presence of lensed sources does not explain the \vvmax\ discrepancy between
XBLs and RBLs, as removing all suspected microlensing candidates (i.e., RBLs with
intervening absorption systems and/or evidence of a foreground galaxy) from the 1Jy
sample does not significantly change the \vvmax\ value (\vvmax\ $= 0.610 \pm
0.08$).  Thus, while gravitational lensing might be a mechanism for
introducing a few BL Lacs into the 1Jy sample, the frequency of sources so affected is
unknown and probably small. 
High-resolution radio and optical imaging of high-$z$ RBLs in the 1Jy sample, which
is currently underway, will address this issue further.

\section{Conclusions}

We have presented new optical and near-IR spectroscopy as well as new high dynamic range,
arcsecond-resolution VLA maps of BL Lacs from the updated 1Jy RBL sample of S91. 
Redshift information is now available for all but six of the 37 BL Lacs
in the sample.  An additional four have only a lower limit on redshift due to
absorption systems; and four more have only a tentative redshift based upon the
detection a single emission line.  All but three of the RBLs have been observed with
the VLA at 20cm in A-array to a high dynamic range ($> 1000:1$); and extended radio flux
has been detected in all but three of those observed.  Thus accurate arcsecond-scale,
20cm extended radio luminosities, or upper limits, are now available for nearly all of
the sample.

While further observations are still necessary to determine or confirm emission-line
redshifts for fourteen RBLs, spectral information for the 1Jy sample is largely
complete.  With this new redshift information the \vvmax\ for the 1Jy sample is
raised slightly to $0.614\pm0.047$, which is 4$\sigma$ higher than the complete EMSS
XBL sample (\vvmax\ $= 0.399\pm 0.057$ for the M91 sample; R00).  Thus the \vvmax\
discrepancy for these two samples cannot be explained as a result of incomplete redshift
information.  Combined with the results of R00, a correlation between \fxfr\ and \vvmax\
is observed, from negative evolution in the more extreme HBLs to positive evolution in
the more extreme LBLs.  However, this correlation should be confirmed in a
uniformly-selected BL Lac sample which spans the entire range of observed \fxfr\ values
(e.g., the ROSAT-Green Bank sample; Laurent-Muehleisen et al. 1999).   

Several other important differences are also found in the properties of RBLs and
XBLs.  In the unified model the, presumably unbeamed and unobscured, emission-line and
extended radio luminosities should be similar for XBLs, RBLs and FR--1s.  However,
whereas XBLs and FR--1s show weak or no emission lines, most RBLs possess
moderately luminous emission lines.  The lines observed in RBLs are statistically less
luminous than FSRQs; however there is significant overlap in the luminosity distributions
of the two classes.  Over half of the 1Jy RBLs have extended radio powers too
luminous to be beamed FR--1 radio galaxies, which disagrees with the unification model
in its simplest form.  This differs  from XBLs, all of which have extended power levels
consistent with FR--1s (R00).  The extended radio luminosities of RBLs are also
statistically lower than FSRQs.  These major observational differences
strongly suggest that RBLs and XBLs are not simply related, and that RBLs are more
closely related to FSRQs; although an explanation for why RBLs statistically have
lower-luminosity emission-lines and lower-power extended radio lobes than FSRQs is
necessary.  One possibility is that of selection effects due to redshift, wherein many
RBLs are simply FSRQs at too low of redshift ($z < 1$) for their strong UV lines to be
seen into the optical.

Unlike the EMSS XBL sample which is almost entirely consistent with a beamed
FR--1 population, the 1Jy sample is a heterogeneous sample in which we find evidence
for, and examples of, three distinct mechanisms for creating the BL Lac phenomenon:
beamed FR--1s, beamed FR--2s and possibly a few gravitationally-lensed quasars. 
More than half of the BL Lacs in the 1Jy sample have extended radio powers and optical
emission line luminosities which are FR--2-like.  The number of gravitationally lensed
sources in the 1Jy sample is not known but likely small; and the microlensed sources are
most likely amongst the FR--2-like RBLs.  But overall, the number of FR--2-like RBLs in
the 1Jy sample can be readily understood as a subclass of beamed FR--2s.  The presence of
FR--2s and gravitationally lensed objects in the 1Jy sample does not explain the
\vvmax\ or the radio core-dominance discrepancy between XBLs and RBLs.  Nor can the
1Jy sample be simply described as beamed FR--1, XBL-like objects contaminated with a
population of high-$z$ FR--2s with low-\wlam\ optical emission lines.  The different
means of producing BL Lac-like objects seen in the 1Jy sample therefore exclude simple
unification models such as the orientation model.  
Because the 1Jy is a very shallow survey, its BL Lac objects may be too small in number
to allow firm conclusions to be drawn about the full RBL population.
Deeper samples of RBLs are required to address the questions
raised but not answered by this work.

\acknowledgments

Research on BL Lac objects at the University of Colorado was supported by NASA grant
NAGW-2675.

\begin{figure}
\plottwo{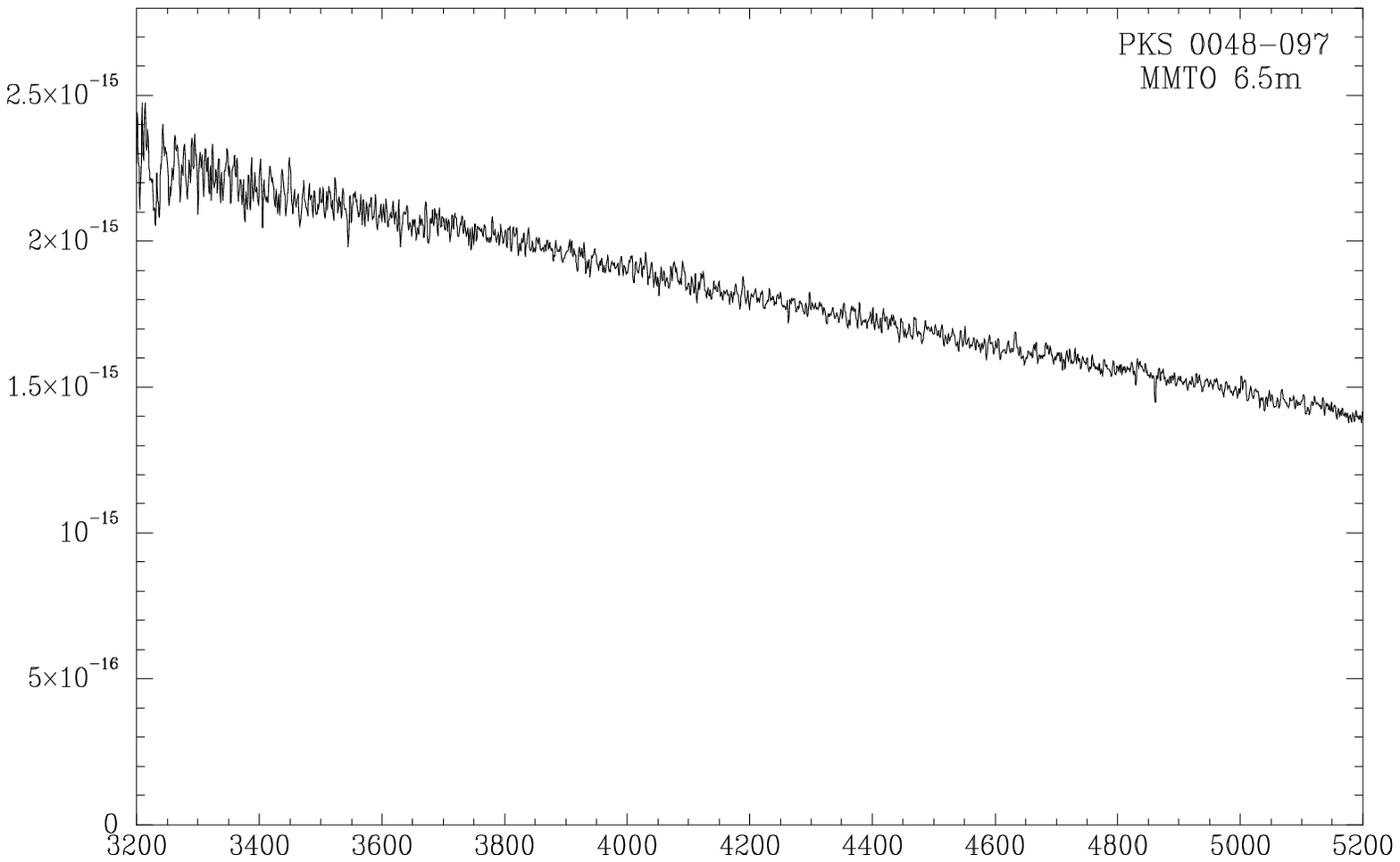}{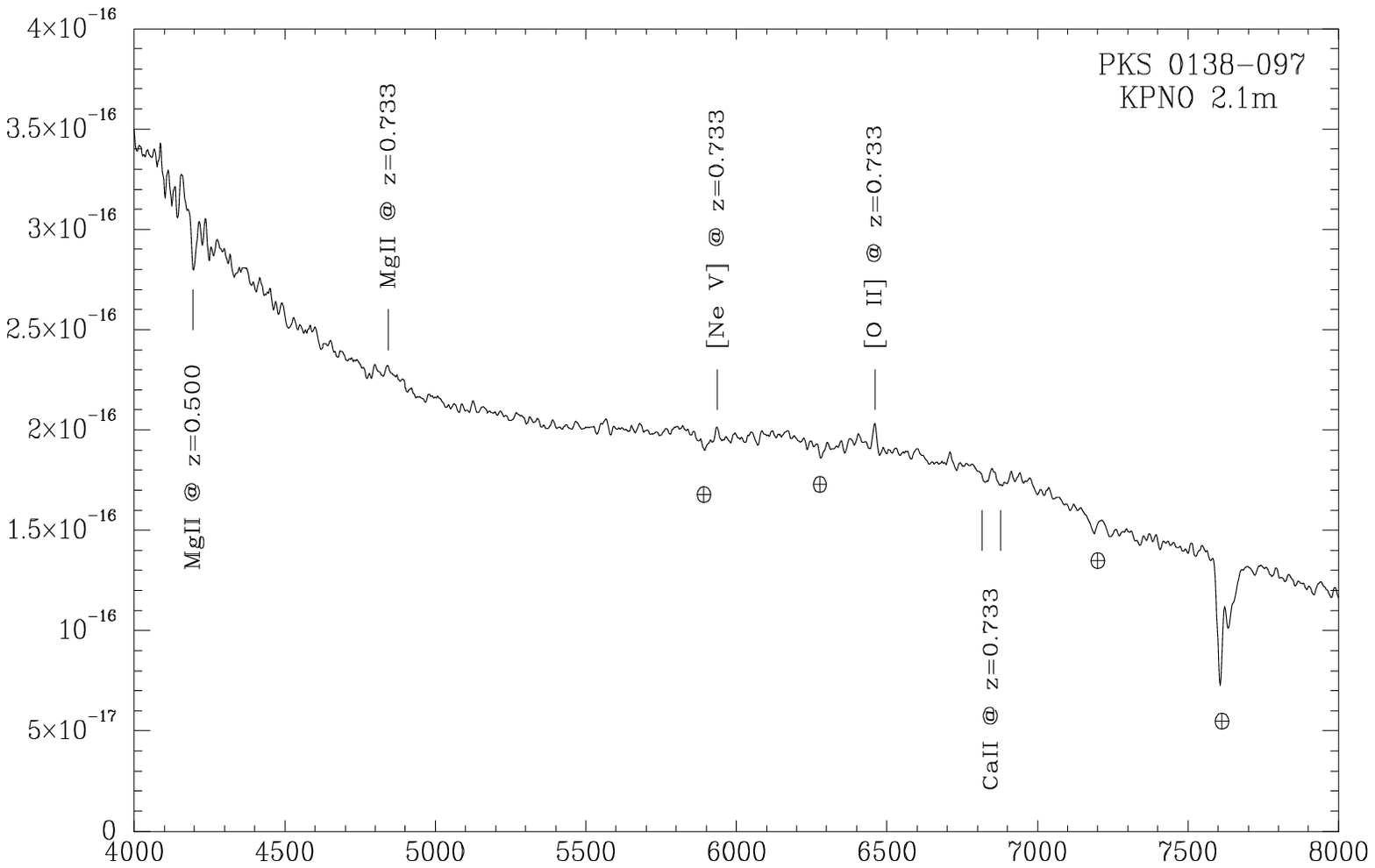}
\end{figure}
\begin{figure}
\plottwo{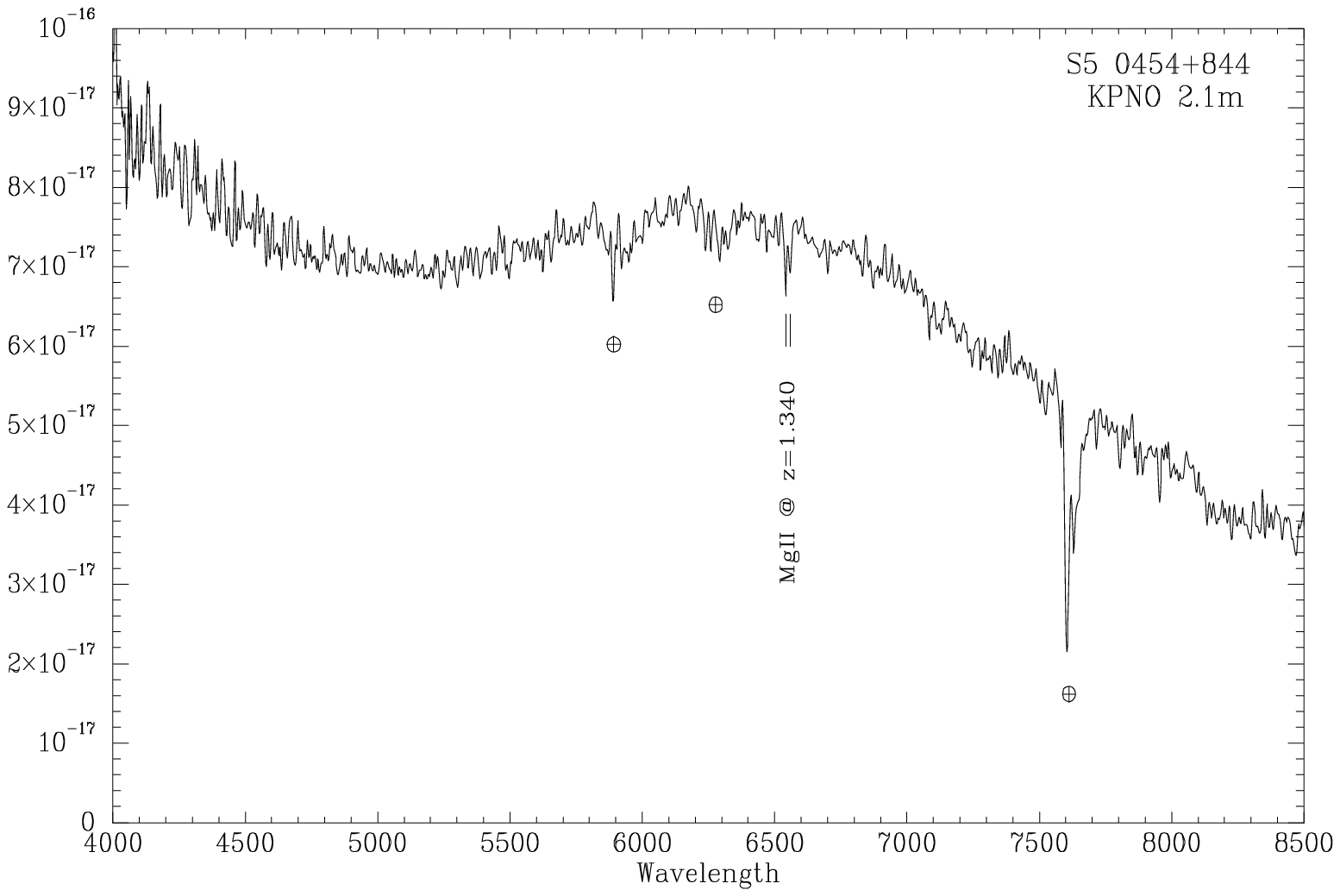}{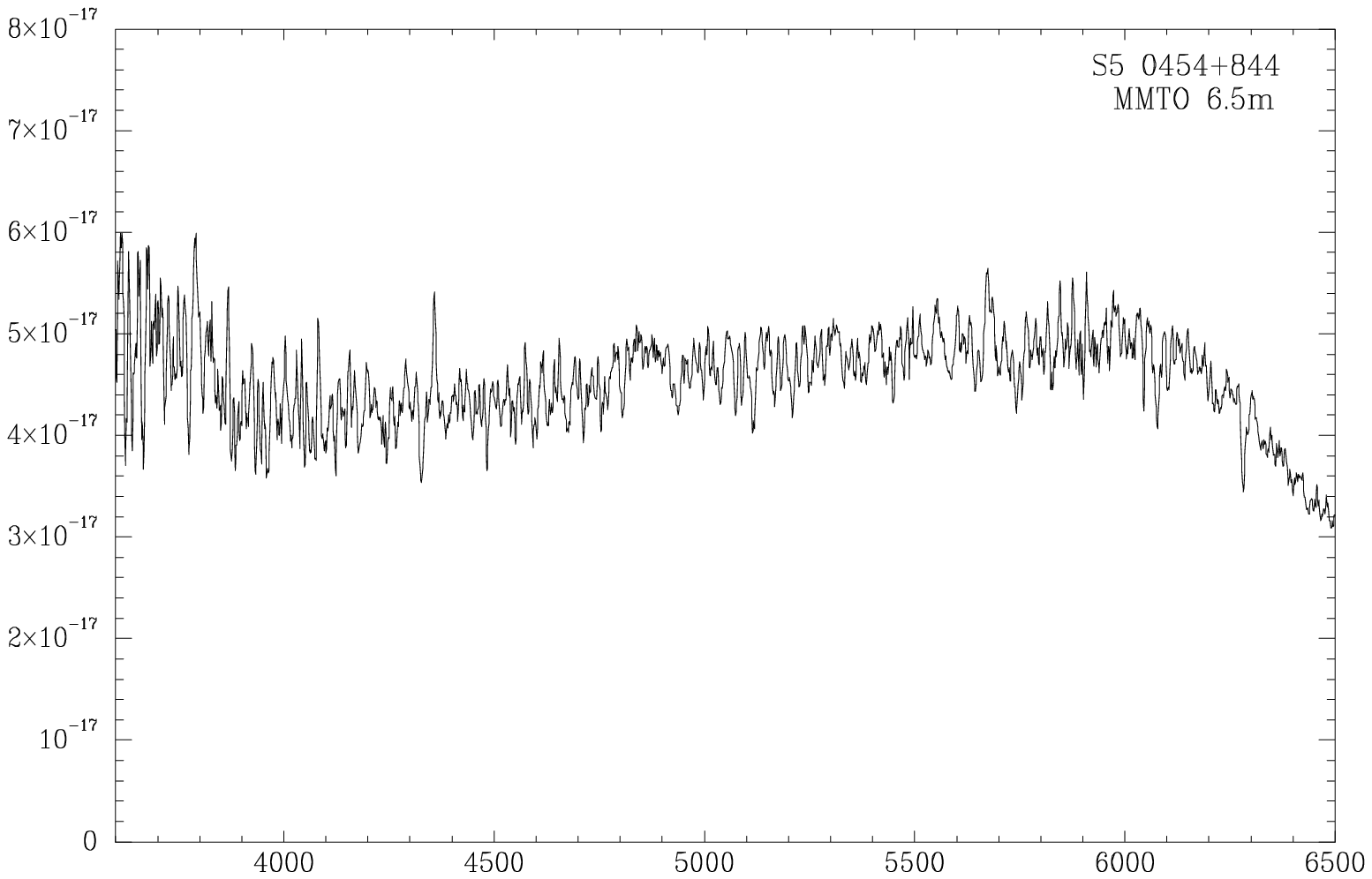}
\end{figure}

\begin{figure}
\plottwo{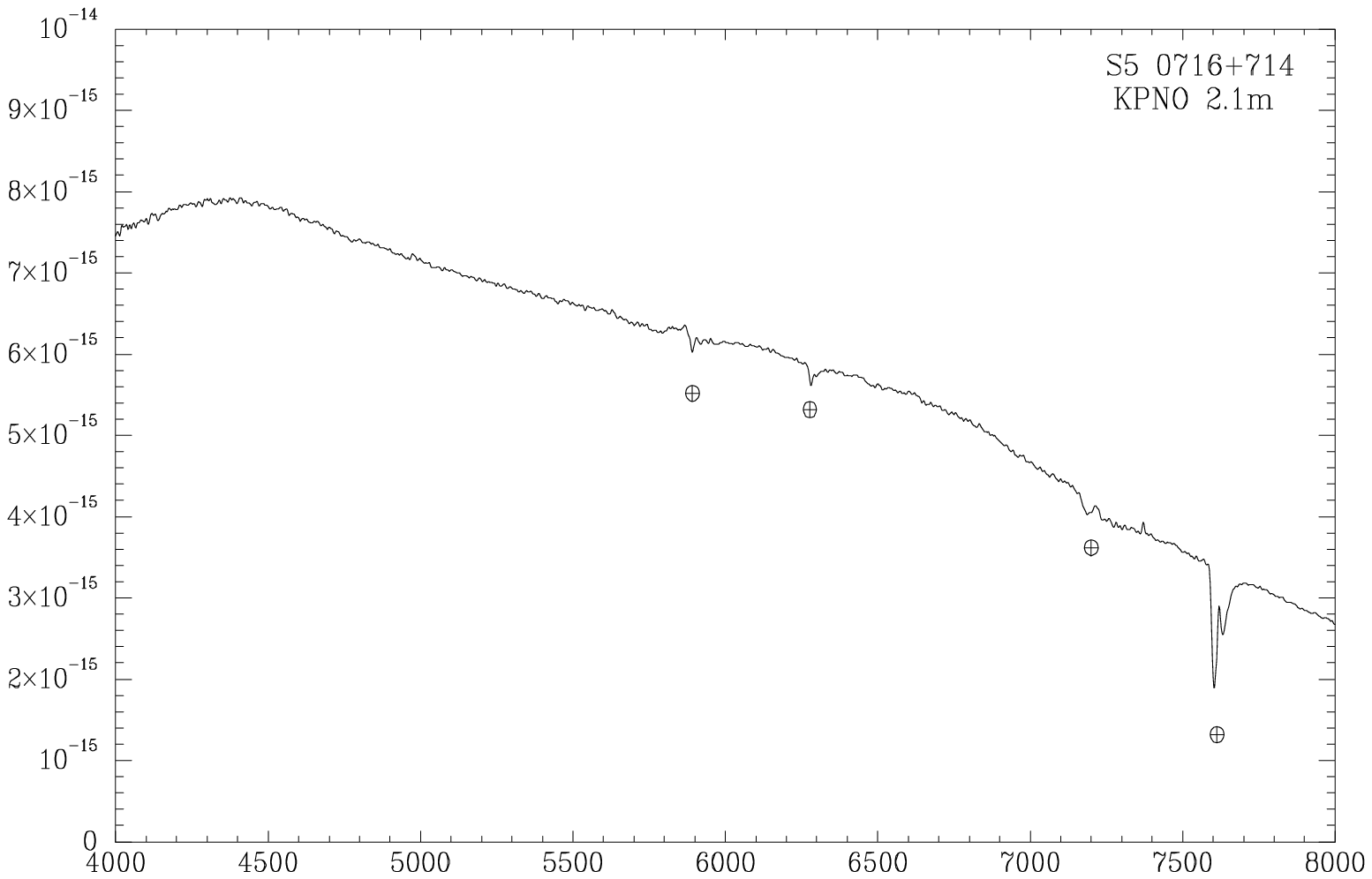}{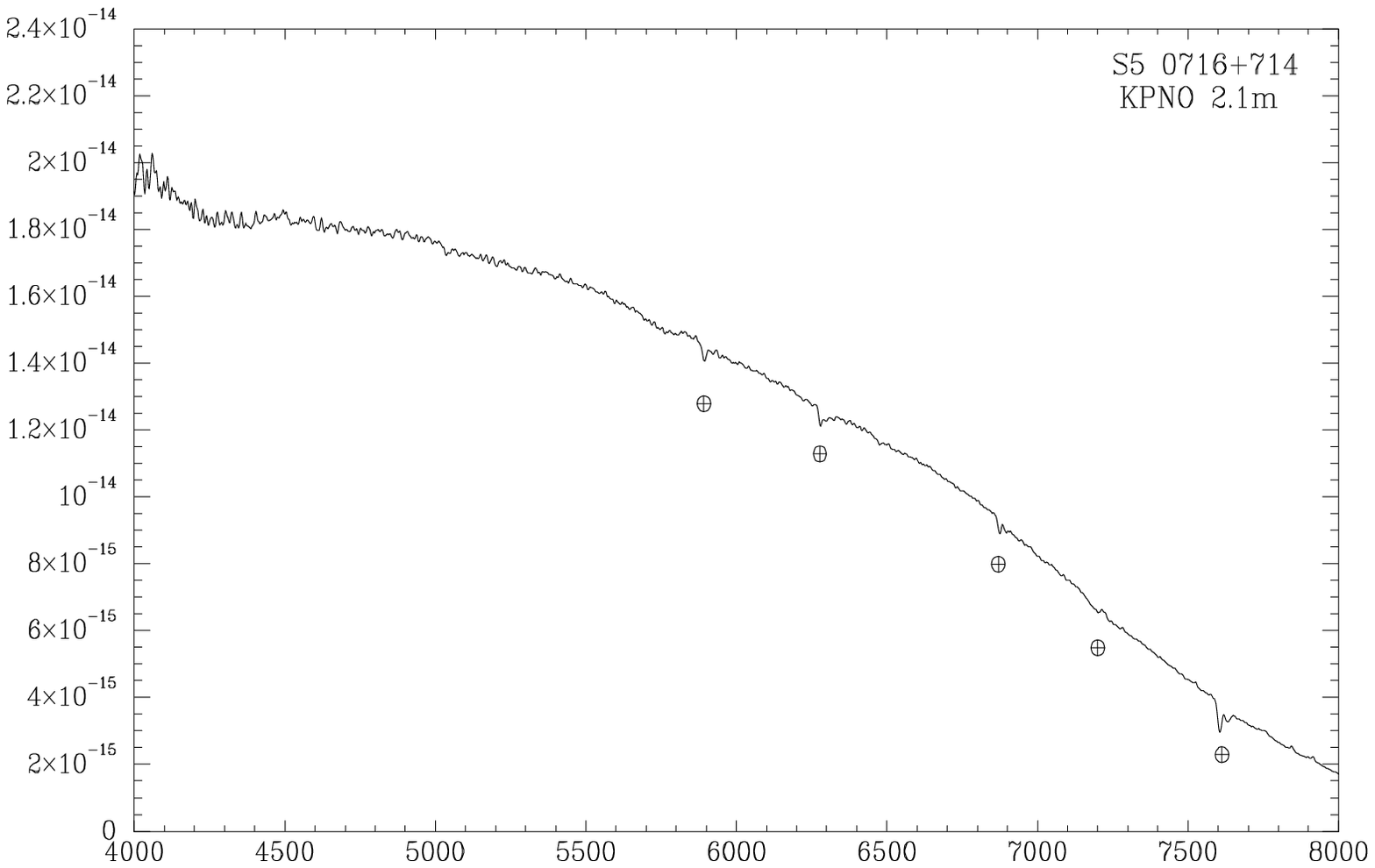}
\end{figure}
\begin{figure}
\plottwo{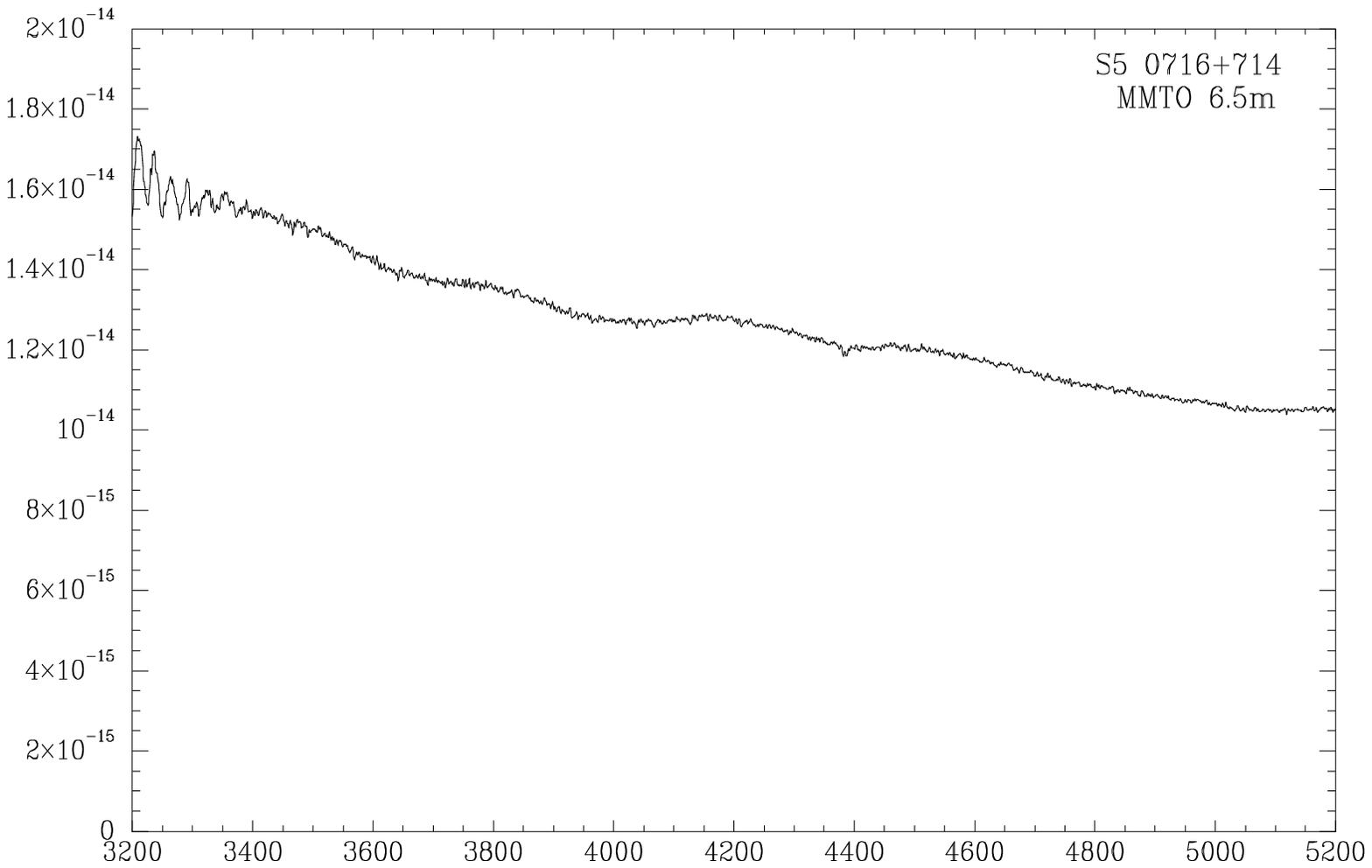}{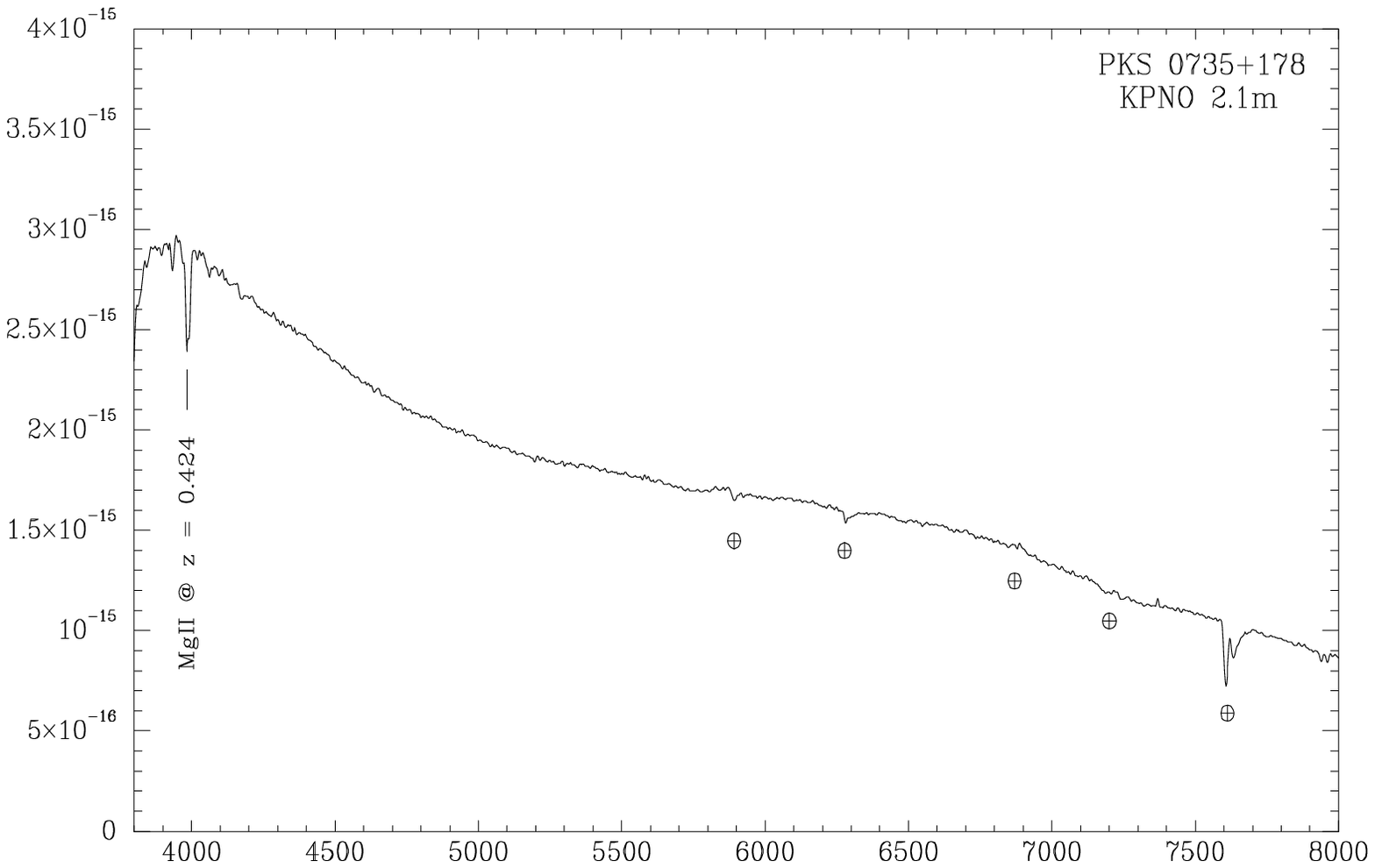}
\end{figure}
\begin{figure}
\plottwo{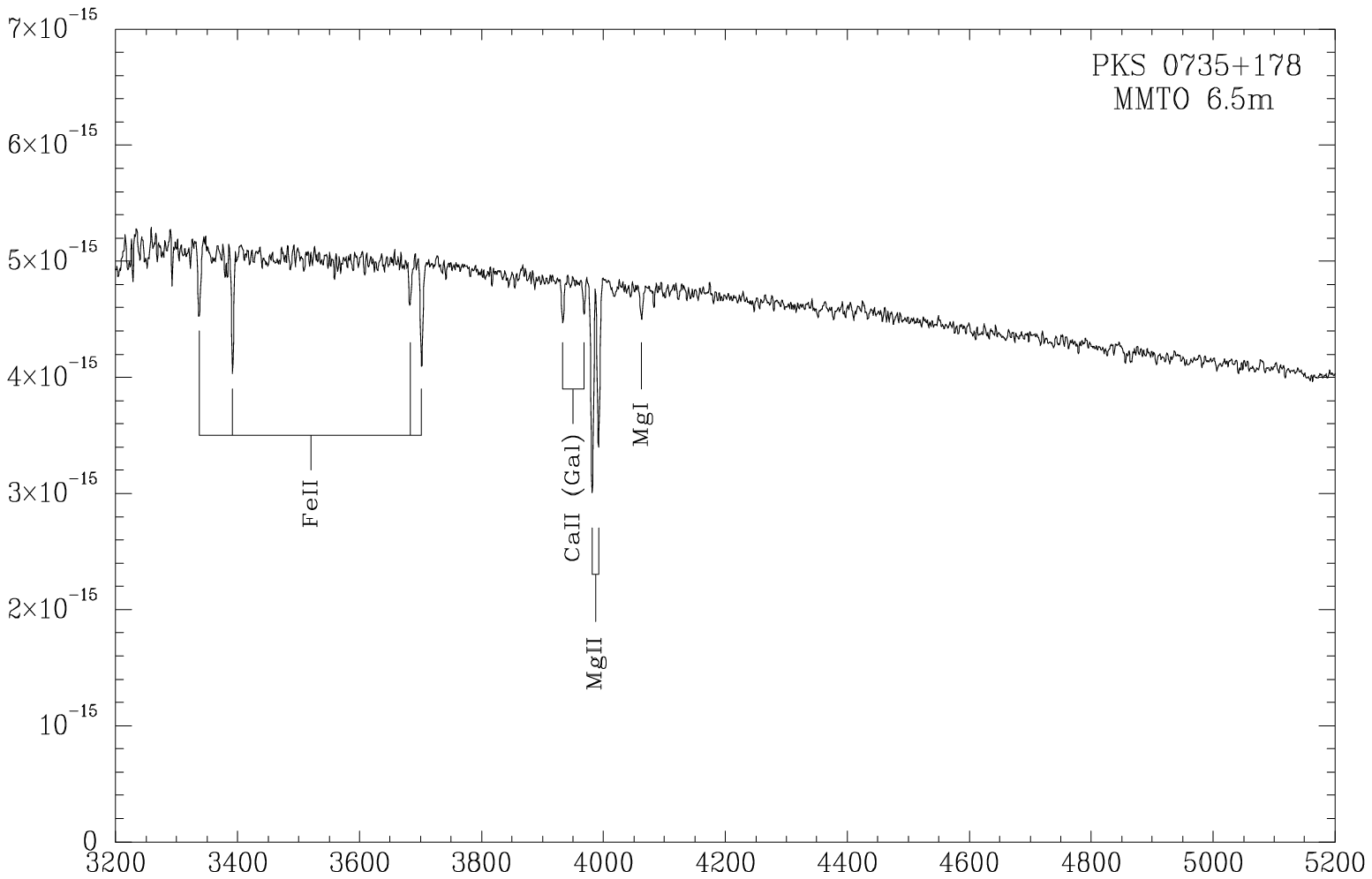}{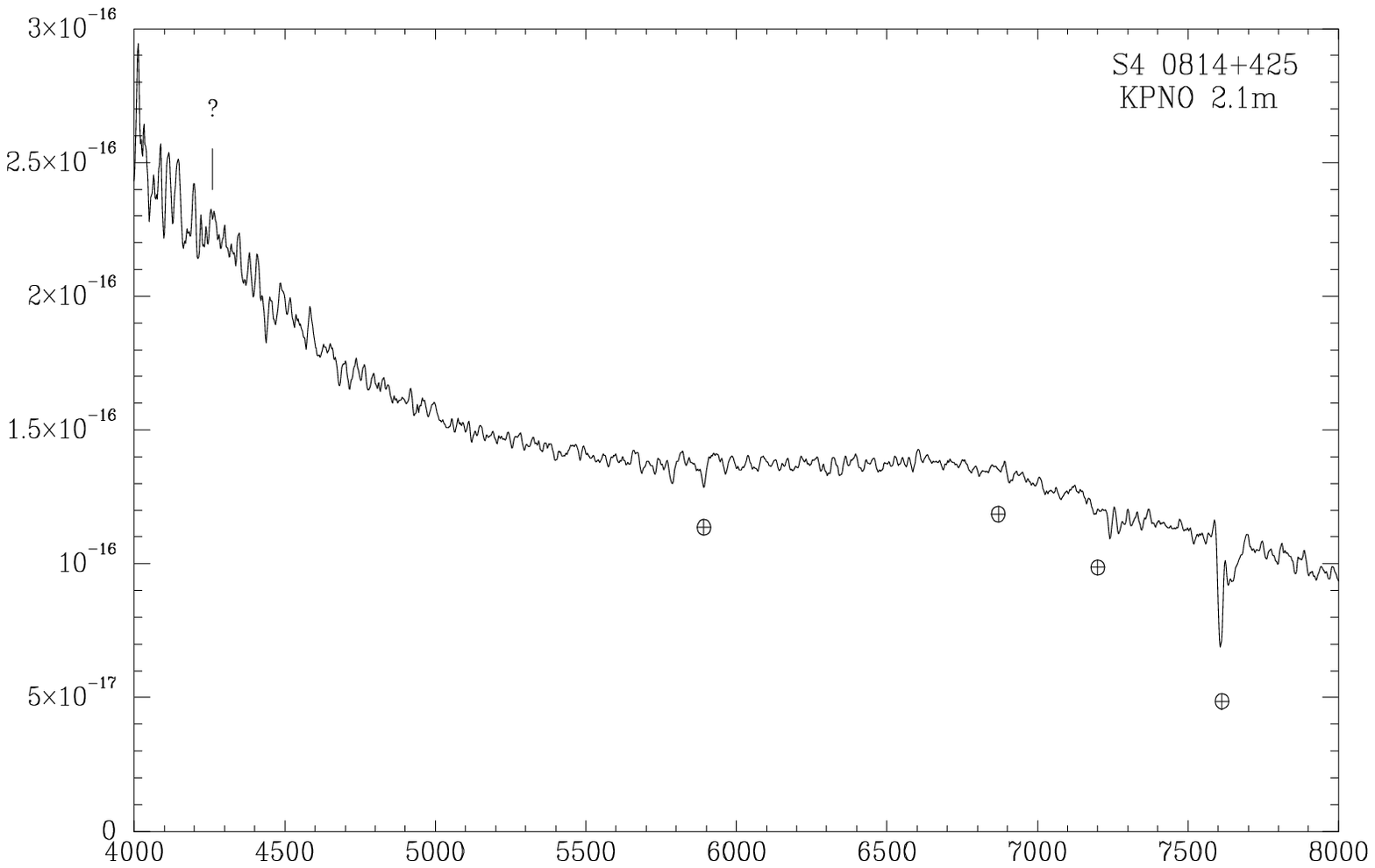}
\end{figure}
\begin{figure}
\plottwo{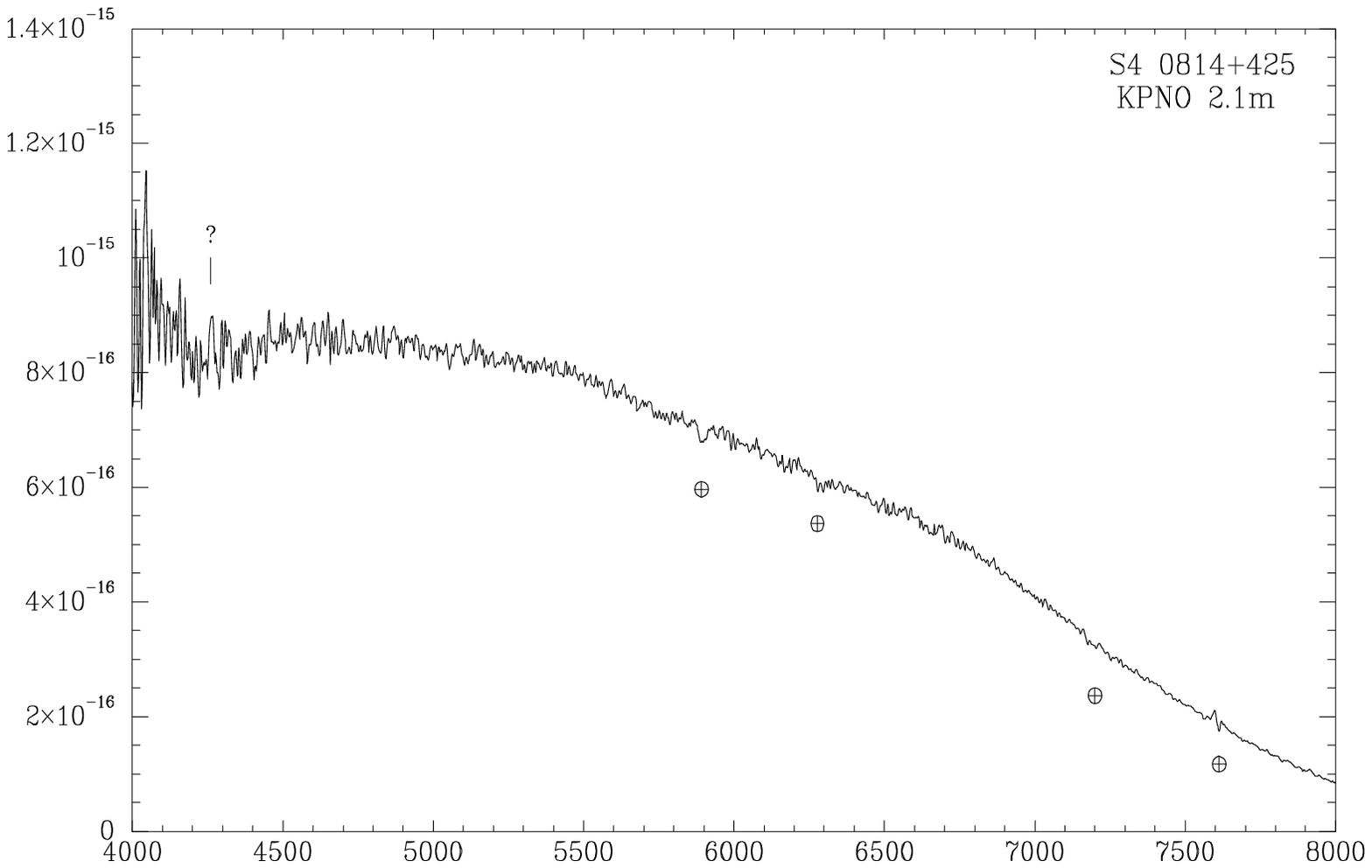}{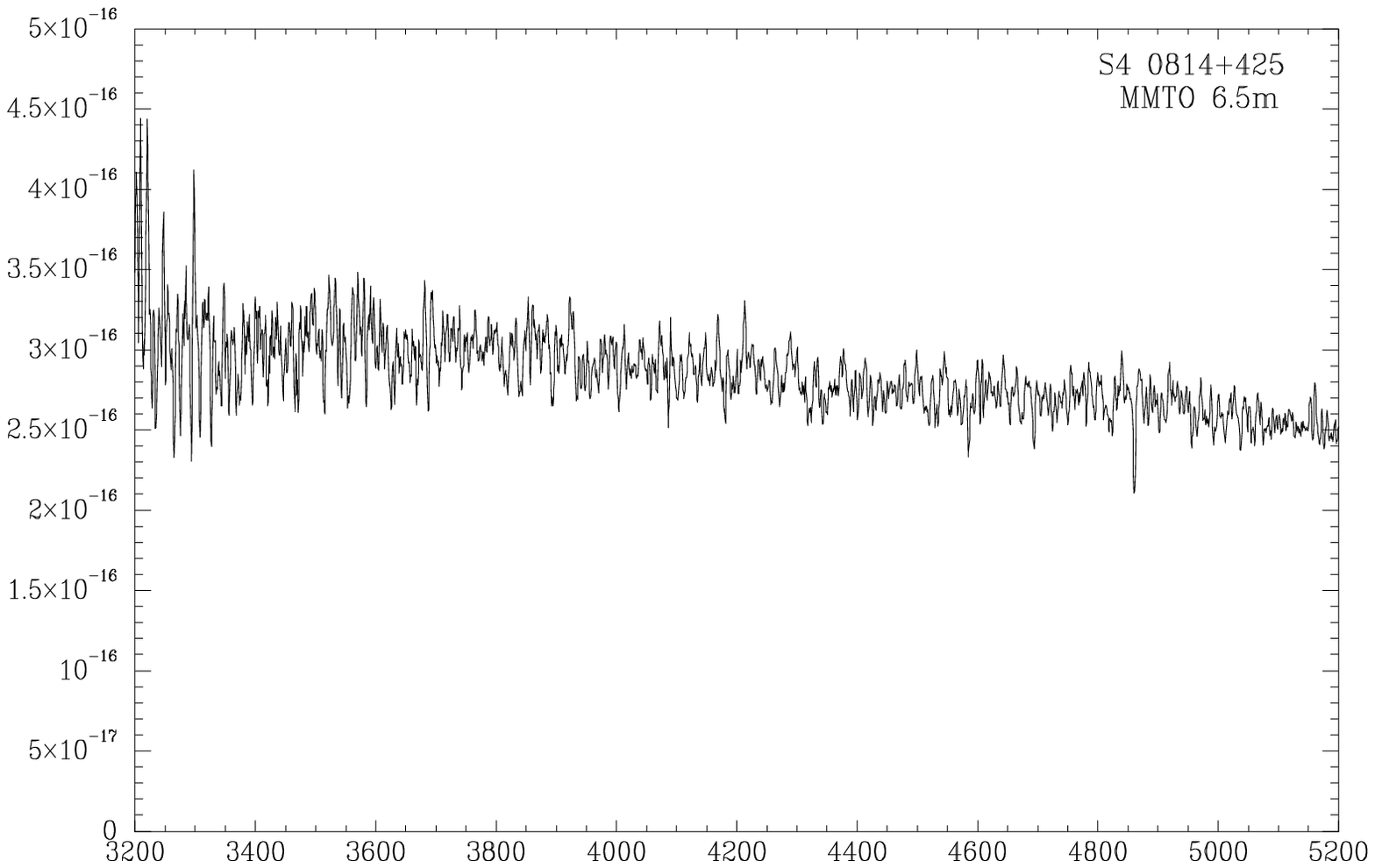}
\end{figure}

\begin{figure}
\plottwo{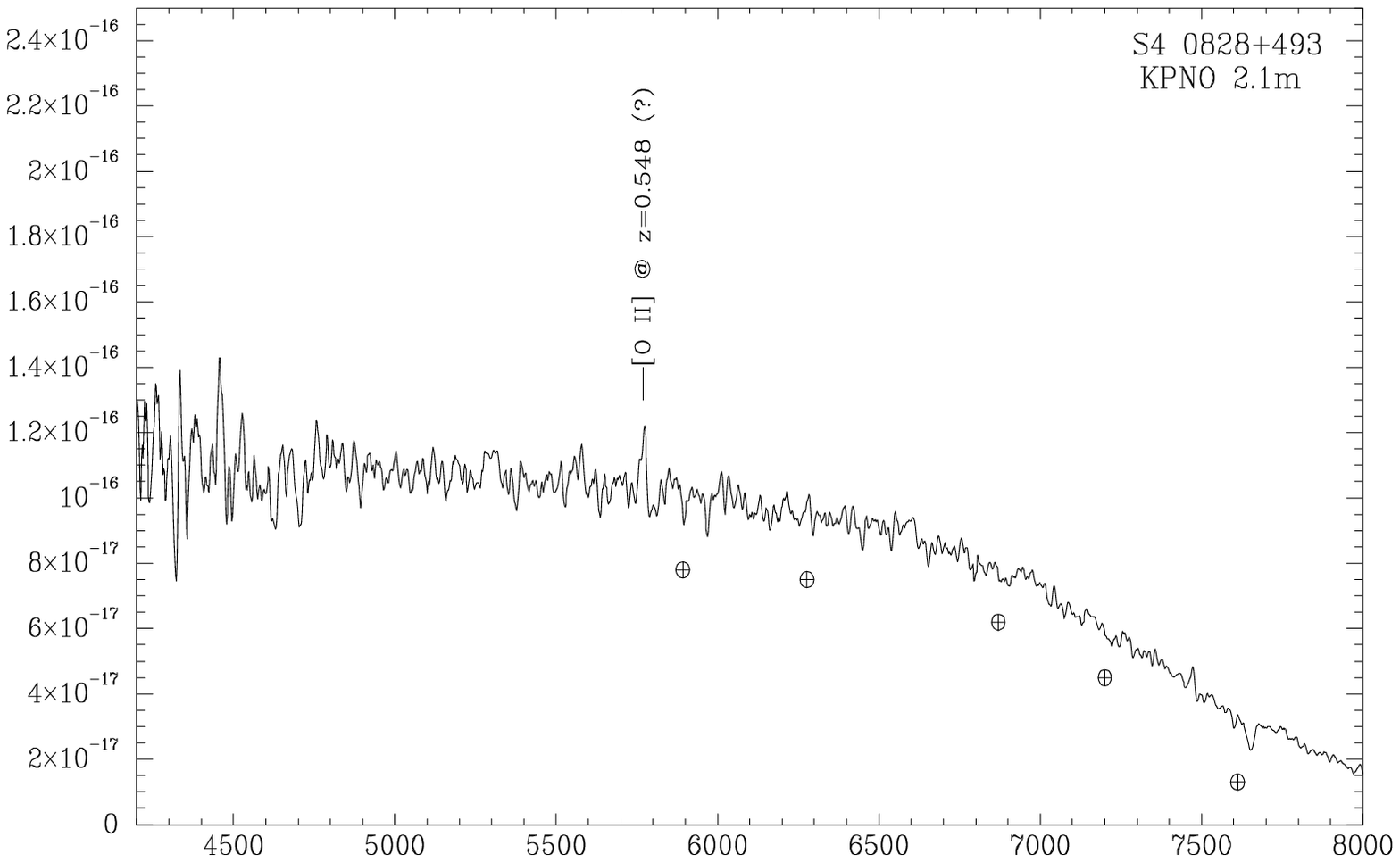}{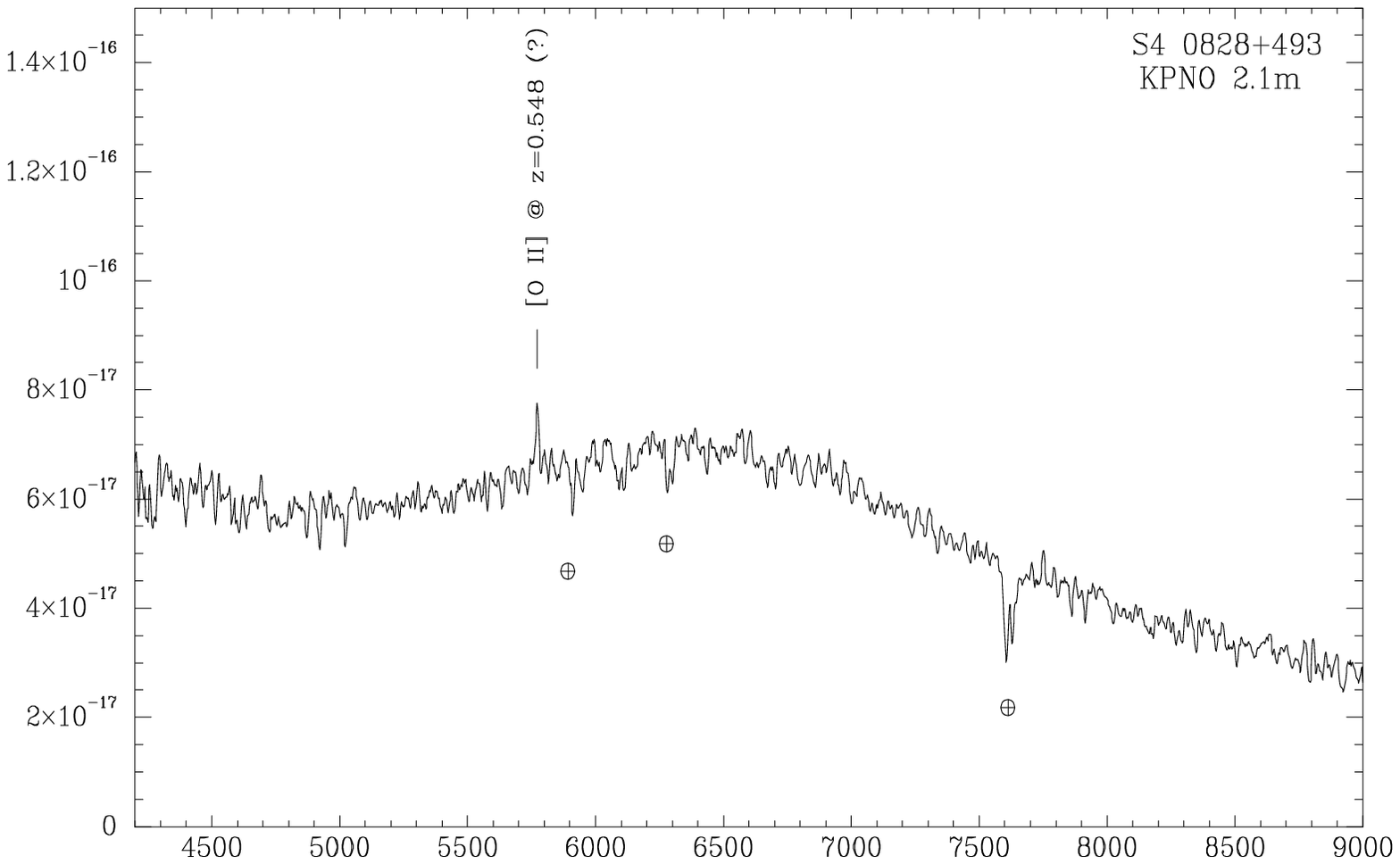}
\end{figure}
\begin{figure}
\plottwo{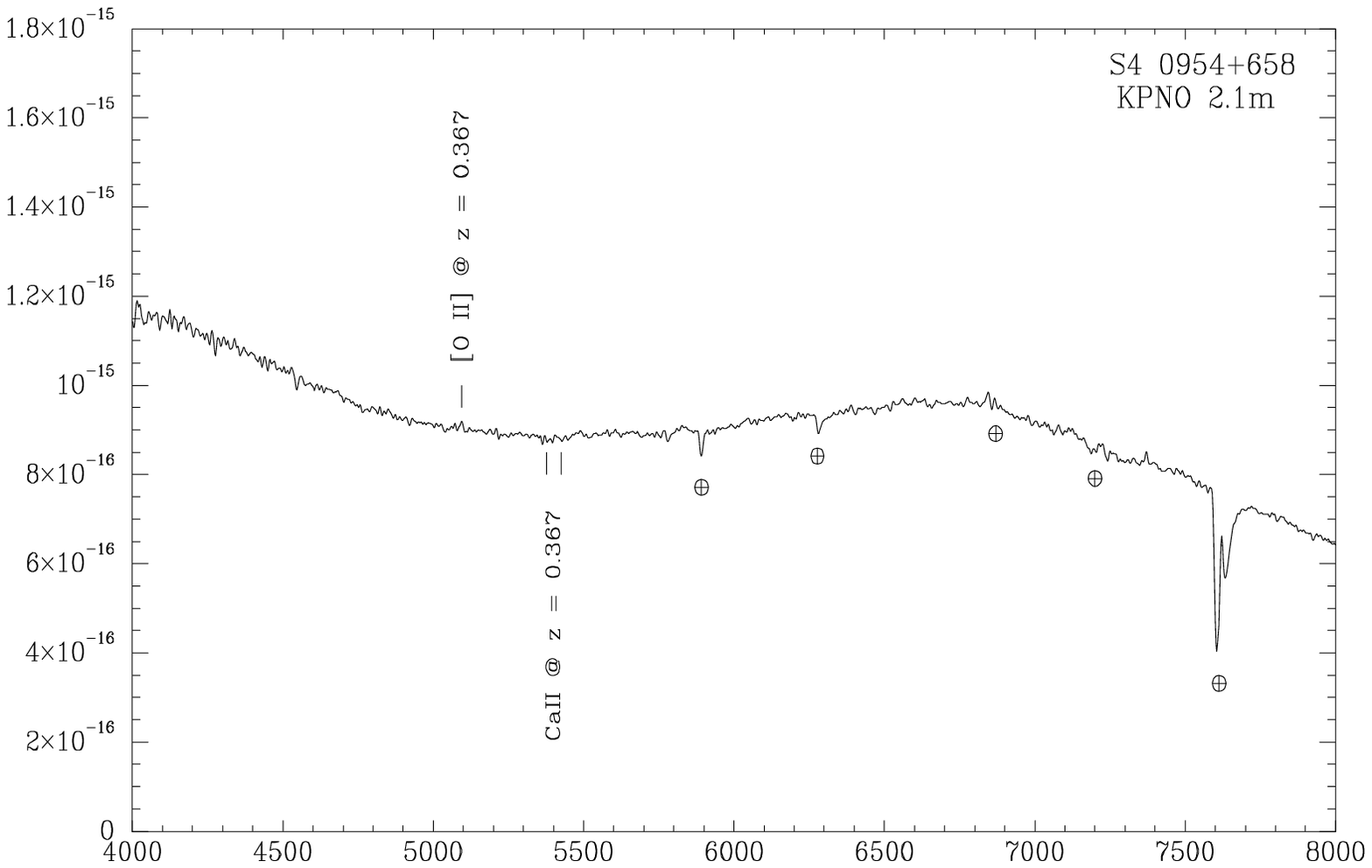}{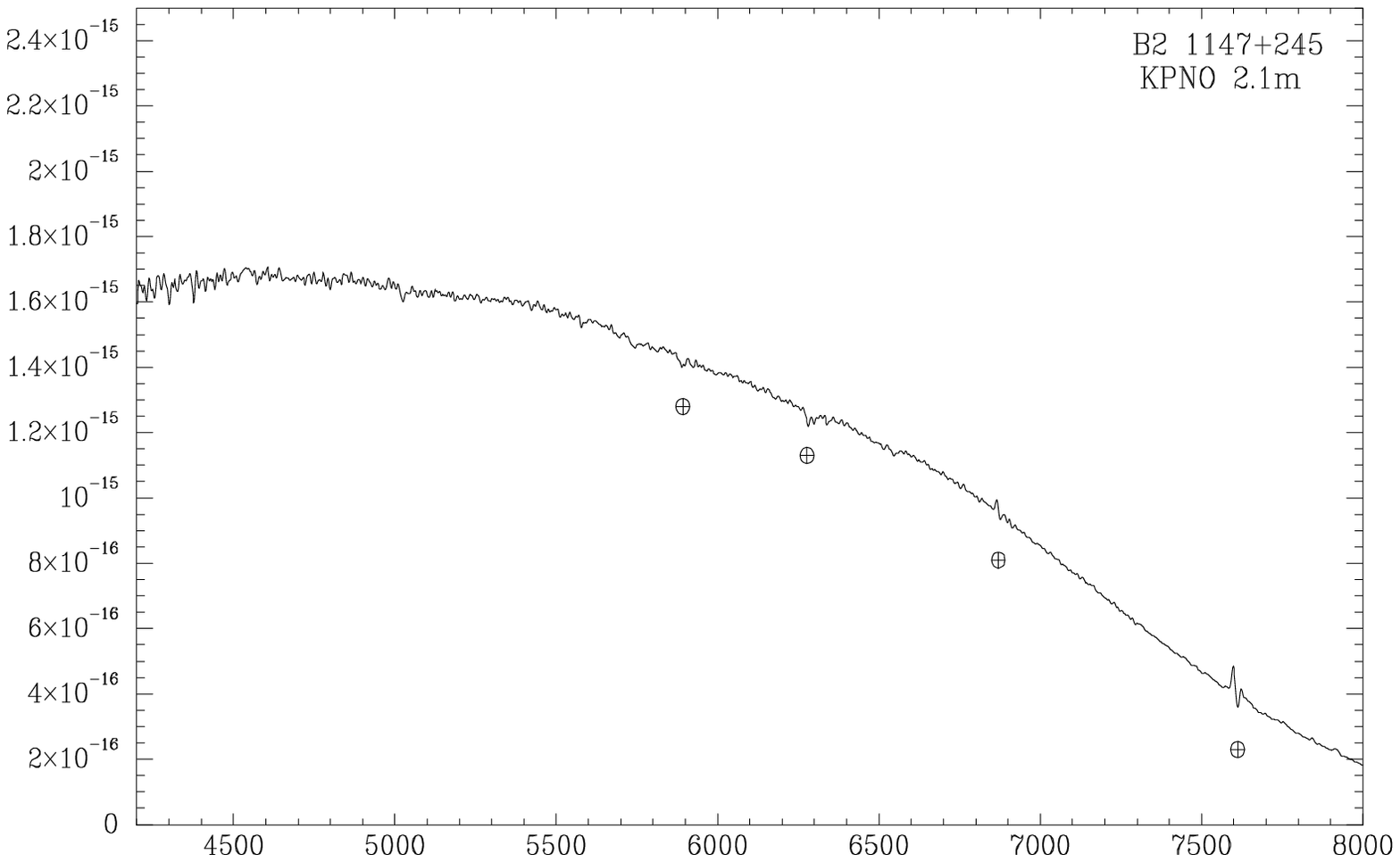}
\end{figure}
\begin{figure}
\plottwo{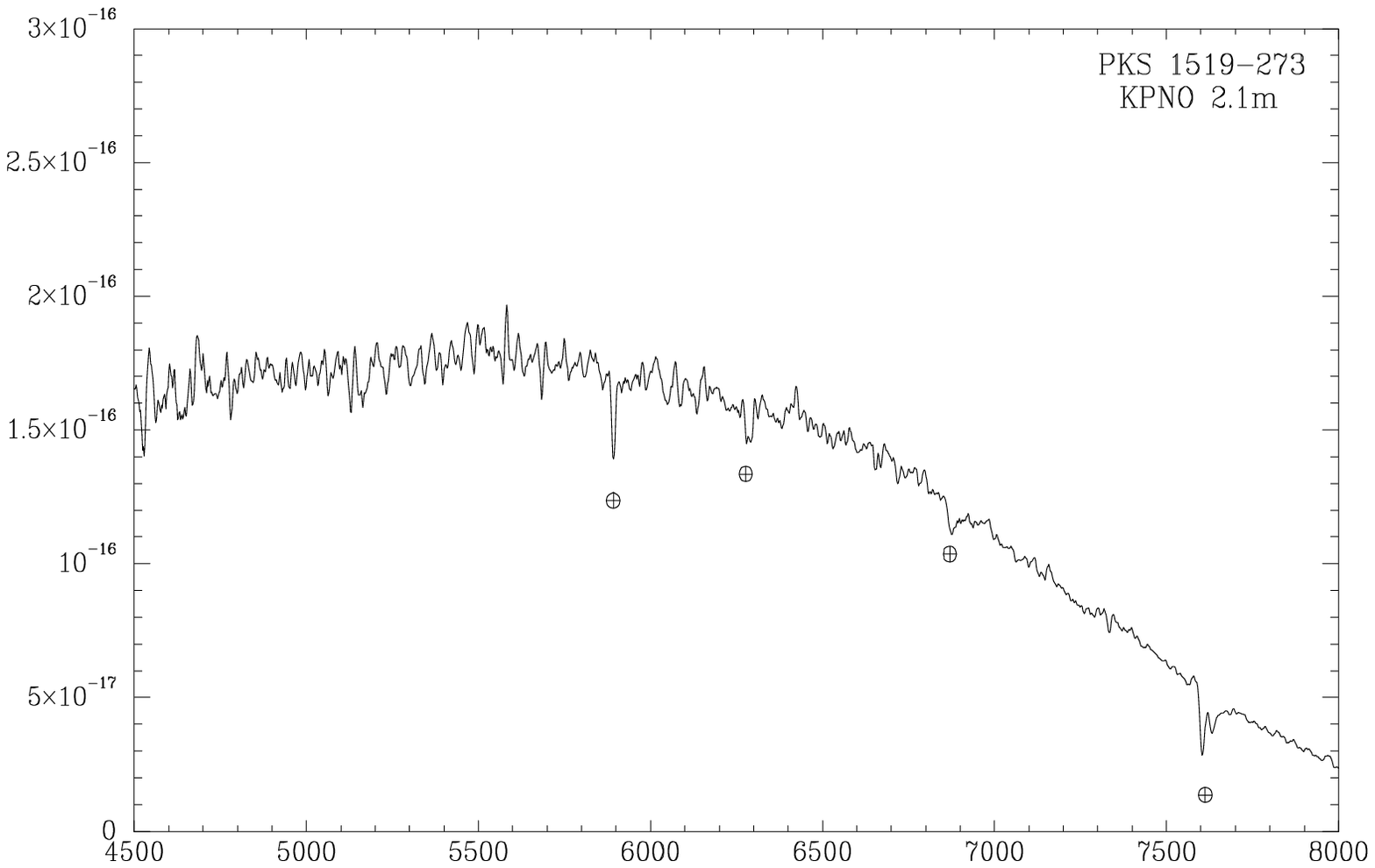}{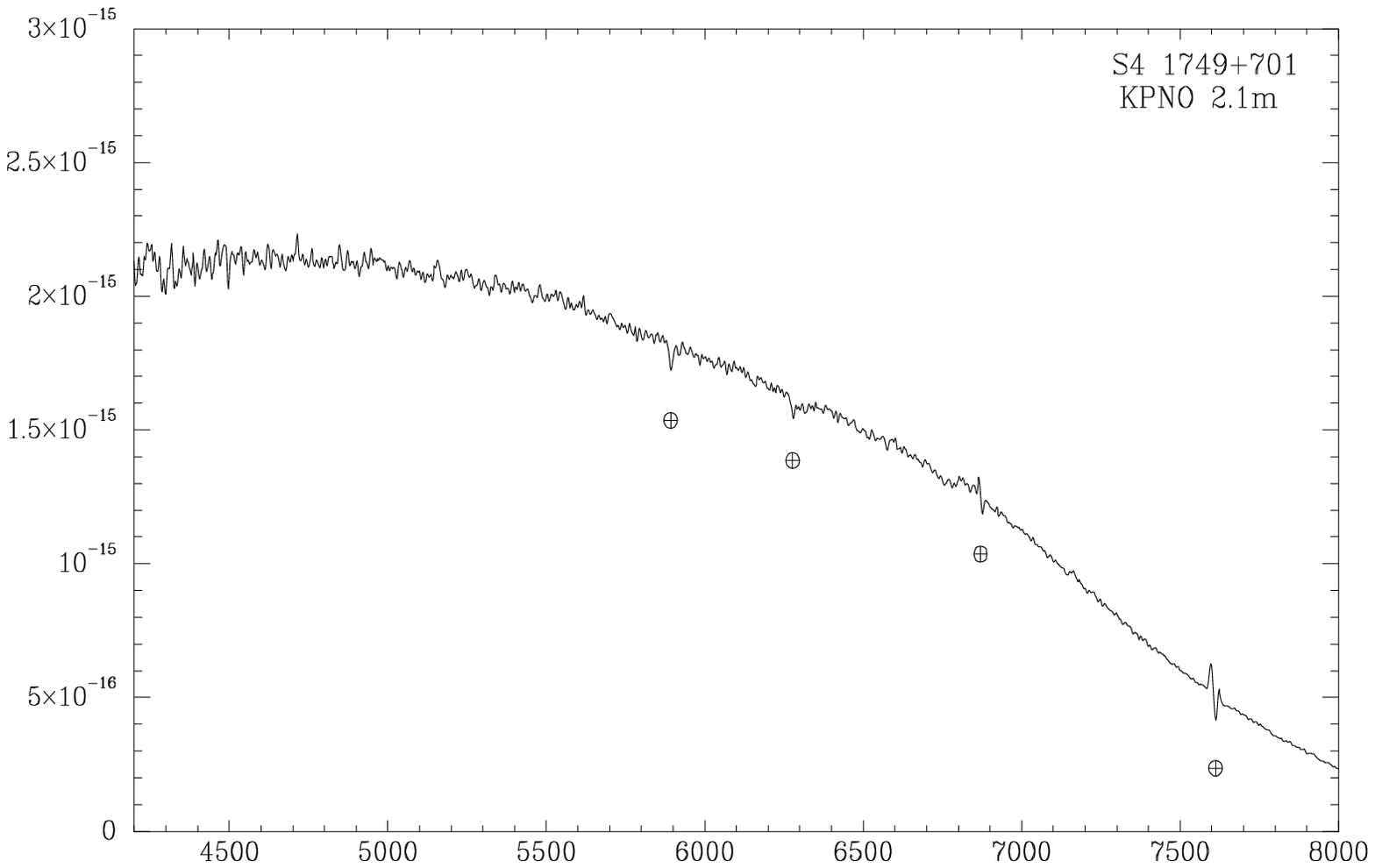}
\end{figure}
\begin{figure}
\plottwo{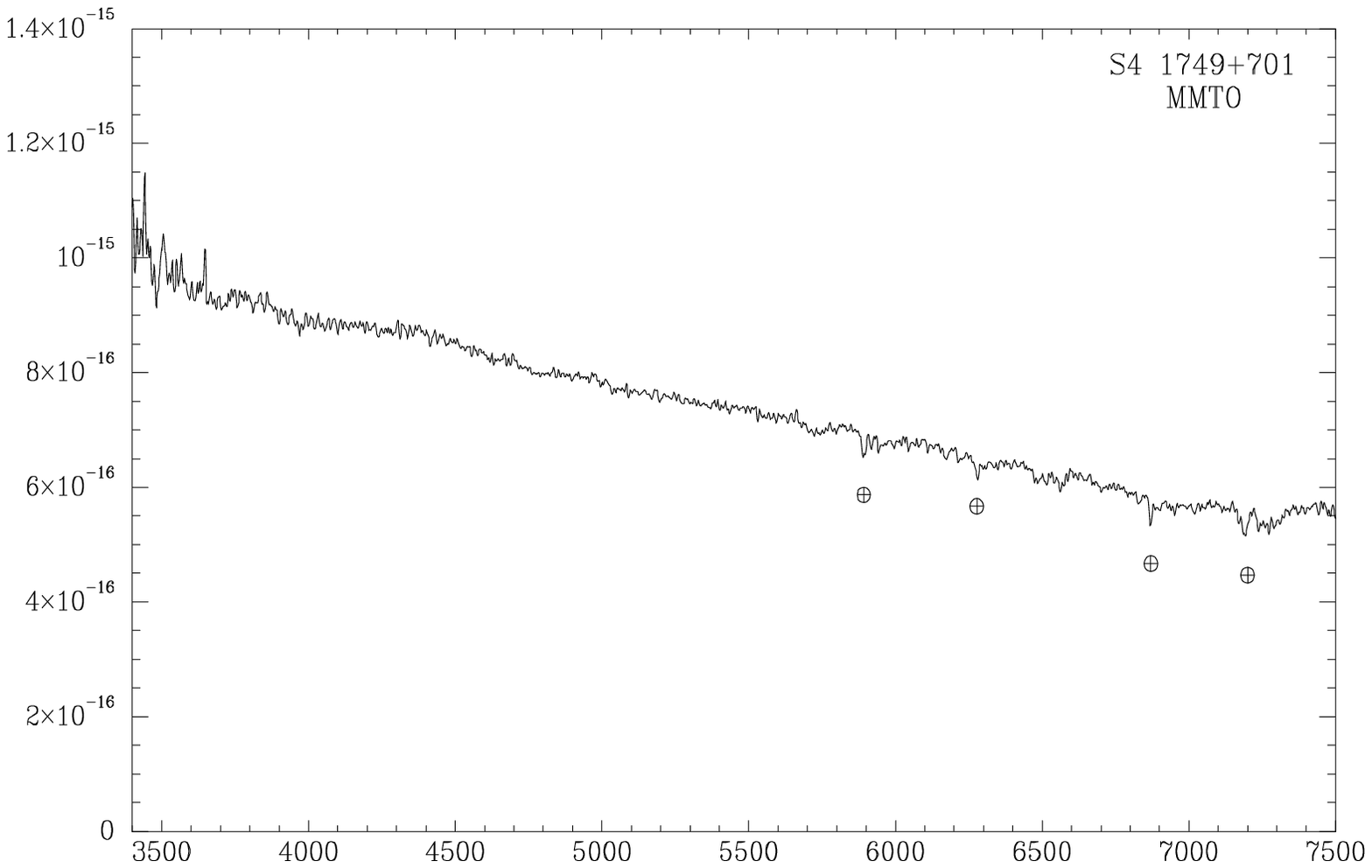}{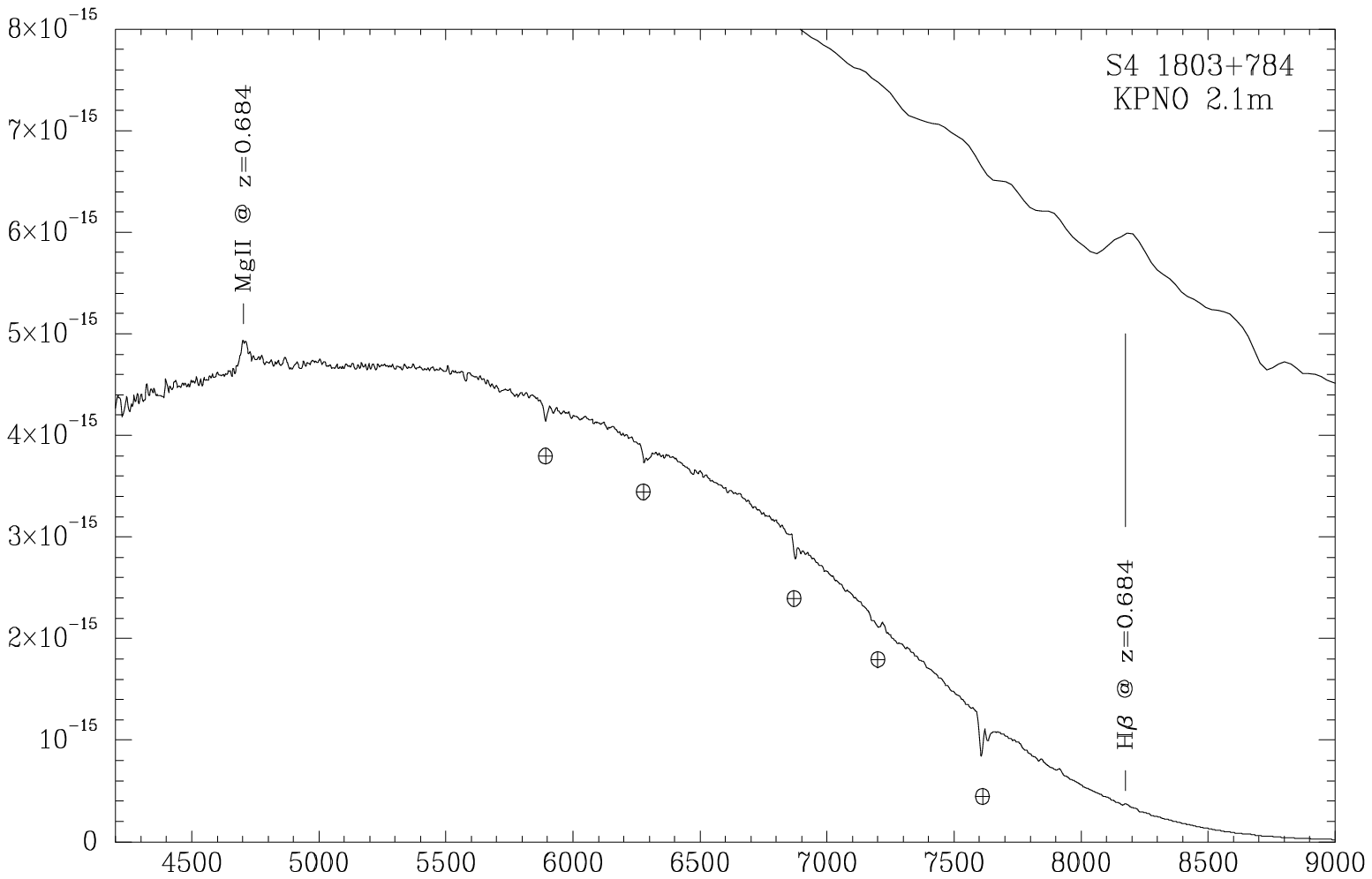}
\end{figure}

\begin{figure}
\plottwo{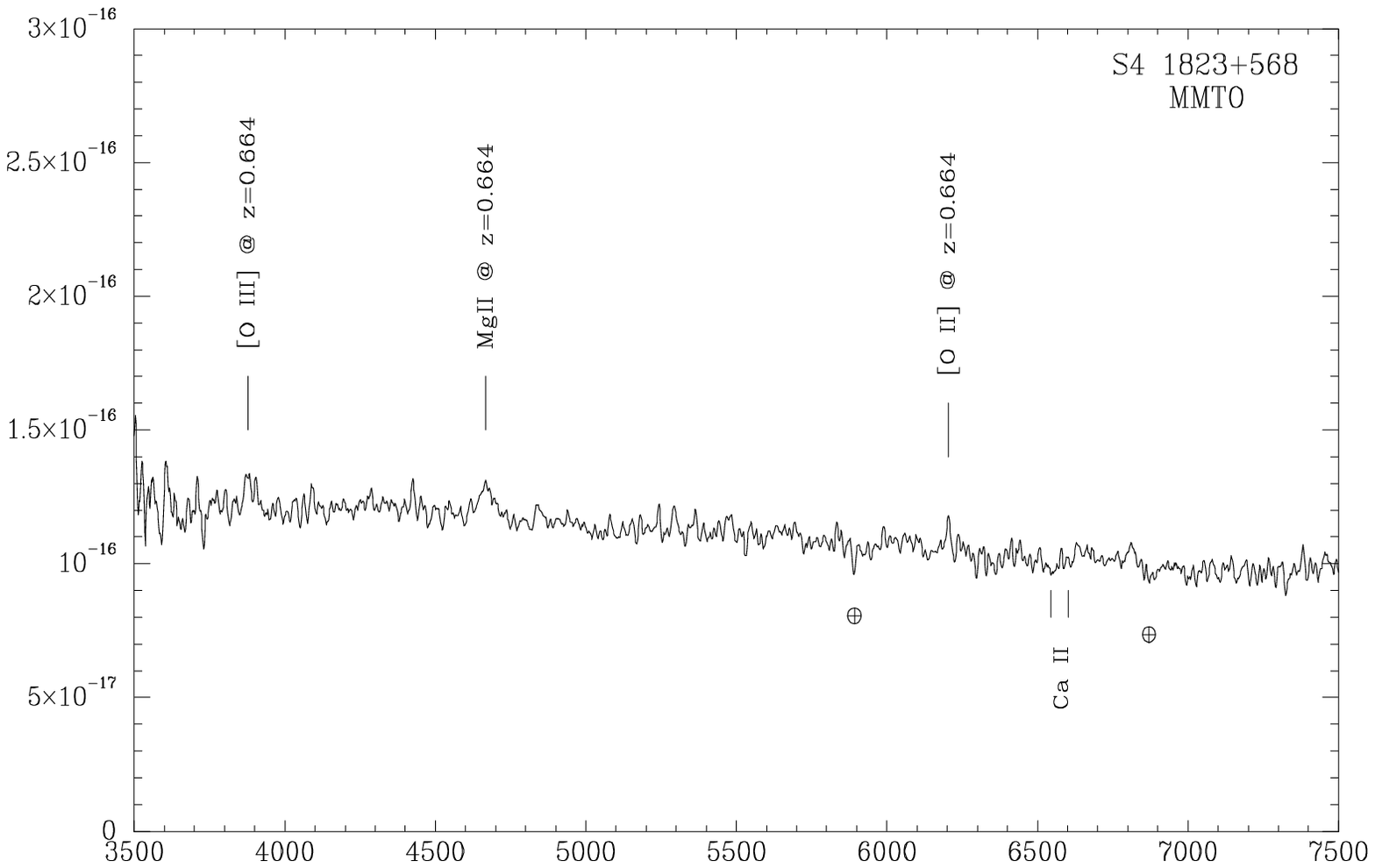}{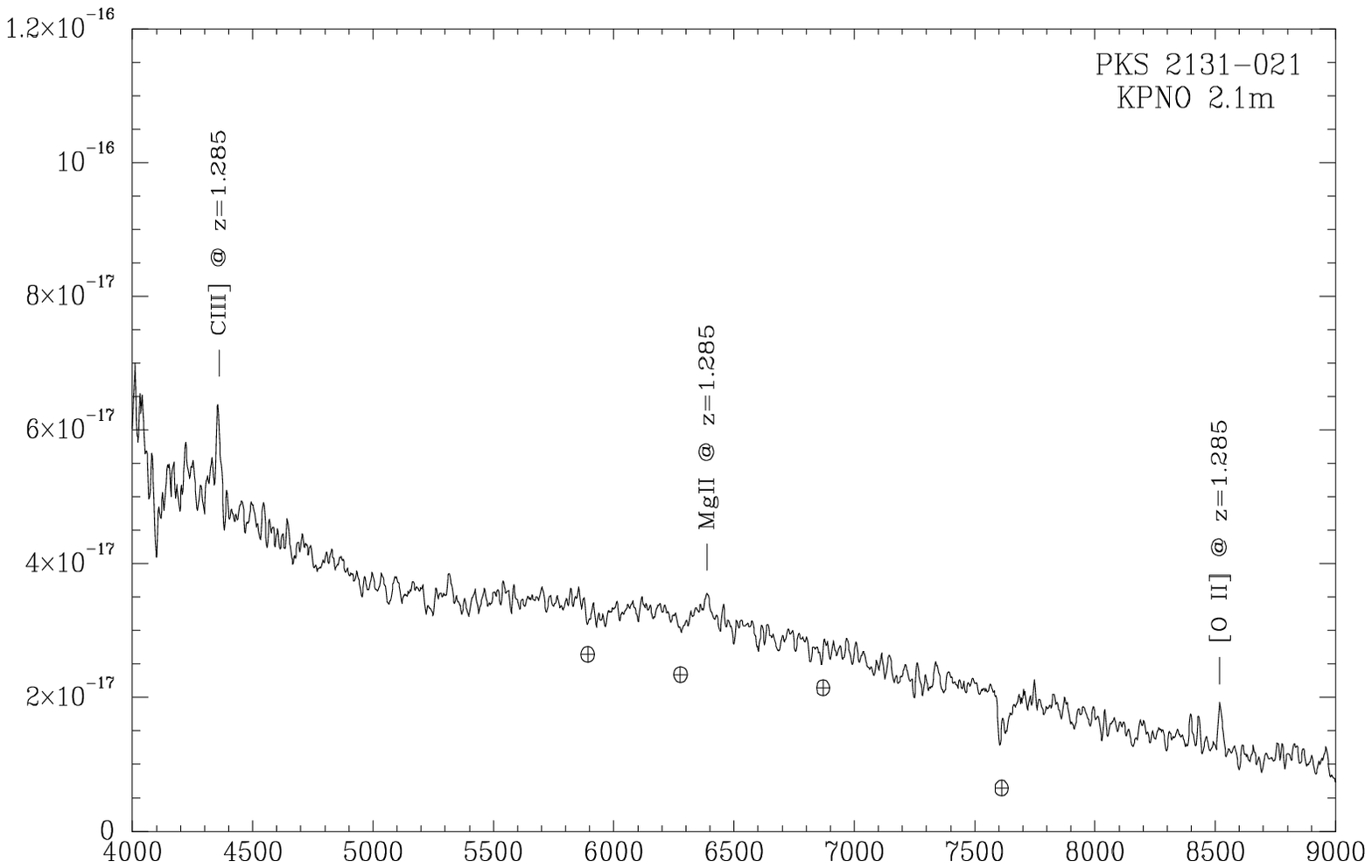}
\end{figure}
\begin{figure}
\plottwo{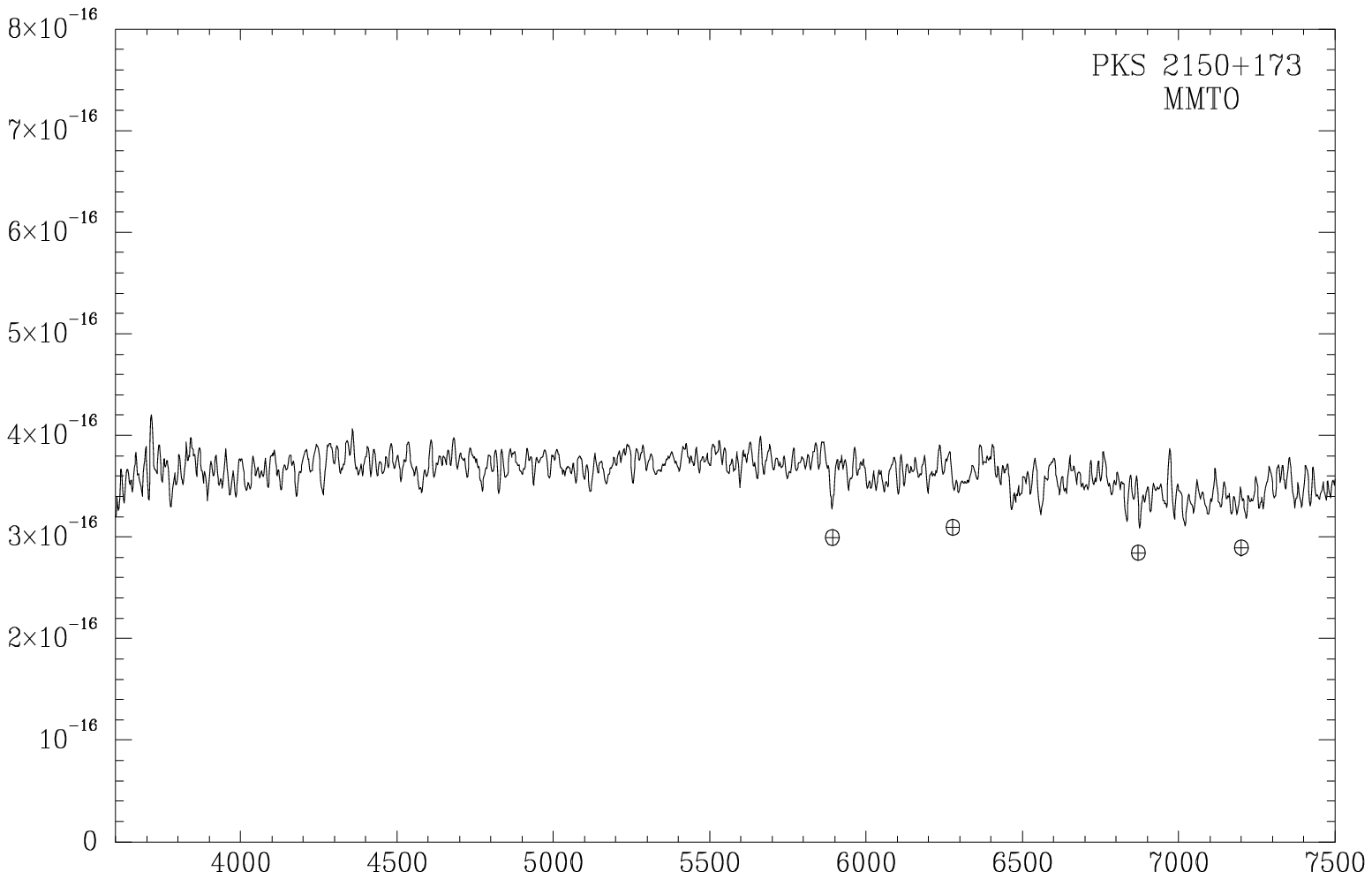}{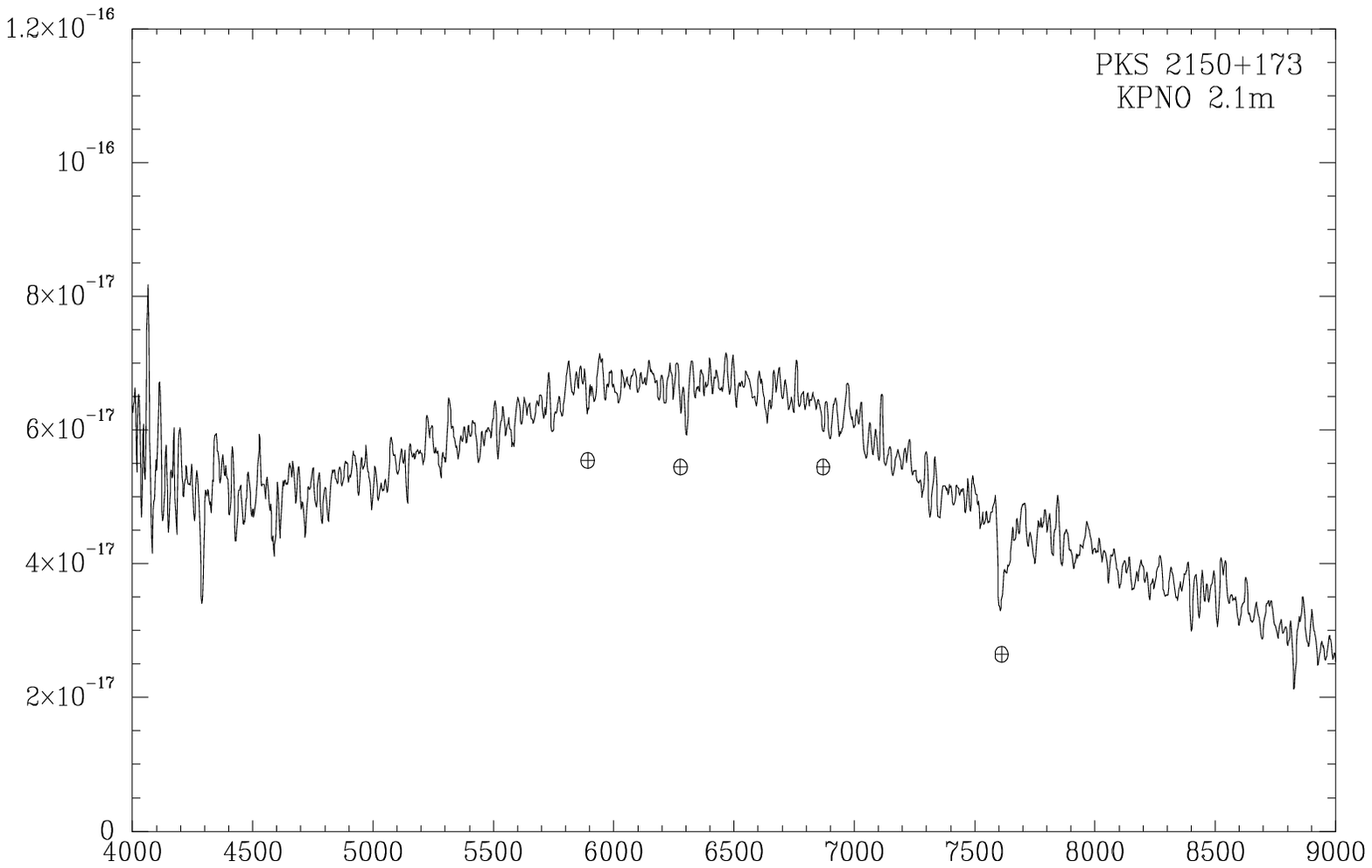}
\end{figure}

\clearpage
\begin{figure}
\plottwo{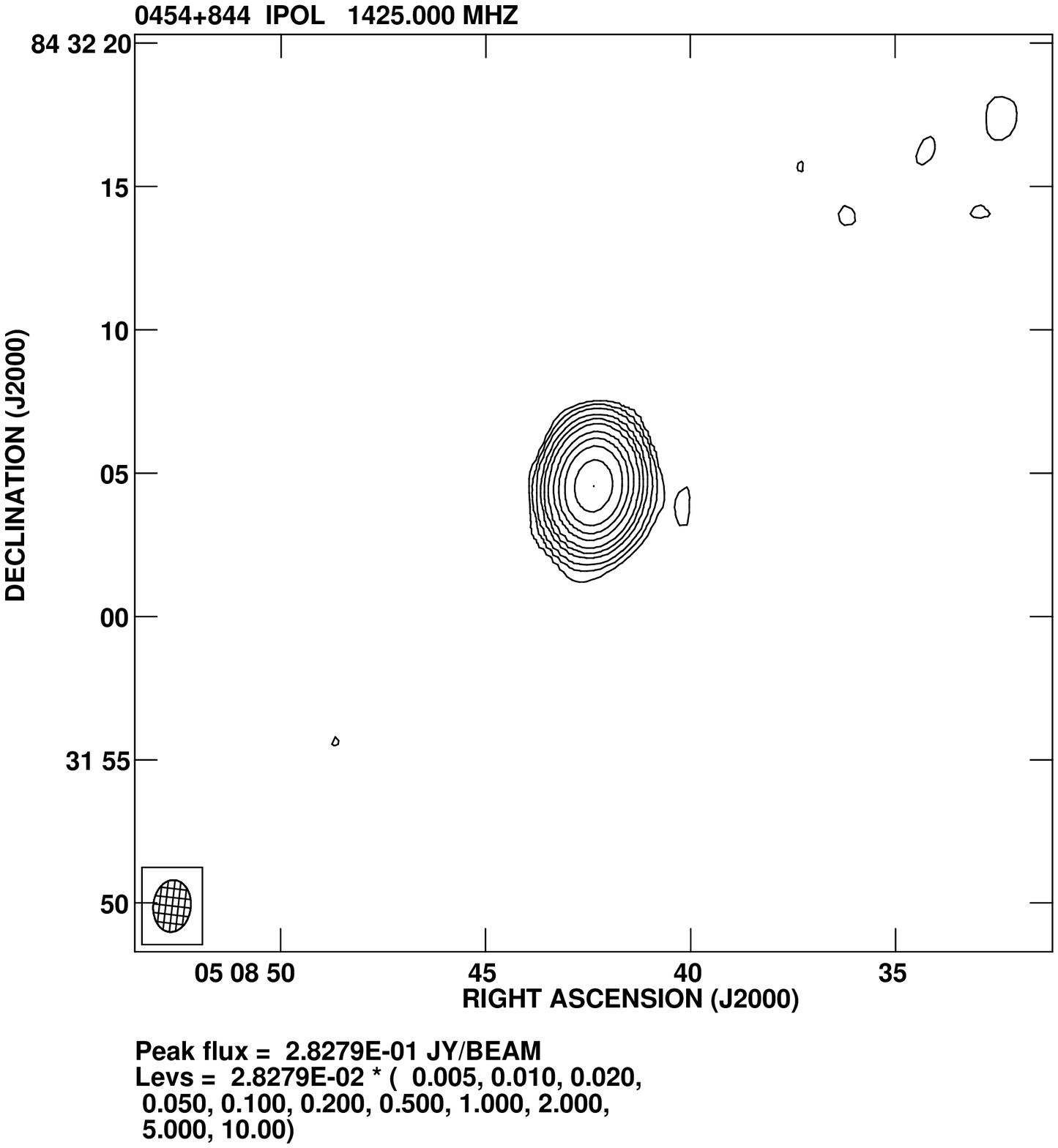}{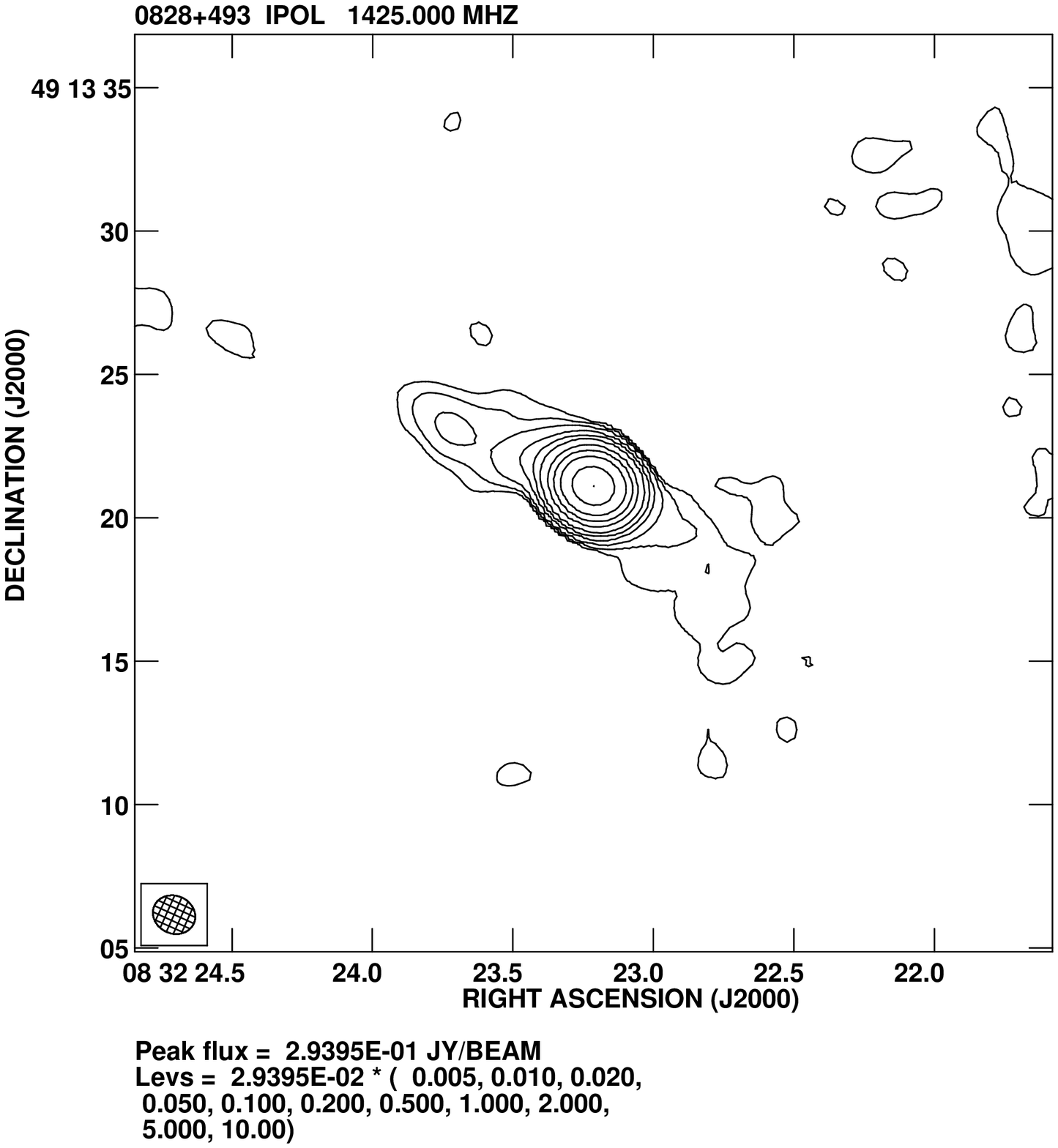}
\end{figure}
\begin{figure}
\plottwo{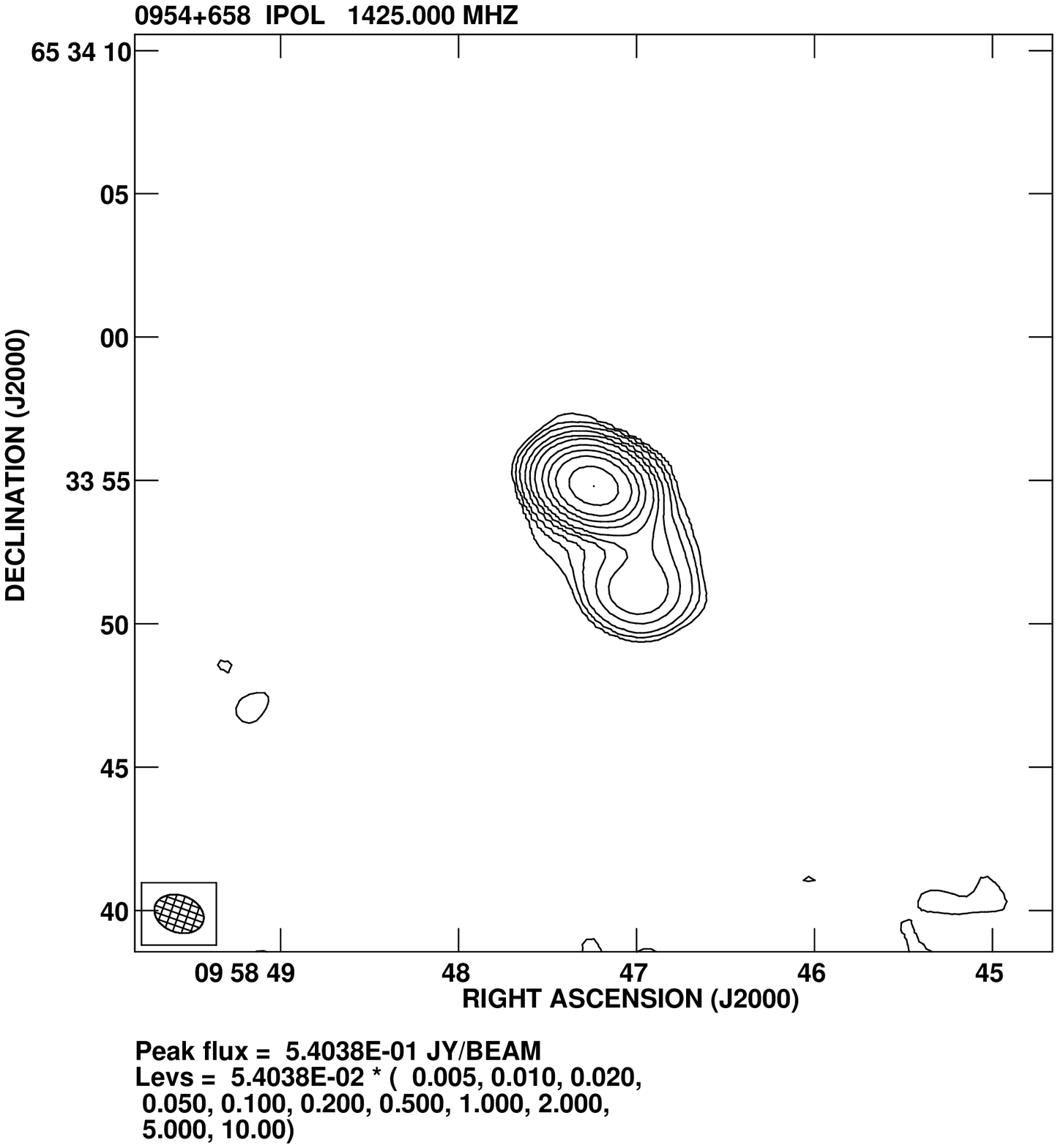}{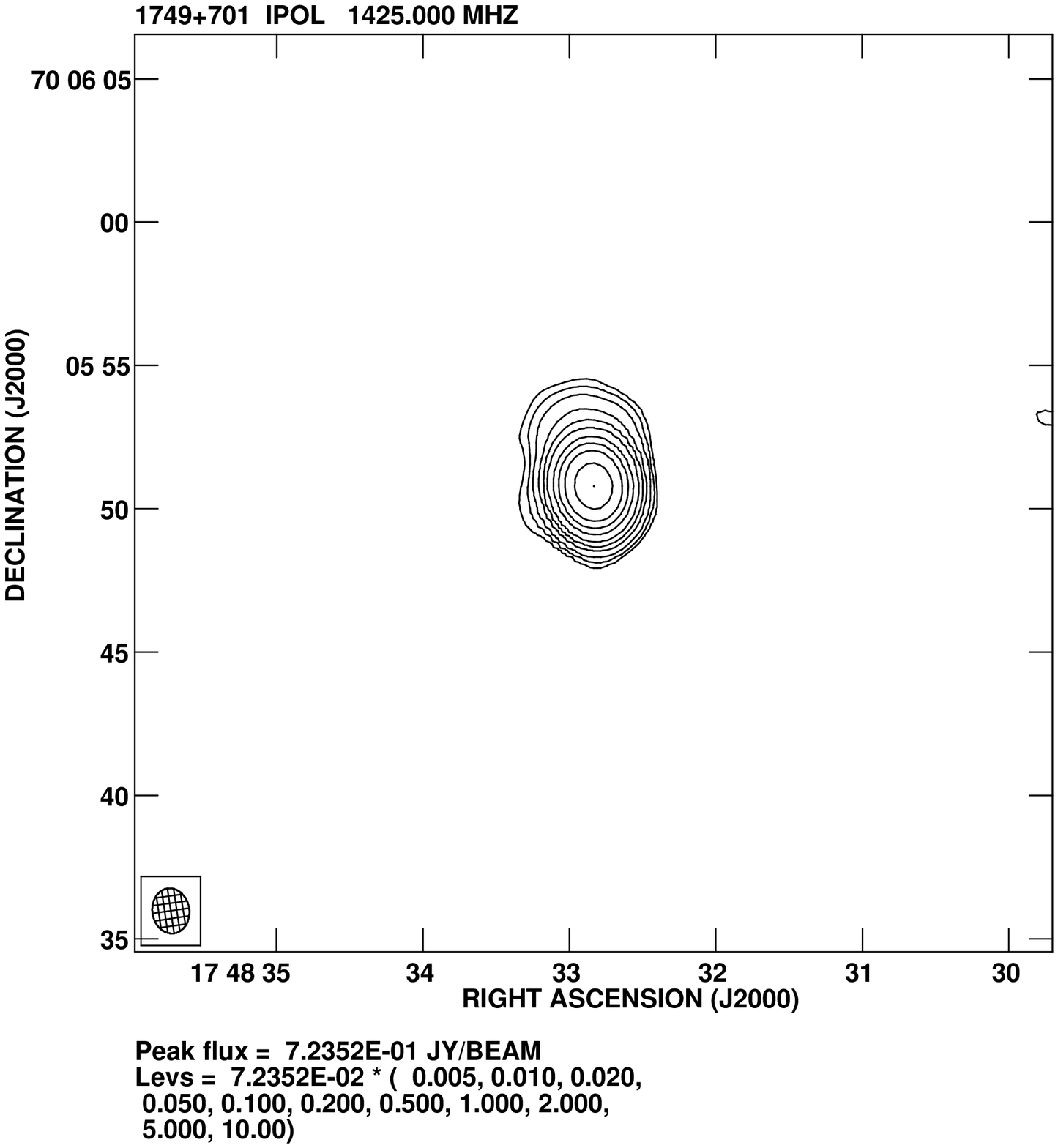}
\end{figure}

\begin{figure}
\plottwo{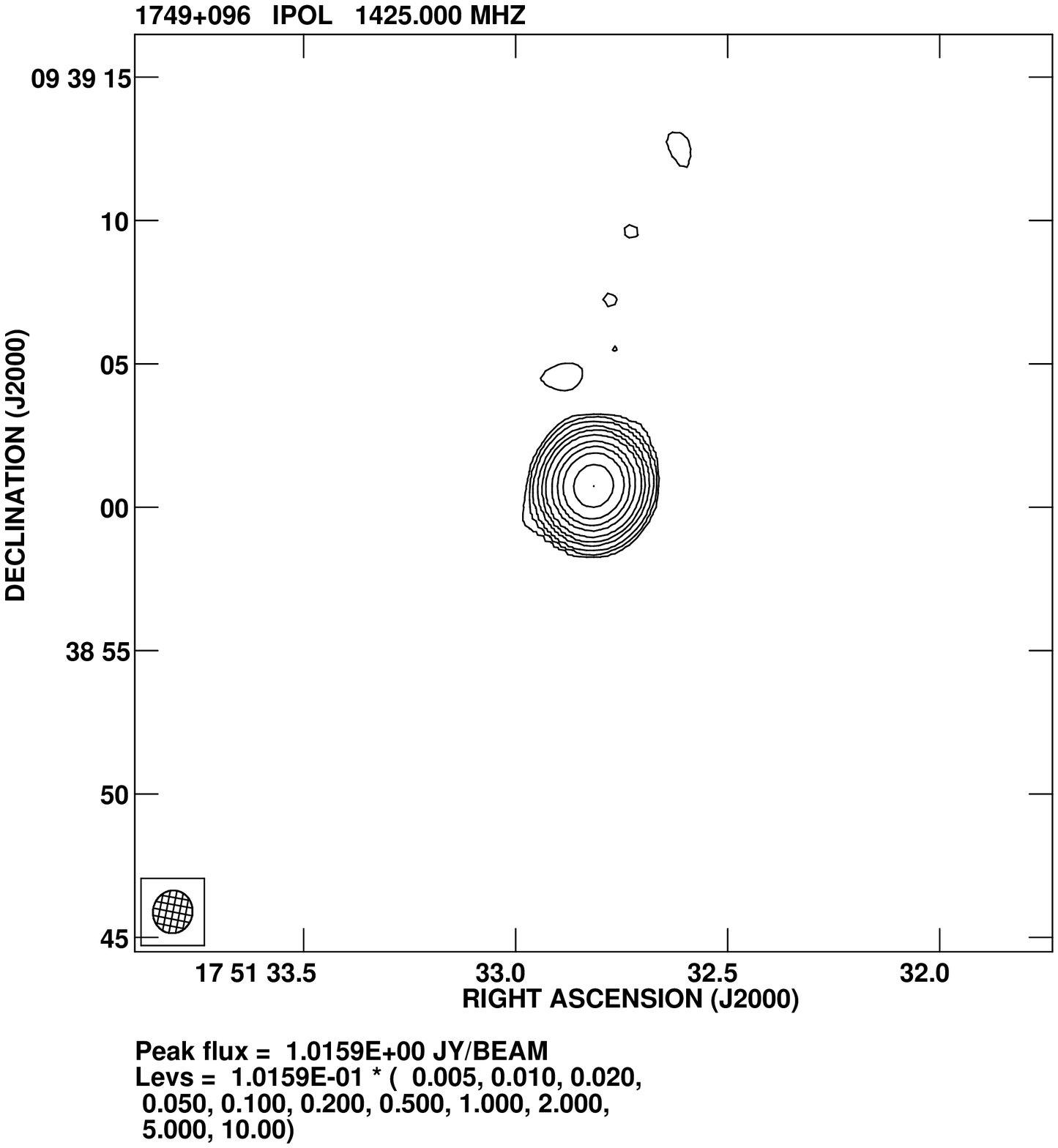}{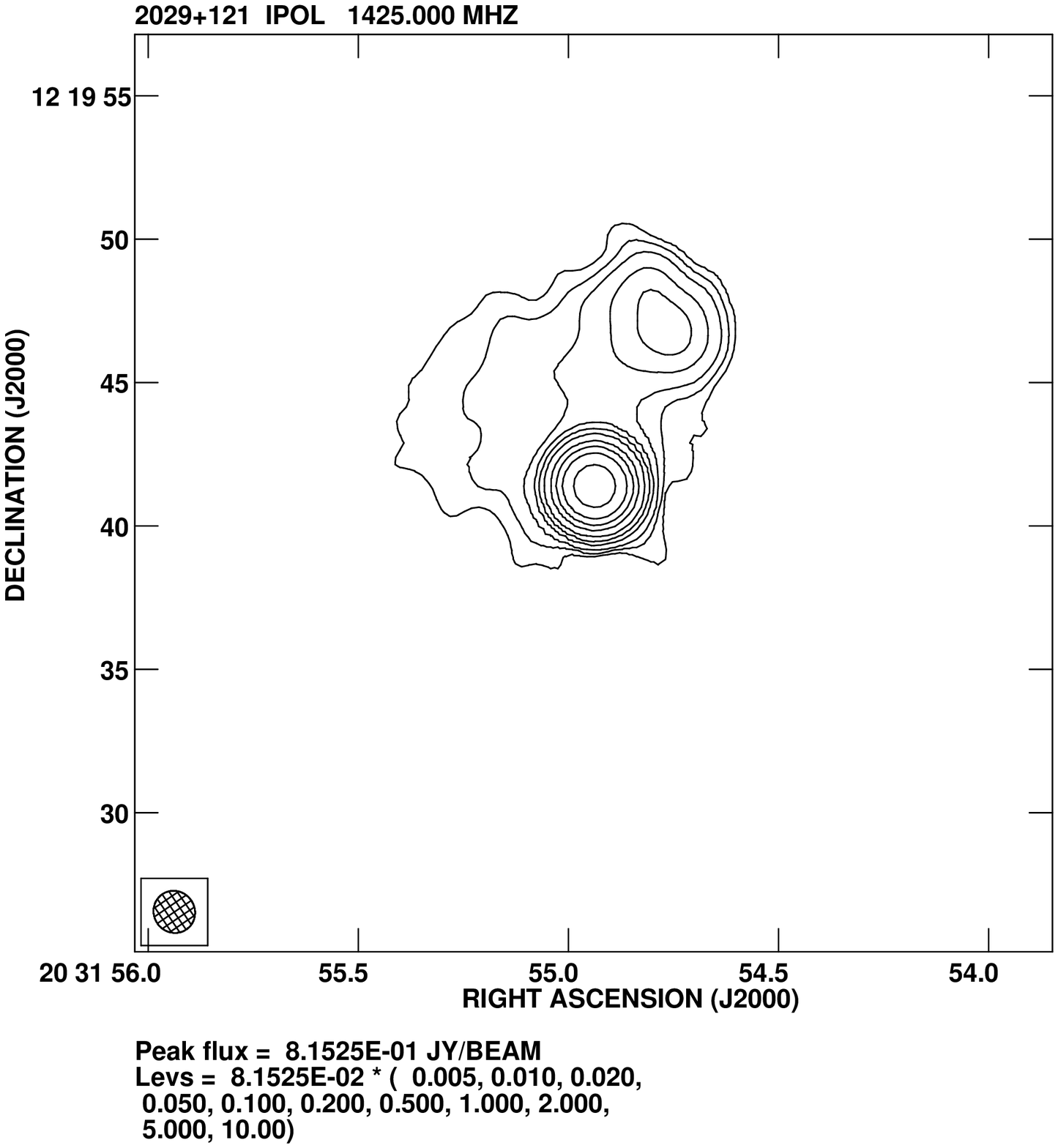}
\end{figure}
\begin{figure}
\plottwo{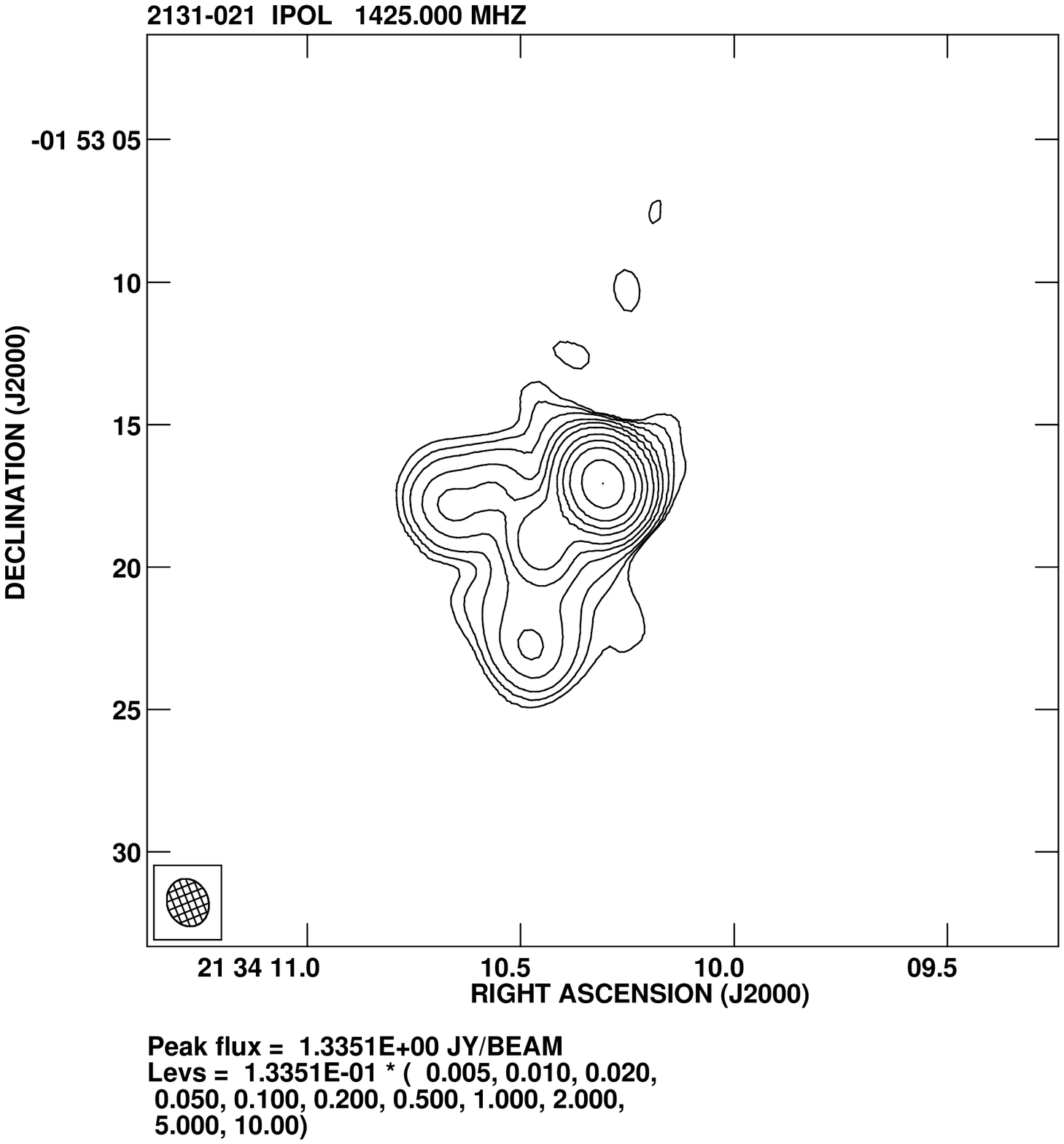}{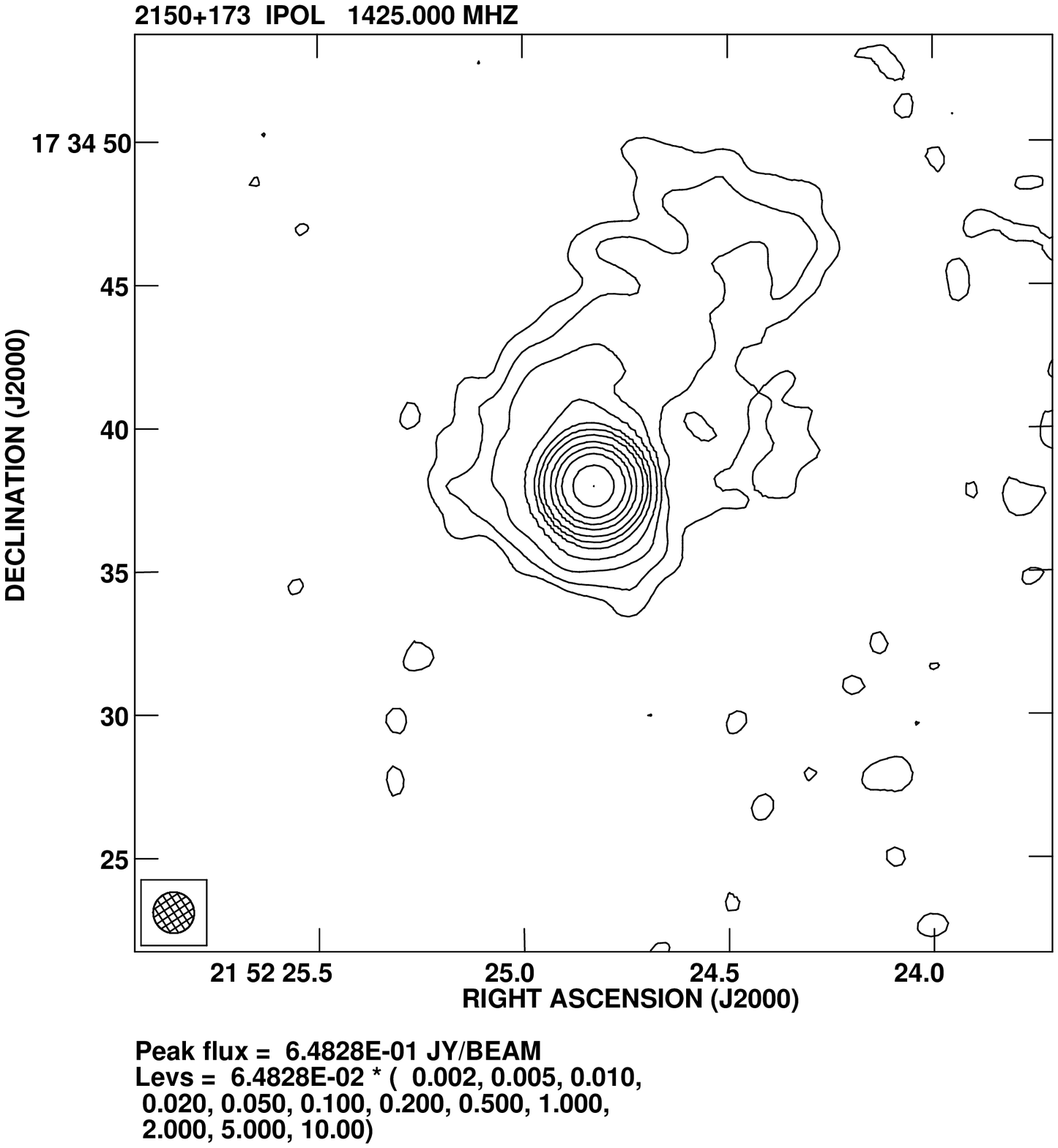}
\end{figure}
\clearpage

\clearpage

\begin{figure}
\plotone{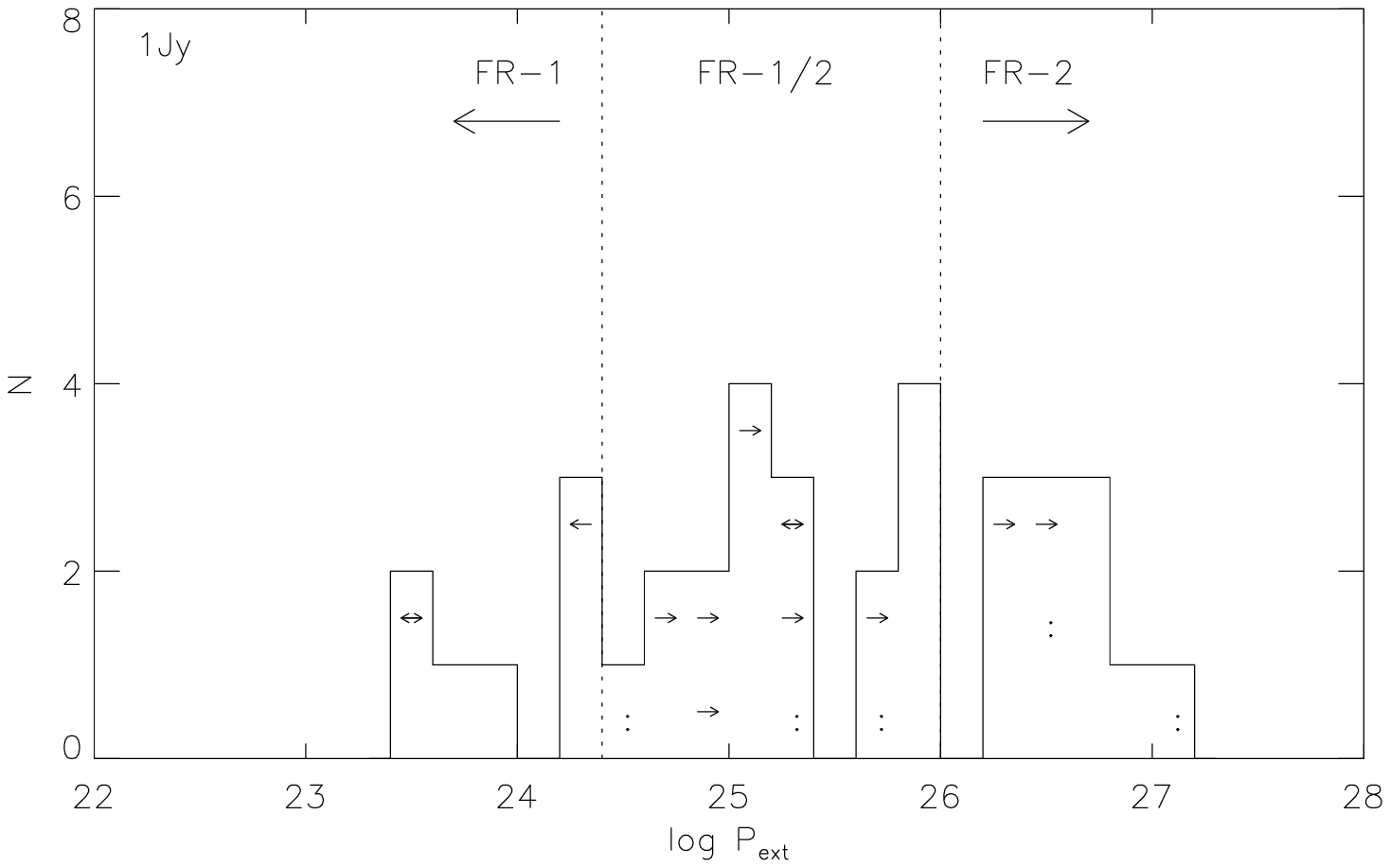}
\end{figure}
\clearpage

\begin{figure}
\plotone{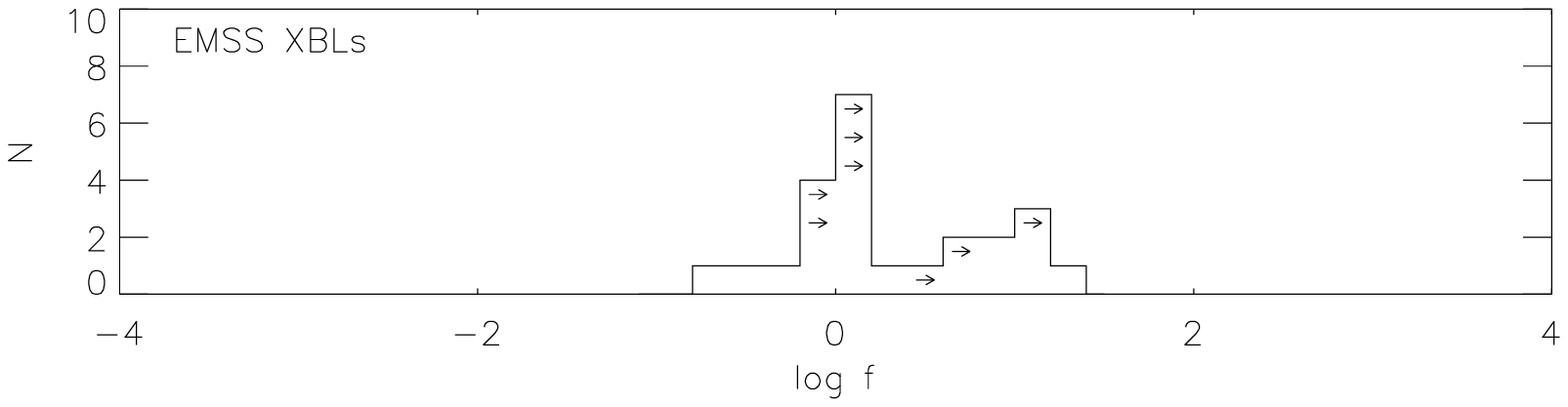}
\end{figure}
\begin{figure}
\plotone{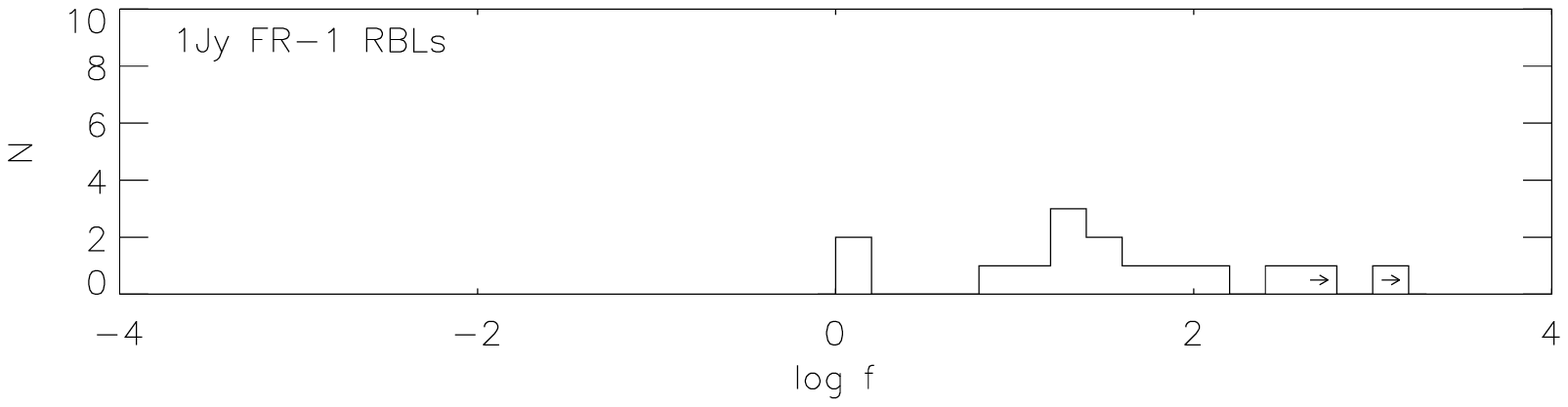}
\end{figure}
\clearpage

\begin{figure}
\plotone{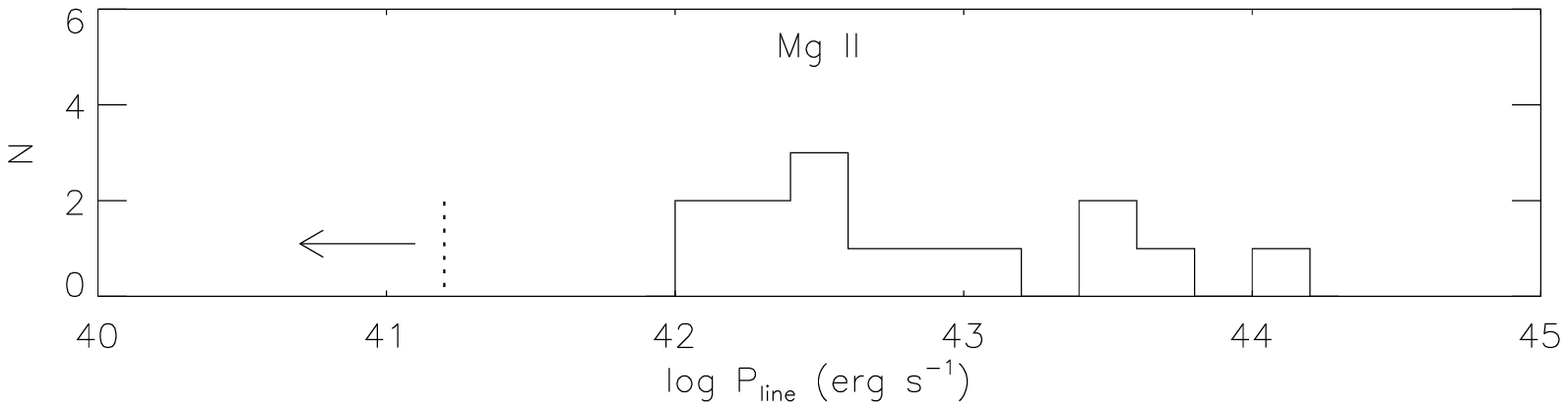}
\end{figure}
\begin{figure}
\plotone{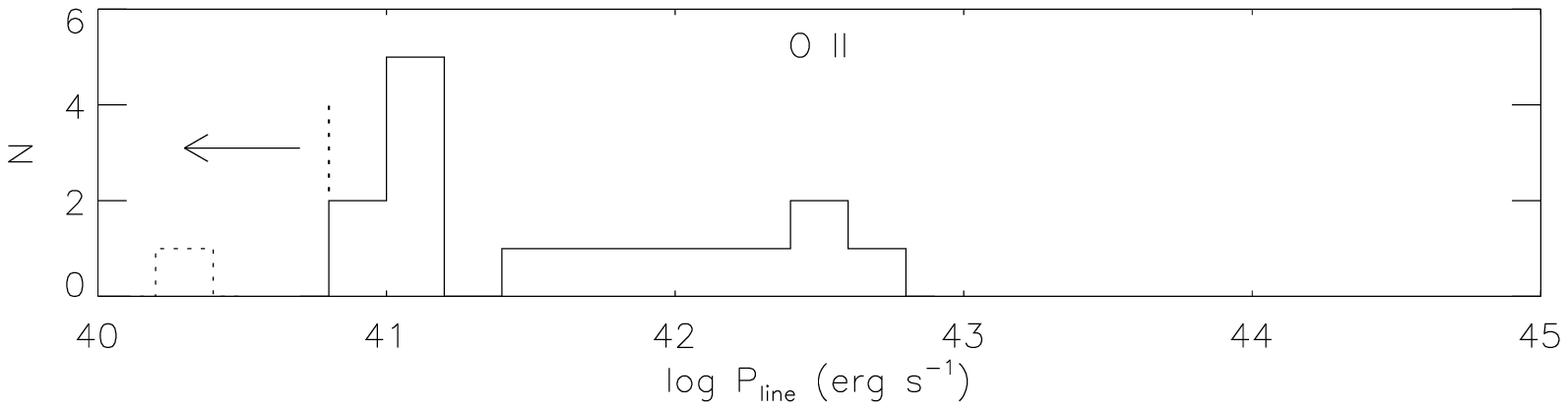}
\end{figure}
\begin{figure}
\plotone{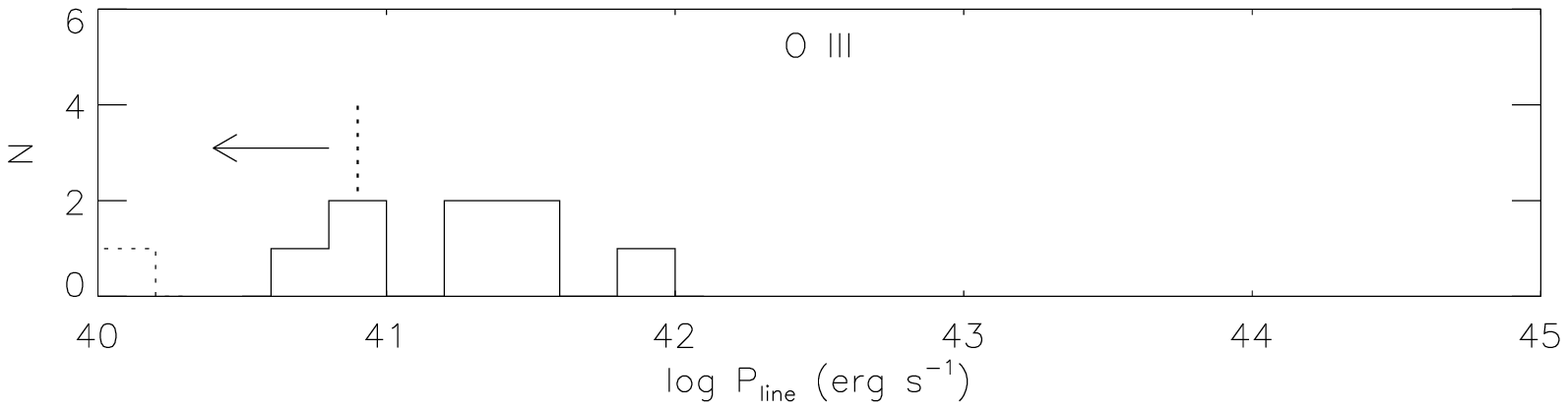}
\end{figure}

\clearpage



\makeatletter
\def\jnl@aj{AJ}
\ifx\revtex@jnl\jnl@aj\let\tablebreak=\\\fi
\makeatother


\begin{deluxetable}{lll}
\tablecaption{Log of Optical Observations \label{tbl-1}}
\tablewidth{0pt}
\tablehead{\colhead{} & \colhead{Dates of} & \colhead{SNR} \\
\colhead{Object} & \colhead{Observation} & \colhead{($1\sigma$)} }
\startdata
0048--097 &              18,20 Nov 95, 14 Aug 96, 23 Nov 00 & 200, 100, 100 \\
0138--097 &              19,21 Nov 95, 15 Aug 96 & 80       \\
0454+844 &   7 Apr 95,   18-21 Nov 95, 23 Nov 00 & 40, 20       \\
0716+714 &   5 Apr 95,      20 Nov 95, 22 Nov 00 & 250, 300, 300 \\
0735+178 & 4,6 Apr 95,      20 Nov 95, 23 Nov 00 & 200, 100      \\
0814+425 &   4 Apr 95,      21 Nov 95, 22 Nov 00 & 50, 40, 30   \\
0828+493 & 6,7 Apr 95,      19 Nov 95            & 20       \\
0954+658 & 5,7 Apr 95,      21 Nov 95            & 250      \\
1147+245 & 4,6 Apr 95                            & 200      \\
1519--273 & 4-8 Apr 95                            & 30       \\
1749+701 &   4 Apr 95,                 15 Aug 96 & 100, 100 \\
1803+784 &   5 Apr 95,                 15 Aug 96 & 200      \\
1823+568 & 6,7 Apr 95,                 15 Aug 96 & 20       \\
2029+121 &               18,19 Nov 95, 15 Aug 96 & 100      \\
2131--021 &               20,21 Nov 95            & 20       \\
2150+173 &               18,19 Nov 95, 15 Aug 96 & 20, 25   \\
\enddata
\end{deluxetable}


\begin{deluxetable}{lllcc}
\tablecaption{Log of Near-IR Observations \label{tbl-5}}
\tablewidth{0pt}
\tablehead{\colhead{} & \colhead{Dates of} & \colhead{IR} & \colhead{SNR} &
\colhead{Min $W_{\lambda}$} \\
\colhead{Object} & \colhead{Observation} & \colhead{Band(s)} &
\colhead{($1\sigma$)} &
\colhead{($3\sigma$)}}
\startdata
0454+844 & 26 Nov 96     & $I,J$     & 20 & 20 \\
0716+714 & 24, 25 Nov 96 & $I,J,H,K$ & 30 & 15 \\
         & 16 Feb 97     & $J,H,K$   & 30 & 15 \\
0735+178 & 24 Nov 96     & $I,J,H,K$ & 10 & 40 \\
0814+425 & 15 Feb 97     & $J,H,K$   & 10 & 50 \\
0851+202 & 25 Nov 96     & $I,J,K$   & 80 & 7  \\
1147+245 & 15, 16 Feb 97 & $J,H,K$   & 10 & 50 \\
1749+701 & 26 Nov 96     & $I$       & 30 & 13 \\
2200+420 & 24 Nov 96     & $J,H,K$   & 40 & 10 \\
2254+074 & 25 Nov 96     & $K$       & 15 & 50 \\
\enddata
\end{deluxetable}

\begin{deluxetable}{llll}
\tablecaption{Measured Redshifts of the 1 Jy BL Lac Sample \label{tbl-2}}
\tablewidth{0pt}
\tablehead{\colhead{}       & \colhead{}         & \colhead{}          & \colhead{}  \\
           \colhead{Object\tablenotemark{1}} & \colhead{$z_{em}$} & \colhead{$z_{abs}$} &
\colhead{References\tablenotemark{2}} }
\startdata
0048--097 & \nodata    & \nodata    &              \\
0118--272 & \nodata    & 0.559      & F91          \\
0138--097 & 0.733      & 0.501      & RS01, SR97   \\
0218+357 & 0.940:     & 0.686      & B93          \\
0235+164 & 0.940      & 0.852      & SFK93        \\ 
         &            & 0.524                     \\
0426-380 & \nodata    & 1.030      & SFK93        \\
0454+844 & \nodata    & 1.340      & SR97         \\
0537--441\tablenotemark{*} & 0.896      & \nodata    & SFK93        \\
0716+714 & \nodata    & \nodata    &              \\
0735+178 & \nodata    & 0.424      & C74          \\
0814+425 & \nodata    & \nodata    & RS01         \\
0820+225 & 0.951:     & \nodata    & SFK93        \\
0823+033 & 0.506:     & \nodata    & SFK93        \\
0828+493 & 0.548:     & \nodata    & RS01, SFK93  \\
0851+202\tablenotemark{*} & 0.306      & \nodata    & SFK93        \\
0954+658 & 0.367      & \nodata    & RS01, L96    \\
1144--379 & 1.048      & \nodata    & SFK93        \\
1147+245 & \nodata    & \nodata    &              \\
1308+326\tablenotemark{*} & 0.997      & \nodata    & SFK93        \\
1418+546 & 0.152      & \nodata    & SFK93        \\
1514--241 & 0.049      & \nodata    & M78          \\
1519--273 & \nodata    & \nodata    &              \\
1538+149 & 0.605      & \nodata    & SFK93        \\
1652+398 & 0.033      & \nodata    & SFK93        \\
1749+701 & 0.770      & \nodata    & L96          \\
1749+096\tablenotemark{*} & 0.320      & \nodata    & SFK93        \\
1803+784 & 0.684      & \nodata    & RS01, L96    \\
1807+698 & 0.051      & \nodata    & SFK93        \\
1823+568 & 0.664      & \nodata    & L96          \\
2007+777 & 0.342      & \nodata    & SFK93        \\
2005--409 & 0.072      & \nodata    & F87          \\
2029+121\tablenotemark{*} & 1.215      & 1.117      & SR97, SK93   \\
2131--021\tablenotemark{*} & 1.285      & \nodata    & RS01, D97    \\
2150+173 & \nodata    & \nodata    & RS01         \\
2200+420\tablenotemark{*} & 0.069      & \nodata    & SFK93        \\
2240--260 & 0.774      & \nodata    & SFK93        \\
2254+074 & 0.190      & \nodata    & SFK93        \\
\tablenotetext{1}{Objects marked with a star have exhibited emission line(s) with rest $W_{\lambda} \geq 5$\AA\ in one
or more epochs.}  
\tablenotetext{2}{References: This paper (RS01); 
Browne et al. 1993 (B93); 
Carswell et al. 1974 (C74); 
Drinkwater et al. 1997 (D97);
Falomo et al. 1987 (F87); 
Falomo 1991 (F91); 
Lawrence et al. 1996 (L96); 
Miller et al. 1978 (M78); 
Stickel, Fried \& K\"uhr 1993 (SFK93); 
Stocke \& Rector 1997 (SR97)}
\enddata
\end{deluxetable}


\begin{deluxetable}{lrlcccrrcc}
\tablecaption{Observed Spectral Line Properties\label{tbl-4}}
\tablewidth{0pt}
\tablehead{\colhead{} & \colhead{} & \colhead{Line} & \colhead{$\lambda_0$} & \colhead{$\lambda_{obs}$} &
\colhead{} & \colhead{FWHM} & \colhead{$W_{\lambda}$} & \colhead{$f_{line}$ (x $10^{-16}$)} & \colhead{$L_{line}$ (x
$10^{41}$)} \\
\colhead{Object} & \colhead{$\langle z \rangle$} & \colhead{ID} & \colhead{(\AA)} & \colhead{(\AA)} & \colhead{$z_{line}$} &
\colhead{(km s$^{-1}$)} & \colhead{(\AA)} & \colhead{(erg s$^{-1}$ cm$^{-2}$)} & \colhead{(erg s$^{-1}$)} }
\startdata
0048--097\tablenotemark{1} & \nodata  & \nodata & \nodata & 6092 & \nodata & 960 & -0.3 & 2.7 & $ \geq 0.6$\\
                                                                        \\
0138--097 & 0.733    & [O II]  & 3727 & 6460 & 0.733 & 492  & -0.9 & 1.7 & 7.4 \\
         &          & [Ne V]  & 3426 & 5936 & 0.733 & 561  & -0.6 & 1.1 & 4.8 \\
         &          & MgII    & 2798 & 4857 & 0.736 & 4842 & -2.2 & 4.8 & 20.9\\
         &          & CaII    & 3933 & 6824 & 0.735 & 1433 &  1.5 &     \\
         &          & CaII    & 3968 & 6883 & 0.734 & 1852 &  2.3 &     \\
         & 0.500    & MgII    & 2798 & 4197 & 0.500 & 1029 &  1.5 &     \\
                                                                        \\

0454+844 &	1.340	&	MgII	& 2796	&	6542	&	1.340	&	323	&	1.03	\\
&	& &	2803	&	6558	&	1.340	&	374	&	0.75	\\
 \\

0735+178 & 0.424	&	FeII	&	2344	&	3337	&	0.424	&	454	&	0.65	\\
&	&		&	2587	&	3683	&	0.424	&	376	&	0.37	\\
&	&		&	2383	&	3392	&	0.424	&	342	&	0.86	\\
&	&	 &	2600	&	3702	&	0.424	&	376	&	0.91	\\
&	&	MgII	&	2796	&	3982	&	0.424	&	364	&	1.98	\\
&	&	&	2803	&	3992	&	0.424	&	380	&	1.64	\\
&	&	MgI	&	2852	&	4063	&	0.424	&	370	&	0.31	\\

& 0.000	&	CaII	&	3934	&	3933	&	0.000	&	294	&	0.30	\\
&	&	&	3968	&	3968	&	0.000	&	283	&	0.24	\\
                                                                        \\
0828+493\tablenotemark{1} & 0.548:   & [O II]: & 3727 & 5770 & 0.548 & 1320 & -6.5 & 6.2 & 13.1: \\
                                                                        \\
0954+658 & 0.367    & [O II]  & 3727 & 5095 & 0.367 & 712 & -0.3 & 2.6 & 2.1 \\
         &          & CaII    & 3933 & 5387 & 0.370 & 1687 &  0.4 &     \\
         &          & CaII    & 3968 & 5439 & 0.370 & 2493 &  0.5 &     \\
                                                                        \\
1803+784 & 0.684    & MgII    & 2798 & 4710 & 0.684 & 3082 & -2.8 & 133 & 487  \\
                                                                        \\
1823+568 & 0.664    & C II]   & 2326 & 3882 & 0.669 & 3377 & -4.2 & 5.1 & 17.3 \\
         &          & MgII    & 2798 & 4668 & 0.668 & 3952 & -6.1 & 7.1 & 24.2 \\
         &          & [O II]  & 3727 & 6204 & 0.664 & 652  & -1.7 & 1.8 &  6.1 \\
                                                                        \\
2029+121 & 1.215    & C IV    & 1549 & 3430 & 1.214 & 2921 & -34.8 & 8.6 & 143   \\
									&          & C III]  & 1909 & 4216 & 1.208 & 1216 & -3.6  & 1.2 & 19.7  \\
									&          & MgII    & 2798 & 6188 & 1.212 & 7757 & -9.2  & 9.2 & 152   \\
									&          & O III   & 3133 & 6946 & 1.217 & 808  & -5.0  & 1.4 & 24.6  \\
									&          \\
2131--021 & 1.285    & C III]  & 1909 & 4356 & 1.281 & 1818 & -11.1 & 4.5 & 87.2 \\
         &          & MgII    & 2798 & 6378 & 1.280 & 3602 & -9.3  & 2.9 & 56.2 \\
         &          & [O II]  & 3727 & 8521 & 1.286 & 690  & -11.8 & 1.4 & 27.1 \\
\tablenotetext{1}{See discussion of this source in \S 3.}
\enddata
\end{deluxetable}


\begin{deluxetable}{lrrrrrrlllc}
\tablecaption{20cm Arcsecond-Scale Radio Properties of the 1 Jy BL Lac Sample \label{tbl-3}}
\tablewidth{0pt}
\tablehead{\colhead{} & \colhead{$S_{core}$} & \colhead{$S_{ext}$} & \colhead{log $P_{core}$} &
\colhead{log $P_{ext}$} & \colhead{} & \colhead{LLS} & \multicolumn{3}{c}{FR Class\tablenotemark{1}} & \colhead{} \\
\colhead{Object}& \colhead{(mJy)} & \colhead{(mJy)} & \colhead{(W Hz$^{-1}$)} & \colhead{(W Hz$^{-1}$)}
& \colhead{$f$} & \colhead{(kpc)} & \colhead{(R)} & \colhead{(O)} & \colhead{(F)} & \colhead{References\tablenotemark{1}}}
\startdata
0048--097 & 537    & 139.0 & $> 25.95$ & $> 25.32$ & 3.9  & $> 72.0$     & 2  & 1: & 2: & RS01     \\
0118--272 & 742    & 168 & $\geq 26.97$ & $\geq 26.23$ & 4.4 & $\geq 133$ & 2 & 2 & 2 & C99    \\
0138--097 & 457    & 41.5  & 26.99     & 25.83     & 11.0 & 205.5        & 2  & 2 & 2 & RS01     \\
0218+357\tablenotemark{2} & 800    & 120   & 27.44:    & 26.48:    &  6.7 & 253.7:       & 2: & 2: & 2: & O92     \\
0235+164 & 972    & 31.4  & 27.53     & 25.98     & 31.0 & 129.3        & 2  & 2 & 2 & M93     \\
0426--380 & 624 & 86 & $\geq 27.42$ & $\geq 26.41$ &7.3 & $\geq 51$ & 2 & 2 & 2 & C99    \\
0454+844 & 283    & $< 5.1$ & $\geq 27.31$ & (25.37) & $>55.5$  & (34.6) & 2: &  & 2: & RS01     \\
0537--441 & 3010 & 220 & 27.98 & 26.71 & 13.7 & 109 & 2 & 2 & 2 & C99 \\
0716+714 & 315    & 316   & $> 25.72$ & $> 25.68$ &  1.0 & $> 76.2$     & 1: &    & 1: & A86a    \\
0735+178 & 2773   & 12    & $\geq 27.30$ & $\geq 24.86$ &  231 & $\geq 29.7$     & 2: &    & 2: & PS94    \\
0814+425 & 1571   & 65.6  & $> 26.42$ & $> 25.00$ & 23.9 & $> 116.5$    & 2: &    & 2: & M93     \\
0820+225 & 1606   & 602.6 & 27.76:    & 27.19:    & 2.67 & 109.6:        & 2  & 2  & 2  & M93     \\
0823+033 & 1325   & 4.1:   & 27.13:    & 24.53:    & 323  & 151.7:       & 1: & 2  & 2  & M93     \\
0828+493 & 294    & 23.4  & 26.55:    & 25.35:    & 12.6 & 113.8:       & 1: & 1: & 1: & RS01     \\
0851+202 & 1719   & 17    & 26.82     & 24.76     &  101 & 175.6        & 1  & 2  & 1: & PS94    \\
0954+658 & 540    & 28.6  & 26.47     & 25.13     & 18.9 &  38.5        & 1  & 1  & 1  & RS01     \\
1144--379 & 1892  & 20:   & 27.92     & 25.79:    & 94.6: & 8:          & 2: & 2  & 2  & C99 \\
1147+245\tablenotemark{3} & 796    & 50    & $> 26.12$ & $> 24.88$ & 15.9 & $> 144$ & 2: &    & 2: & C99    \\
1308+326 & 1577   & 67.9  & 27.79     & 26.28     & 23.2 & 211.7        & 2  & 2  & 2 & M93     \\
1418+546 & 1189   & 22.0  & 26.06     & 24.30     &   54 & 122.7        & 1  & 1  & 1 & M93     \\
1514--241\tablenotemark{3} & 1652   & 210    & 25.22     & 24.32     & 7.7 &  354        & 1  & 1  & 1 & C99    \\
1519--273 & 1690 & $< 3.0$ & $> 26.45$ & (23.56) & $> 560$ & (4.0) & 1: & 1  & 1 &  C99 \\
1538+149 & 1330   & 202.0 & 27.29     & 26.37     & 6.6 &  29.6        & 2  & 2  & 2 & M93     \\
1652+398 & 1376   & 67    & 24.80     & 23.49     & 20.5 &  26.7        & 1  & 1  & 1 & UJW83   \\
1749+701 & 720    & 45.7  & 27.23     & 25.91     & 15.8 &  34.2        & 2  & 1  & 1: & RS01     \\
1749+096 & 1016   & $< 1.0$  & 26.59 & $< 24.40$ & $> 145$ & $< 13.8$ & 1: & 1  & 1 & RS01 \\ 
1803+784 & 1557   & 68.0  & 27.46     & 25.99     & 22.9 & 445.0        & 2  & 2  & 2  & M93      \\
1807+698 & 1350   & 990   & 25.18     & 25.03     & 1.36 &  92.7        & 1  & 1  & 1 & AU85,WL90    \\
1823+568 & 858    & 452   & 27.18     & 26.78     & 1.90 & 195.0        & 2  & 2  & 2 & M93     \\
2005--489 & \nodata & \nodata & \nodata & \nodata & \nodata & \nodata   &    & 1  & 1 & \nodata \\
2007+777 & 823    & 28.9  & 26.59     & 25.07     & 28.5 & 217.2        & 1  & 1  & 1 & M93     \\
2029+121 & 815    & 84.2  & 27.68     & 26.52     &  9.7 & 100.1        & 2  & 2  & 2 & RS01     \\ 
2131--021 & 1335   & 194.0 & 27.95     & 26.93     &  6.9 & 181.4        & 2  & 2  & 2 & RS01     \\ 
2150+173 & 648    & 36.7  & $> 26.03$ & $> 24.74$ & 17.7 & $> 21.4$  & 1: &    & 1: & RS01     \\
2200+420\tablenotemark{4}  & 3310   & 40    & 25.83     & 23.90     & 82.8 &  86.0        & 1 & 1 & 1 & A86b    \\
2240--260 & 813 & 333 & 27.29 & 26.77 & 2.4 & 212 & 2 & 2: & 2 & C99 \\
2254+074 & 454    & 17    & 25.83     & 24.37     & 26.7 &  78.7        & 1  & 1  & 1 & AU85    \\
\tablenotetext{1}{References: 
This paper (RS01); 
Antonucci \& Ulvestad 1985 (AU85);
Antonucci et al. 1986 (A86a);
Antonucci 1986 (A86b);
Cassaro et al. 1999 (C99); 
Murphy et al. 1993 (M93); 
O'Dea et al. 1992 (O92); 
Perlman \& Stocke 1994 (PS94);
Ulvestad, Johnston, \& Weiler 1983 (UJW83);
Wrobel \& Lind 1990 (WL90)}
\tablenotetext{2}{A+C array}
\tablenotetext{3}{A+B array}
\tablenotetext{4}{B array}
\enddata
\end{deluxetable}



\makeatletter
\def\jnl@aj{AJ}
\ifx\revtex@jnl\jnl@aj\let\tablebreak=\nl\fi
\makeatother


\def\vvmax{$\langle V/V_{max} \rangle$}
\def\veva{$\langle V_e/V_a \rangle$}
\def\avgz{$\langle z \rangle$}
\def\fxfr{$f_x/f_r$}

\begin{deluxetable}{llrr}
\tablecaption{\vvmax\ for BL Lac Subsamples \label{tbl-7}}
\tablewidth{0pt}
\tablehead{
\colhead{Survey} & \colhead{Subsample} & \colhead{$N$} &
\colhead{\vvmax} }
\startdata

1Jy  & $-8.00 <$ \fxfr\ $ \leq -7.25$ & 10 & $0.655\pm0.091$  \\
1Jy  & $-7.25 <$ \fxfr\ $ \leq -6.75$ & 13 & $0.597\pm0.080$  \\
1Jy  & $-6.75 <$ \fxfr\ $ \leq -5.00$ & 11 & $0.497\pm0.087$  \\
\\
EMSS & $-6.00 <$ \fxfr\ $ \leq -5.00$ & 13 & $0.550\pm0.080$  \\
EMSS & $-5.00 <$ \fxfr\ $ \leq -4.50$ & 14 & $0.470\pm0.077$  \\
EMSS & $-4.50 <$ \fxfr\ $ \leq -3.90$ & 14 & $0.271\pm0.077$  \\

\enddata
\end{deluxetable}

\end{document}